\documentclass[%
superscriptaddress,
preprint,
showpacs,preprintnumbers,
nofootinbib,
amsmath,amssymb,
]{revtex4-2}

\usepackage{graphicx}
\usepackage{dcolumn}
\usepackage{bm}
\usepackage{epstopdf}
\usepackage{color}

\begin{document}
	
	
\title{Multiple-relaxation-time discrete Boltzmann modeling of multicomponent mixture with nonequilibrium effects}

\author{Chuandong Lin}
\affiliation{Sino-French Institute of Nuclear Engineering and Technology, Sun Yat-Sen University, Zhuhai 519082, China}
\affiliation{Key Laboratory for Thermal Science and Power Engineering of Ministry of Education, Department of Energy and Power Engineering, Tsinghua University, Beijing 100084, China.}

\author{Kai H. Luo}
\email{K.Luo@ucl.ac.uk}
\affiliation{Department of Mechanical Engineering, University College London, Torrington Place, London WC1E 7JE, United Kingdom}

\author{Aiguo Xu}
\affiliation{Laboratory of Computational Physics, Institute of Applied Physics and Computational Mathematics, P. O. Box 8009-26, Beijing 100088, China}
\affiliation{State Key Laboratory of Explosion Science and Technology, Beijing Institute of Technology, Beijing 100081, China}
\affiliation{Center for Applied Physics and Technology, MOE Key Center for High Energy Density Physics Simulations, College of Engineering, Peking University, Beijing 100871, China}

\author{Yanbiao Gan}
\affiliation{North China Institute of Aerospace Engineering, Langfang 065000, China}

\author{Huilin Lai}
\email{hllai@fjnu.edu.cn}
\affiliation{College of Mathematics and Informatics $\&$ FJKLMAA, Fujian Normal University, Fuzhou 350007, China}

\date{\today}

\begin{abstract}
A multiple-relaxation-time discrete Boltzmann model (DBM) is proposed for multicomponent mixtures, where compressible, hydrodynamic, and thermodynamic nonequilibrium effects are taken into account. It allows the specific heat ratio and the Prandtl number to be adjustable, and is suitable for both low and high speed fluid flows. From the physical side, besides being consistent with the multicomponent Navier-Stokes equations, Fick's law and Stefan-Maxwell diffusion equation in the hydrodynamic limit, the DBM provides more kinetic information about the nonequilibrium effects. The physical capability of DBM to describe the nonequilibrium flows, beyond the Navier-Stokes representation, enables the study of the entropy production mechanism in complex flows, especially in multicomponent mixtures. Moreover, the current kinetic model is employed to investigate nonequilibrium behaviors of the compressible Kelvin-Helmholtz instability (KHI). It is found that, in the dynamic KHI process, the mixing degree and fluid flow are similar for cases with various thermal conductivity and initial temperature configurations. Physically, both heat conduction and temperature exert slight influences on the formation and evolution of the KHI. 
\end{abstract}


\maketitle


\section{Introduction}

Numerical simulations of multicomponent mixtures with essential nonequilibrium characteristics are of great importance in many fields of science and engineering \cite{Cussler2000,Law2006,Bertevas2019POF,Zhao2019SR,Yang2020CF}, as these practical systems are often too complex to be studied by experiment or theory in a simple and intuitive way. A typical case is the spacecraft reentry into the atmosphere under the condition of low air density and high flight speed \cite{Yang1995JCP,Peng2016JCP,Celiberto2016PSST}. Other typical examples include the porous media bio-filtration device, micro electro-mechanical system, microfluidic device, geological storage of nuclear wastes, carbon dioxide sequestration, combustion chamber, and rotating detonation propulsion engine. 
In fact, the thermodynamic nonequilibrium effect (TNE) accompanied by the hydrodynamic nonequilibrium effect (HNE) takes place due to remarkable regional variations, such as around the shock front, rarefaction wave, and material interface \cite{Lin2014PRE}. 	For such phenomena where both TNE and HNE have a significant role due to the small characteristic length and/or sharp physical gradient (noticeable differences exist between the distribution functions and their equilibrium counterparts, the equipartition of energy between different degrees of freedom breaks down), the traditional continuum description may be inadequate \cite{Yang1995JCP,Ivanov1998ARFM,Rapaport2004Book}, and the need for a microscopic or mesoscopic description arises \cite{Rapaport2004Book,Rykov1975FD}. 

To resolve the above issue, the computational kinetic theory is sought as a promising approach. As a central equation in the kinetic theory, the Boltzmann equation has the capability to describe complex fluid flows with both HNE and TNE. However, in practice, it is usually too complicated to be employed for simulations in a straightforward way due to the quadratic nonlinearity of the collision integral dependence of the integrand function on postcollision velocities and high multiplicity of integration \cite{Ivanov1998ARFM}. An alternative approach to the use of the Boltzmann equation is the molecular dynamics (MD) that provides accurate results, but is incapable of simulating large-scale systems owing to prohibitive increasing of the computational cost \cite{Rapaport2004Book,Liu2016FP,Liu2017PRE,Murugesan2019CNF}. As another alternative method, the direct simulation Monte Carlo (DSMC) method is widely used for modeling nonequilibrium systems, including multicomponent flows and chemically reacting flows \cite{Ivanov1998ARFM,Sebastiao2018CNF,White2018CPC,Gimelshein2019PRF}, but the probabilistic nature of DSMC leads to noisy solutions \cite{Mieussens2000JCP}. To overcome these difficulties, various kinetic models based on a simplified Boltzmann equation were proposed \cite{SucciBook,Wu2013JCP,Liu2017CCP,Zhang2018PRE,Xu2012FP,Xu2018Book}.
Such kinetic models have existed for a very long time, starting with the famous Bhatnagar-Gross-Krook (BGK) equation in 1954 \cite{BGK1954}. Later, an ellipsoidal statistical (ES) model \cite{Holway1966POF} and a Shakhov model \cite{Shakhov1968FD} were proposed. An overview of their properties can be found in the reference \cite{Struchtrup2005Book}.
Moreover, there were also kinetic models for the gas mixtures \cite{Andries2002JSP,Groppi2004POF} and gas flows consisting of molecules with internal degrees of freedom \cite{Rykov1975FD}. 
Numerous effective numerical methods were proposed for solving these equations \cite{Yang1995JCP,Mieussens2000JCP,Titarev2007CF,Morinishi2006CF,Kudryavtsev2013JSC}, and a large number of essentially nonequilibrium problems where the continuum description is inadequate or adequate have been solved. 

To further utilize the simplified Boltzmann equations, a straightforward method is to discretize the time, space, as well as the particle velocity. In fact, the idea of using a finite set of discrete speeds appeared early in the seminal work \cite{Broadwell1964POF}, where the discrete velocity model was constructed for the Boltzmann equation, and the collision integral was expressed as a nonlinear quadratic term. In recent three decades, the lattice Boltzmann method (LBM), developed from the lattice gas method and originally based on the discrete simplified Boltzmann equations \cite{SucciBook,GuoBook2013,Qian1992EL,Zhang2011PRE,Zhang2011MN,Zhang2013,Qin2015}, has been successfully used as an alternative tool of various partial differential equations for complex systems with multi-phase \cite{Fakhari2013PRE,Liang2016PRE,Qin2018POF,Chen2018PRE,Fei2019POF,Wang2019POF}, multi-component \cite{Makhija2012CAF,Chai2012AMS,Liu2016JCP}, mass diffusion \cite{Chai2019PRE,Hosseini2019IJHMT}, external force \cite{Fei2018IJHMT}, and/or chemical reactions \cite{Chen2015IJHMT,Feng2018CNF}, etc. 
In the evolution of the discrete Boltzmann equation, the particle velocity space is dicretized besides the discretization in physical space. The physical variables are calculated from the discrete distribution functions whose evolution is obtained with proper numerical methods. 
To be specific, there are two stages (i.e., ``collision" + ``propagation") in the procedure of the standard LBM. In the stage of collision, the lattice distribution functions evolve under the control of the artificial relaxation time. In the phase of propagation, the artifical particle population transfers from one node of the square grid in physical space to exactly one of the neighboring nodes. To meet the requirement that the time step, space step and discrete velocities are coupled, the discrete speeds should be chosen in a particular way (such as D1Q5, D2Q9, D3Q27, etc.), which is one of the characteristic features of standard LBMs \cite{SucciBook}.
Although the LBM has achieved great success in replacing traditional continuum governing equations, few of the lattice Boltzmann models go beyond the continuum equations to provide various significant thermodynamic nonequilibrium information. 

To address this problem, one possible method is to modify the discrete Boltzmann equation by introducing an artificial discrete equilibrium distribution function that satisfies higher order kinetic moments \cite{Kang2014PRE}. However, the artifical term becomes particularly complicated with increasing kinetic moments required \cite{Kang2014PRE}. In fact, a more direct way is to invoke a novel methodology, the discrete Boltzmann method (DBM), which is regarded as a modern variant of the standard LBM \cite{Gan2015SM,Zhang2019SM,Lin2015PRE,Lin2016CNF,Zhang2016CNF,Lin2017SR,Lin2018CNF,Lin2014PRE,Lai2016PRE,Lin2017PRE,Chen2018POF,Lin2018CAF,Lin2019PRE}. 
The DBM is based on the discrete Boltzmann equation which can be solved with various numerical approaches. For time discretization, the implicit, explicit or implicit-explicit scheme \cite{Wang2007IJMPC} can be employed. For space discretization, the frequently used schemes include the finite difference, the finite volume, the finite element, and the spectral methods. The numerical flexibility makes it eaiser to perform simulations with desirable robustness, accuracy and efficiency. The numerical scheme for the DBM can be chosen to balance the desired physical fidelity and computational cost.

Actually, the DBM does not belong to the family of classic LBM solvers. Standard LBMs mainly serve as solvers of (incompressible or slightly compressible) Navier-Stokes (NS) equations or other partial differential equations and aim to be loyal to these original equations. The DBM is equivalent to a modified hydrodynamic model plus a coarse-grained model of the thermodynamic nonequilibrium behaviors \cite{Gan2015SM,Zhang2019SM,Lin2015PRE,Lin2016CNF,Zhang2016CNF,Lin2017SR,Lin2018CNF,Lin2014PRE,Lai2016PRE,Lin2017PRE,Chen2018POF,Lin2018CAF,Lin2019PRE}.
In other words, the DBM kinetic modeling goes beyond traditional macroscopic governing equations in terms of physics recovered. 
To be specific, the DBM provides two tools to describe the TNE: One is to employ the viscous stresses and heat fluxes derived via Chapman-Enskog multi-scale analysis; The other is to use kinetic moments of the differences between the distribution functions and its equilibrium counterparts \cite{Gan2015SM,Zhang2019SM,Lin2015PRE,Lin2016CNF,Zhang2016CNF,Lin2017SR,Lin2018CNF,Lin2014PRE,Lai2016PRE,Lin2017PRE,Chen2018POF,Lin2018CAF,Lin2019PRE}. The former simply describes the TNE upon the evolution of macroscopic fluid behaviors, while the latter provides a detailed description of the specific nonequilibrium degree.
The study of TNE based on the DBM is helpful to deepening the understanding of the linear and nonlinear constitutive relations in hydrodynamic fluid models from a more fundamental point of view \cite{Zhang2016CNF}. 
It is convenient to use the DBM to probe the relationship between the nonequilibrium quantities and other concerned physical variables (e.g., entropy), and identify the correlation and similarity between different nonequilibrium states or processes \cite{Chen2018POF}.

Due to its solid physical foundation, the DBM has been applied to investigate various complex fluid flows and gained some new physical insights into the corresponding systems, including multiphase flows \cite{Gan2015SM,Zhang2019SM}, reactive flows \cite{Lin2016CNF,Zhang2016CNF,Lin2017SR,Lin2018CNF}, and fluid instabilities \cite{Lin2014PRE,Lai2016PRE,Lin2017PRE,Chen2018POF,Gan2019FOP,Ye2020Entropy}. Besides by theoretical analyses and experimental data \cite{Lin2017SR}, DBM results have been confirmed and supplemented by numerical solutions of MD \cite{Liu2016FP,Liu2017PRE}, DSMC \cite{Zhang2019CPC}, etc. Generally, in terms of relaxation time, the DBM can be divided into two classes, single-relaxation-time (SRT) DBM \cite{Gan2015SM,Lai2016PRE,Lin2018CNF,Zhang2019CPC} and MRT DBM \cite{Lin2015PRE,Chen2018POF,Lin2018CAF,Lin2019PRE}. From the perspective of fluid species, it can be classified into two categories, single-component DBM \cite{Lin2014PRE,Gan2015SM} and multi-component DBM \cite{Lin2015PRE,Lin2017PRE,Lin2017SR,Lin2018CNF}. Now, we propose a first MRT DBM for multicomponent flows. Compared with SRT DBMs where there is only one relaxation time and a fixed Prandtl number $\Pr = 1$ \cite{Gan2015SM,Lin2018CNF,Zhang2019SM}, the MRT DBM has various relaxation times for different nonequilibrium processes and a flexible $\Pr$. 
In contrast to single-component DBMs \cite{Lin2014PRE,Gan2015SM}, $N$-component DBM describes each chemical species by an individual distribution function, and consequently presents a much finer treatment of the flow system, for example, each component has its own particle mass, density, flow velocity, temperature, viscosity, heat conductivity, etc. 
As a preliminary application, the current model is used to study the nonequlibrium mixing process induced by the Kelvin-Helmholtz instability (KHI) in this work. 

The KHI is a fundamental interfacial instability in fluid mechanics \cite{Batchelor2000,Umeda2020PP,Hoshoudy2020EPJP}. It occurs when there is velocity shear across a wrinkled interface in a fluid system, and leads to the formation of vortices and turbulence \cite{Batchelor2000}. KHI phenomena are ubiquitous in nature and are of considerable interest in scientific and engineering fields \cite{Awasthi2014IJHMT,Liu2015IJHMT,Wang2009EPL,Wang2010POP,Gan2019FOP,Lin2019CTP}. 
Although the KHI has been investigated extensively, there are still some open problems, such as the effect of heat conduction or ablation, on which the conclusion is highly controversial \cite{Awasthi2014IJHMT,Liu2015IJHMT,Wang2009EPL,Wang2010POP,Gan2019FOP}. Viscous potential flow analysis of the KHI around an liquid-vapor interface suggests that heat transfer (resulting in mass transfer) tends to enhance the unstable process of a fluid system \cite{Awasthi2014IJHMT,Liu2015IJHMT}. 
On the contrary, comparison of numerical results between the classical and ablative KHIs indicates that thermal conduction (with dissipative nature) stabilizes the flow by impeding the linear growth rate and frequency, suppressing the perturbation transmission and fine structures, but it promotes the vortex pairing process and large-scale structures \cite{Wang2009EPL,Wang2010POP}. 
Very recently, Gan et al. proposed an easily implementable DBM for the KHI with flexible specific-heat ratio and Prandtl number, and found that the thermal conduction firstly restrains then strengthens the KHI afterwards \cite{Gan2019FOP} because it extends both density and velocity transition layers simultaneously.

However, the aforementioned studies on KHI are based on numerical models only applicable to single-component fluids. 
These models have the following constraints: 
(i) The fluid within the same chemical species can be studied, while the interaction between different components is beyond its capability. 
(ii) To set the pressure invariant across a material interface in an initial configuration, a heavy (light) medium on one side of the interface should have a low (high) temperature, because the pressure is a linear function of the concentration and temperature. 
(iii) As changes of density and temperature are strongly coupled due to the equation of state, the heat transfer always results in mass transfer, and vice versa. In other words, the independent impact of either density or temperature (i.e., mass or heat transfer) can not be accurately probed. For example, the effect of the Atwood number is bound to the influence of temperature differences.	
For the sake of investigating an independent thermal effect (or impact of temperature variation) on KHI, it is necessary to adopt a two-component (or multicomponent) physical model suitable for the practical situation where the changes of mass density and temperature are not combined together \cite{Lin2017PRE}. 
For instance, the multicomponent model is applicable to fluid systems where the Atwood number is constant and the component temperature is variable.  
In fact, it is one reason why we develop the MRT DBM for multicomponent mixtures and apply it to the thermal KHI in this research. The rest of the paper is organized as follows. Details of our DBM are described in Sec. \ref{SecII}. In Sec. \ref{SecIII}, the model is validated by three benchmarks, i.e., the three-component diffusion, the thermal Couette flow, and the Sod tube shock. Then, the DBM is employed to investigate the compressible nonequilibrium KHI with various initial temperature and thermal conductivity in Sec. \ref{SecIV}. Finally, Sec. \ref{SecV} gives conclusions and discussions.

\section{Discrete Boltzmann model}\label{SecII}

In nonequilibrium statistical physics, the system is described by the particle velocity distribution functions that are equivalent to all their kinetic moments (from zero to infinite orders). In theory, the main features of the distribution function can be captured by the initial parts of its kinetic moments (with relatively low orders) \cite{Struchtrup2005Book}. More kinetic moments are needed to describe the nonequilibrium behaviors with increasing deviation from the equilibrium state. 

In the constructing process of the DBM, there are three main stages: (i) simplification of the collision term, (ii) discretization of the velocity space, and (iii) description of meaningful nonequilibrium information. The first two phases belong to coarse-grained physical modeling, where the concerned physical variables (including conserved quantities and some nonconserved ones) should remain unchanged during the simplification and discretization process. The last step is actually the core and main purpose of DBM, where the nonequilibrium effects can be measured by using the high-order kinetic moments of the differences between the discrete distribution functions and their equilibrium counterparts. 

Note that the DBM is a special discretization of the Boltzmann equation in particle velocity space. First of all, let us introduce symbols $f_{i}^{\sigma }$ and $\hat{f}_{i}^{\sigma }$ which denote the discrete distribution functions in the velocity and moment spaces, respectively, see Eq. (\ref{Moment_f}). Here the subscript $i$ ($ = 1$, $2$, $\dots$, $N$) represents the number of discrete velocities $v_{i \alpha}^{\sigma }$, and the total number is $N$ ($ = 16$) in this work, see Eq. (\ref{DVM}). The superscript $\sigma$ stands for the chemical species in a fluid mixture. 

The individual mass density ${\rho}^{\sigma }$, molar number density ${n}^{\sigma }$, momentum ${J}^{\sigma }_{\alpha}$, and velocity ${u}^{\sigma }_{\alpha}$ are obtained from the following relations,
\begin{equation}
	{{\rho }^{\sigma }}={{m}^{\sigma }}{{n}^{\sigma }}={{m}^{\sigma }}\sum\nolimits_{i}{f_{i}^{\sigma }}
	\tt{,}
\end{equation}
\begin{equation}
	J_{\alpha }^{\sigma }={{\rho }^{\sigma }}u_{\alpha }^{\sigma }={{m}^{\sigma }}\sum\nolimits_{i}{f_{i}^{\sigma }v_{i\alpha }^{\sigma }}
	\tt{,}
\end{equation}
with the molar mass $m^{\sigma}$. The mixing mass density ${\rho}$, number density ${n}$, momentum ${J}_{\alpha}$, and velocity ${u}_{\alpha}$ are given by 
\begin{equation}
	\rho =\sum\nolimits_{\sigma }{{{\rho }^{\sigma }}}
	\tt{,}
\end{equation}
\begin{equation}
	n=\sum\nolimits_{\sigma }{{{n}^{\sigma }}}
	\tt{,}
\end{equation}
\begin{equation}
	{{J}_{\alpha }}=\rho {{u}_{\alpha }}=\sum\nolimits_{\sigma }{J_{\alpha }^{\sigma }}
	\tt{.}
\end{equation}
The individual and mixing energies are, respectively,
\begin{equation}
	{{E}^{\sigma }}=\frac{1}{2} m^{\sigma} \sum\nolimits_{i}{f_{i}^{\sigma}\left( v_{i}^{\sigma 2}+\eta _{i}^{\sigma 2} \right)}
	\tt{,}
\end{equation}
\begin{equation}
	E=\sum\nolimits_{\sigma }{{{E}^{\sigma }}}
	\tt{,}
\end{equation}
where $\eta _{i}$ is used to describe extra energies corresponding to molecular rotation, vibration, and a third translational motion hidden by the two-dimensional DBM. The remaining degrees of freedom serve to adjust the heat capacity ratio to a desired value. Results of the two-dimensional model are helpful in understanding the real processes of energy exchange between translational, rotational and vibrational degrees of freedom of molecules, which play an important role in nonequilibrium flows.

The individual temperature relative to the mixing velocity $u_{\alpha}$  and average temperature are, respectively,
\begin{equation}
	{{T}^{\sigma *}}=\frac{2{{E}^{\sigma }}-{{\rho }^{\sigma }}{{u}^{2}}}{\left( D+{{I}^{\sigma }} \right){{n}^{\sigma }}}
	\label{TemperatureI}
	\tt{,}
\end{equation}
\begin{equation}
	T=\frac{2E-\rho {{u}^{2}}}{\sum\nolimits_{\sigma }{\left( D+{{I}^{\sigma }} \right){{n}^{\sigma }}}}
	\tt{,}
\end{equation}
where $D = 2$ and $I^{\sigma}$ indicates extra degrees of freedom. Different from the definition (\ref{TemperatureI}), 
\begin{equation}
	{{T}^{\sigma }}=\frac{2{{E}^{\sigma }}-{{\rho }^{\sigma }}{{u}^{\sigma 2}}}{\left( D+{{I}^{\sigma }} \right){{n}^{\sigma }}}
	\label{TemperatureII}
	\tt{,}
\end{equation}
denotes the individual temperature relative to the individual velocity $u_{\alpha}^{\sigma}$

The individual and mixing pressures take the form
\begin{equation}
	{p}^{\sigma *} = n^{\sigma} T^{\sigma *}
	\tt{,}
	\label{Pressure_sigma_asterisk}
\end{equation}
\begin{equation}
	p=\sum\nolimits_{\sigma }{{{p}^{\sigma *}}}
	\tt{,}
\end{equation}
respectively. Corresponding to Eq. (\ref{TemperatureII}), the definition
\begin{equation}
	{p}^{\sigma } = n^{\sigma} T^{\sigma }
	\tt{,}
	\label{Pressure_sigma}
\end{equation}
is introduced as well.

Furthermore, let us introduce two kinds of discrete equilibrium distribution functions in the velocity and moment spaces, respectively. The first sets are $f_{i}^{\sigma eq}$ and $\hat{f}_{i}^{\sigma eq}$ which are functions of (${n}^{\sigma }$, ${u}_{\alpha}$, $T$) (see Appendix \ref{APPENDIXA}). 
The second ones are $f_{i}^{\sigma seq}$ and $\hat{f}_{i}^{\sigma seq}$ that depend upon (${n}^{\sigma }$, ${u}^{\sigma }_{\alpha}$, $T^{\sigma }$), and their expression are given by substituting (${n}^{\sigma }$, ${u}^{\sigma }_{\alpha}$, $T^{\sigma }$) for (${n}^{\sigma }$, ${u}_{\alpha}$, $T$) in formulas of $f_{i}^{\sigma eq}$ and $\hat{f}_{i}^{\sigma eq}$, respectively (see Appendix \ref{APPENDIXA}). The projection of discrete (equilibrium) distribution functions from velocity onto moment spaces is
\begin{equation}
	{{\mathbf{\hat{f}}}^{\sigma}}={{\mathbf{M}}^{\sigma }}{{\mathbf{f}}^{\sigma}}
	\label{Moment_f}
	\tt{,}
\end{equation}
\begin{equation}
	{{\mathbf{\hat{f}}}^{\sigma eq}}={{\mathbf{M}}^{\sigma }}{{\mathbf{f}}^{\sigma eq}}
	\label{Moment_feq}
	\tt{,}
\end{equation}
\begin{equation}
	{{\mathbf{\hat{f}}}^{\sigma seq}}={{\mathbf{M}}^{\sigma }}{{\mathbf{f}}^{\sigma seq}}
	\label{Moment_fseq}
	\tt{,}
\end{equation}
in terms of the column matrices
\begin{equation}
	\mathbf{f}^{\sigma}=\left(\begin{array}{ccc}{f_{1}^{\sigma}} \ {f_{2}^{\sigma}} \ {\dots} \ {f_{N}^{\sigma}}\end{array}\right)^{\mathrm{T}}
	\label{Moment_f_R}
	\tt{,}
\end{equation}
\begin{equation}
	\hat{\mathbf{f}}^{\sigma}=\left(\begin{array}{ccc}{\hat{f}_{1}^{\sigma}} \ {\hat{f}_{2}^{\sigma}} \ {\dots} \ {\hat{f}_{N}^{\sigma}}\end{array}\right)^{\mathrm{T}}
	\label{Moment_f_L}
	\tt{,}
\end{equation}
\begin{equation}
	\mathbf{f}^{\sigma eq}=\left(\begin{array}{ccc}{f_{1}^{\sigma eq}} \ {f_{2}^{\sigma eq}} \ {\dots} \ {f_{N}^{\sigma eq}}\end{array}\right)^{\mathrm{T}}
	\label{Moment_feq_R}
	\tt{,}
\end{equation}
\begin{equation}
	\hat{\mathbf{f}}^{\sigma eq}=\left(\begin{array}{ccc}{\hat{f}_{1}^{\sigma eq}} \ {\hat{f}_{2}^{\sigma eq}} \ {\dots} \ {\hat{f}_{N}^{\sigma eq}}\end{array}\right)^{\mathrm{T}}
	\label{Moment_feq_L}
	\tt{,}
\end{equation}
\begin{equation}
	\mathbf{f}^{\sigma seq}=\left(\begin{array}{ccc}{f_{1}^{\sigma seq}} \ {f_{2}^{\sigma seq}} \ {\dots} \ {f_{N}^{\sigma seq}}\end{array}\right)^{\mathrm{T}}
	\label{Moment_fseq_R}
	\tt{,}
\end{equation}
\begin{equation}
	\hat{\mathbf{f}}^{\sigma seq}=\left(\begin{array}{ccc}{\hat{f}_{1}^{\sigma seq} \ \hat{f}_{2}^{\sigma seq} \ \cdots \ \hat{f}_{N}^{\sigma seq}}\end{array}\right)^{\mathrm{T}}
	\label{Moment_fseq_L}
	\tt{.}
\end{equation}

The discrete Boltzmann equations take the form,
\begin{equation}
	\partial _t f_{i}^{\sigma } + v_{i\alpha }^{\sigma } \partial _{\alpha } f_{i}^{\sigma } = - {{M_{il}^{\sigma }}^{-1}} S_{lk}^{\sigma } \left( \hat{f}_{k}^{\sigma } - \hat{f}_{k}^{\sigma eq} \right) + {A}_{i}^{\sigma }
	\label{DBEquation}
	\tt{.} 
\end{equation}
On the left-hand side, $t$ is the time, and $\alpha = x$, $y$ the physical space for a $2$-D system. On the right-hand side, ${S_{lk}^{\sigma }}$ is the element of a diagonal matrix $\mathbf{S}^{\sigma } = \mathrm{diag}\left(S_{1}^{\sigma} \ S_{2}^{\sigma} \ \cdots \ S_{N}^{\sigma}\right)$, and the parameter $S^{\sigma }_{i}$ controls the relaxation speed of $\hat{f}^{\sigma }_{i}$ approaching $\hat{f}_{i}^{\sigma eq}$. ${{M_{il}^{\sigma }}^{-1}}$ is the element of the square matrix ${\mathbf{M}^{\sigma }}^{-1}$ which is the inverse of $\mathbf{M}^{\sigma}$ with the element $M^{\sigma}_{il}$ (see Appendix \ref{APPENDIXA}). ${A}_{i}^{\sigma }$ is an additional term expressed by Eqs. (\ref{A_8}) and (\ref{A_9}). 

Actually, Eq. (\ref{DBEquation}) is a reduced form of 
\begin{equation}
	{{\partial }_{t}}f_{i}^{\sigma }+v_{i\alpha }^{\sigma }{{\partial }_{\alpha }}f_{i}^{\sigma }=-{{M_{il}^{\sigma }}^{-1}}\left[ S_{lk}^{\sigma s}\left( \hat{f}_{k}^{\sigma }-\hat{f}_{k}^{\sigma seq} \right)+S_{lk}^{\sigma }\left( \hat{f}_{k}^{\sigma seq}-\hat{f}_{k}^{\sigma eq} \right) \right]+A_{i}^{\sigma }
	\label{DBEquationFull}
	\tt{,} 
\end{equation}
where $S_{lk}^{\sigma s} = S_{lk}^{\sigma}$. Equations (\ref{DBEquation}) and (\ref{DBEquationFull}) are regarded as one- and two-step relaxation models, respectively. 
During the thermodynamic process described by Eq. (\ref{DBEquation}), $\hat{f}_{k}^{\sigma }$ tends towards $\hat{f}_{k}^{\sigma eq}$ with a speed controlled by the relaxation parameter $S_{lk}^{\sigma }$ in a straightforward way. 
While in Eq. (\ref{DBEquationFull}), $\hat{f}_{k}^{\sigma }$ firstly relaxes to $\hat{f}_{k}^{\sigma seq}$ at a speed controlled by the parameter $S_{lk}^{\sigma s }$, then $\hat{f}_{k}^{\sigma seq}$ to $\hat{f}_{k}^{\sigma eq}$ with $S_{lk}^{\sigma }$. (Further study on the latter equation is beyond this work.)

\begin{figure}
	\begin{center}
		\includegraphics[bbllx=0pt,bblly=0pt,bburx=186pt,bbury=181pt,width=0.4\textwidth]{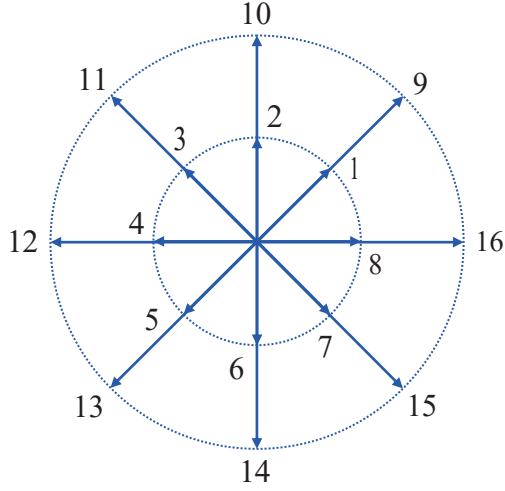}
	\end{center}
	\caption{Sketch of the discrete velocities.}
	\label{Fig01}
\end{figure}

As shown in Fig. \ref{Fig01}, there are two groups of discrete velocities whose magnitudes are $v_{a}^{\sigma }$ and $v_{b}^{\sigma }$, respectively. The expression of the discrete velocities reads
\begin{equation}
	\left(v_{i x}^{\sigma}, v_{i y}^{\sigma}\right)=\left\{\begin{array}{l}{v_{a}^{\sigma}\left(\cos \frac{i \pi}{4}, \sin \frac{i \pi}{4}\right) \ \mathrm { for } \ 1 \leq i \leq 8} , \\ [6pt]  
		{v_{b}^{\sigma}\left(\cos \frac{i \pi}{4}, \sin \frac{i \pi}{4}\right) \ \mathrm { for } \ 9 \leq i \leq N}  \tt{.} \end{array}\right.
	\label{DVM}
\end{equation}
Obviously, there is a good isotropy in the velocity space. 
Besides, we define $\eta _{i}^{\sigma }=\eta _{a}^{\sigma }$ for $1\le i\le 4$, and $\eta _{i}^{\sigma }=\eta _{b}^{\sigma }$ for $9\le i\le 12$, otherwise, $\eta _{i}^{\sigma }=0$. Here $v_{a}^{\sigma }$, $v_{b}^{\sigma }$, $\eta _{a}^{\sigma }$, and $\eta _{b}^{\sigma }$ are flexible parameters. It is worth mentioning that these parameters can be adjusted to optimize the DBM properties. 
(I) The conditions $v_{a}^{\sigma } \ne v_{b}^{\sigma } \ne 0$ and $\eta _{a}^{\sigma } \ne \eta _{b}^{\sigma } \ne 0$ should be satisfied to ensure the matrix $\mathbf{M}^{\sigma}$ invertible.
(II) For the sake of numerical stability, the sizes of $v_{a}^{\sigma }$ and $v_{b}^{\sigma }$ should be given by reference to the values of flow velocity $\mathbf{u}^{\sigma }$ and sound speed $v_s^{\sigma } = \sqrt{{\gamma^{\sigma}} {T^{\sigma}}/{m^{\sigma }}}$, where $\gamma^{\sigma}$ denotes the specific heat ratio. For example, $v_{a}^{\sigma }$ is less than $v_s^{\sigma }$, $v_{b}^{\sigma }$ is greater than $v_s^{\sigma }$, and one of them is around $\mathbf{u}^{\sigma }$. However, in general, both the flow velocity and sound speed take variable values in different fluid areas and also vary with time. Further, in a slow flow, the real molecular velocities still remain large and there are no two distinct groups of molecules, one of which moves at low speeds, and the other at high speeds. The complexities of practical systems often challenge the numerical robustness of the current model. One feasible method to solve this problem is to use adaptive discrete velocities which are functions of $\mathbf{u}^{\sigma }$ and sound speed $v_s^{\sigma }$. For simplicity, the discrete velocities are constant in this work. 
(III) The values of $\eta _{a}^{\sigma }$ and $\eta _{b}^{\sigma }$ should be set by reference to the value of $\bar{\eta} = \sqrt{I^{\sigma} {T^{\sigma}} / m^{\sigma}}$, because the extra internal energy is $\frac{1}{2} m^{\sigma } {\bar{\eta}}^{2}=\frac{1}{2} I^{\sigma} {T^{\sigma}}$ in the local thermodynamic equilibrium according to the equipartition of energy theorem. One of them should be less than $\bar{\eta}$ if the other is greater than $\bar{\eta}$, and vice versa. For instance, $\eta _{a}^{\sigma } < \bar{\eta}$ and $\eta _{b}^{\sigma } > \bar{\eta}$.
Moreover, for convenience, ($v_{a}^{\sigma}$, $v_{b}^{\sigma}$, ${\eta}_{a}^{\sigma}$, ${\eta}_{a}^{\sigma}$) can be given the same values for various chemical species $\sigma$ if their properties (including $\mathbf{u}^{\sigma }$, ${T}^{\sigma }$, and $v_s^{\sigma }$) do not have remarkable differences in practical simulations. 

It is noteworthy that, to ensure consistency with traditional NS equations in the hydrodynamic limit (see Appendix \ref{APPENDIXB}), an additional term ${A}_{i}^{\sigma }$ is imposed on the right-hand side of Eq. (\ref{DBEquation}). 
Similar to Eqs. (\ref{Moment_f}) - (\ref{Moment_fseq}), the relation between the additional term ${{\mathbf{A}}^{\sigma}}$ and its moment ${{\mathbf{\hat{A}}}^{\sigma}}$ takes the form            
\begin{equation}
	{{\mathbf{\hat{A}}}^{\sigma}} = {{\mathbf{M}}^{\sigma }} {{\mathbf{A}}^{\sigma}}
	\label{Moment_A}
	\tt{,}
\end{equation}
with 
\begin{equation}
	\mathbf{A}^{\sigma}=\left(\begin{array}{ccc}{{A}_{1}^{\sigma}} \ {{A}_{2}^{\sigma}} \ {\dots} \ {{A}_{N}^{\sigma}}\end{array}\right)^{\mathrm{T}}
	\label{Moment_A_R}
	\tt{,}
\end{equation}
\begin{equation}
	\hat{\mathbf{A}}^{\sigma}=\left(\begin{array}{ccc}{\hat{A}_{1}^{\sigma}} \ {\hat{A}_{2}^{\sigma}} \ {\dots} \ {\hat{A}_{N}^{\sigma}}\end{array}\right)^{\mathrm{T}}
	\label{Moment_A_L}
	\tt{,}
\end{equation}
where $\hat{A}_{i}^{\sigma } = 0$ for $1 \le i \le 7$ and $10 \le i \le 16$, and 
\begin{equation}
	\hat{A}_{8}^{\sigma }=2\left( S_{8}^{\sigma }-S_{5}^{\sigma } \right)u_{x}^{\sigma }\Delta _{5}^{\sigma }+2\left( S_{8}^{\sigma }-S_{6}^{\sigma } \right)u_{y}^{\sigma }\Delta _{6}^{\sigma }
	\label{A_8}
	\tt{,}
\end{equation}
\begin{equation}
	\hat{A}_{9}^{\sigma }=2\left( S_{9}^{\sigma }-S_{7}^{\sigma } \right)u_{y}^{\sigma }\Delta _{7}^{\sigma }+2\left( S_{9}^{\sigma }-S_{6}^{\sigma } \right)u_{x}^{\sigma }\Delta _{6}^{\sigma }
	\label{A_9}
	\tt{,}
\end{equation}
in terms of 
\begin{equation}
	\Delta _{5}^{\sigma }=\frac{2{{n}^{\sigma }}{{T}^{\sigma }}}{S_{5}^{\sigma }{{m}^{\sigma }}}\left( \frac{1-D-{{I}^{\sigma }}}{D+{{I}^{\sigma }}}{{\partial }_{x}}u_{x}^{\sigma }+\frac{{{\partial }_{y}}u_{y}^{\sigma }}{D+{{I}^{\sigma }}} \right)
	\tt{,}
\end{equation}
\begin{equation}
	\Delta _{6}^{\sigma }=-\frac{{{n}^{\sigma }}{{T}^{\sigma }}}{S_{6}^{\sigma }{{m}^{\sigma }}}\left( {{\partial }_{y}}u_{x}^{\sigma }+{{\partial }_{x}}u_{y}^{\sigma } \right)
	\tt{,}
\end{equation}
\begin{equation}
	\Delta _{7}^{\sigma }=\frac{2{{n}^{\sigma }}{{T}^{\sigma }}}{S_{7}^{\sigma }{{m}^{\sigma }}}\left( \frac{{{\partial }_{x}}u_{x}^{\sigma }}{D+{{I}^{\sigma }}}+\frac{1-D-{{I}^{\sigma }}}{D+{{I}^{\sigma }}}{{\partial }_{y}}u_{y}^{\sigma } \right)
	\tt{.}
\end{equation}
From Eq. (\ref{Moment_A}), the following formula is derived, 
\begin{equation}
	{{\mathbf{A}}^{\sigma}} = {{{\mathbf{M}}^{\sigma }}^{-1}} {{\mathbf{\hat{A}}}^{\sigma}}
	\label{Exresssion_A}
	\tt{,}
\end{equation}
which is the expression of the additional term.

In addition, the Fick's laws of diffusion and Stefan-Maxwell diffusion equation could also be derived from the multicomponent NS equations under corresponding assumptions (see Appendix \ref{APPENDIXC}). Besides giving the continuum equations, the DBM also provides a set of handy, effective and efficient tools to describe and probe the abundant kinetic information beyond them. Let us define  $\hat{f}_{i}^{\sigma } = \hat{f}_{i}^{\sigma sneq} + \hat{f}_{i}^{\sigma seq}$, with the equilibrium part $\hat{f}_{i}^{\sigma seq}$ and nonequilibrium part $\hat{f}_{i}^{\sigma sneq}$, respectively. In a similar way, we can define $\hat{f}_{i}^{\sigma } = \hat{f}_{i}^{\sigma neq} + \hat{f}_{i}^{\sigma eq}$. Namely, there are two kinds of nonequilibrium physical quantities $\hat{f}_{i}^{\sigma sneq}$ and $\hat{f}_{i}^{\sigma neq}$, which can be obtained in each iterative step and used to investigate the nonequilibrium effects. (It is the key reason why $\hat{f}_{i}^{\sigma seq}$ and $\hat{f}_{i}^{\sigma eq}$ are introduced.) Concretely, $\hat{f}_{i}^{\sigma sneq} = 0$ for $1 \le i \le 4$ in line with conservation laws, as $\hat{f}_{i}^{\sigma } = \hat{f}_{i}^{\sigma seq} = {{n}^{\sigma }}$, ${{J}^{\sigma }_{x}}/{{m}^{\sigma }}$, ${{J}^{\sigma }_{y}}/{{m}^{\sigma }}$, $2{E}^{\sigma}/{{m}^{\sigma }}$ for $i = 1$, $2$, $3$, $4$, respectively. In contrast, the nonequilibrium quantity $\hat{f}_{i}^{\sigma sneq}$ may not equal zero for $5 \le i \le 16$ in a nonequilibrium state. Physically, $\hat{f}_{i}^{\sigma sneq}$ denotes the departure of a kinetic mode $\hat{f}_{i}^{\sigma }$ from its equilibrium counterpart $\hat{f}_{i}^{\sigma seq}$. The speed of relaxation process from $\hat{f}_{i}^{\sigma }$ to $\hat{f}_{i}^{\sigma seq}$ is controlled by the relaxation parameter ${S_{i}^{\sigma }}$, and both $\hat{f}_{i}^{\sigma sneq}$ and ${S_{i}^{\sigma }}$ exert influence on the thermodynamic and hydrodynamic behaviors. Simultaneously, various nonequilibrium effects interplay with each other, and these kinetic modes are coupled as well. For instance, at the NS level,
\begin{eqnarray}{\hat{f}_{5}^{\sigma s n eq}=\Delta_{5}^{\sigma}-n^{\sigma}\left(u_{x}^{\sigma 2}-u_{x}^{2}\right)+\frac{2 S_{2}^{\sigma} n^{\sigma}}{S_{5}^{\sigma}} u_{x}^{\sigma}\left(u_{x}^{\sigma}-u_{x}\right)} ~~~ \nonumber \\ 
	{-\frac{S_{4}^{\sigma} n^{\sigma}}{S_{5}^{\sigma}} \frac{\left(u_{x}^{\sigma}-u_{x}\right)^{2}+\left(u_{y}^{\sigma}-u_{y}\right)^{2}}{D+I^{\sigma}}+\left(S_{4}^{\sigma}-S_{5}^{\sigma}\right) n^{\sigma} \frac{T^{\sigma}-T}{S_{5}^{\sigma} m^{\sigma}}} 
	,
	\label{hatf5_sneq}
\end{eqnarray}
\begin{eqnarray}
	\hat{f}_{6}^{\sigma sneq}=\Delta _{6}^{\sigma }-{{n}^{\sigma }}\left( u_{x}^{\sigma }u_{y}^{\sigma }-{{u}_{x}}{{u}_{y}} \right) ~~~~~  \nonumber \\
	+\frac{S_{2}^{\sigma }{{n}^{\sigma }}}{S_{6}^{\sigma }}u_{y}^{\sigma }\left( u_{x}^{\sigma }-{{u}_{x}} \right)+\frac{S_{3}^{\sigma }{{n}^{\sigma }}}{S_{6}^{\sigma }}u_{x}^{\sigma }\left( u_{y}^{\sigma }-{{u}_{y}} \right) ,
	\label{hatf6_sneq}
\end{eqnarray}
\begin{eqnarray}
	\hat{f}_{7}^{\sigma sneq}=\Delta _{7}^{\sigma }-{{n}^{\sigma }}\left( u_{y}^{\sigma 2}-u_{y}^{2} \right)+2\frac{S_{3}^{\sigma }{{n}^{\sigma }}}{S_{7}^{\sigma }}u_{y}^{\sigma }\left( u_{y}^{\sigma }-{{u}_{y}} \right) ~~  \nonumber \\ 
	-\frac{S_{4}^{\sigma }{{n}^{\sigma }}}{S_{7}^{\sigma }}\frac{{{\left( u_{x}^{\sigma }-{{u}_{x}} \right)}^{2}}+{{\left( u_{y}^{\sigma }-{{u}_{y}} \right)}^{2}}}{D+{{I}^{\sigma }}}+\left( S_{4}^{\sigma }-S_{7}^{\sigma } \right){{n}^{\sigma }}\frac{{{T}^{\sigma }}-T}{S_{7}^{\sigma }{{m}^{\sigma }}}
	,
	\label{hatf7_sneq}
\end{eqnarray}
which are derived from the Chapman-Enskog analysis, reduce to $\hat{f}_{5}^{\sigma sneq}=\Delta _{5}^{\sigma }$, $\hat{f}_{6}^{\sigma sneq}=\Delta _{6}^{\sigma }$, and $\hat{f}_{7}^{\sigma sneq}=\Delta _{7}^{\sigma }$ under conditions of $u_{\alpha}^{\sigma } = {u}_{\alpha}$ and ${T}^{\sigma } = T$. 

The above mentioned capability of this DBM makes convenient to study behaviors in the nonequilibrium process, such as the entropy production \cite{Lin2019CTP,Zhang2016CNF,Zhang2019SM}. Especially, with ${{X}^{\sigma }}$ the molar fraction of species $\sigma$, the entropy of mixing, 
\begin{equation}
	S_M=-\sum\nolimits_{\sigma }{{{n}^{\sigma }}\ln {{X}^{\sigma }}}
	\tt{,}
\end{equation}
which is part of the increasing entropy as separate mixable fluids contact and mix, can be obtained in each iterative step.

It should be stressed that kinetic effects are significant and traditional hydrodynamic models are not sufficient for fluid flows with small characteristic scales or large Knudsen numbers \cite{Gan2015SM,Zhang2019SM,Lin2015PRE,Lin2016CNF,Zhang2016CNF,Lin2017SR,Lin2018CNF,Lin2014PRE,Lai2016PRE,Lin2017PRE,Chen2018POF,Lin2018CAF,Lin2019PRE}. 
The TNE becomes crucial and even dominant in the evolution of multicomponent flows due to the existence of various complex material and/or mechanical interfaces \cite{Gan2015SM,Zhang2019SM,Lin2015PRE,Lin2016CNF,Zhang2016CNF,Lin2017SR,Lin2018CNF,Lin2014PRE,Lai2016PRE,Lin2017PRE,Chen2018POF,Lin2018CAF,Lin2019PRE}. 
In such complicated cases, to investigate the TNE is a significant and convenient way to study the fundamental kinetic processes, which is made easy by the discrete Boltzmann modeling. 
The DBM is equivalent to the modified NS equations plus a coarse-grained thermodynamic nonequilibrium model in fluid systems with essential TNE.
In the continuum limit, it reduces to the usual NS equations supplemented with a coarse-grained model for the most relevant thermodynamic nonequilibrium behaviors. In any case, a DBM brings more physical information than a pure hydrodynamic model. Because the hydrodynamic model generally consists of only the evolution of the conserved kinetic moments, i.e., the density, momentum and energy.

In addition, the DBM has the advantage of simplicity for coding and high efficiency of parallel processing, since the set of formulas in Eq. (\ref{DBEquation}) is uniformly linear and the information transfer in DBM is local in both time and space \cite{Lin2019CTP}. Actually, the parallel programming based on the message-passing interface are used for all simulations in this work. Moreover, we adopt the second-order nonoscillatory and nonfree-parameter dissipation difference scheme \cite{Zhang1991NND} to deal with the space derivatives and the second-order Runge-Kutta method to treat the time derivative in Eq. (\ref{DBEquation}). Note that the current Runge-Kutta method is an explicit scheme, so the temporal step should be no greater than the minimum of the relaxation times $\tau_m$ in order to have accurate and robust solutions. To be specific, it is necessary to satisfy the relation $\Delta t \le \tau_m$, where $\tau_m = \mathrm{min} \left({1/{S_{i}^{\sigma}}}
\right)$ is the minimum of the reciprocal of $S_{i}^{\sigma}$, 
and another restriction is on the Courant number: $\Delta t \le \Delta x / \mathrm{max} (v_i^{\sigma})$. 

Remark: The DBM and other discrete ordinate methods are based on special discretization forms of the (simplified) Boltzmann equation in particle velocity space \cite{Broadwell1964POF,Yang1995JCP,Mieussens2000JCP}. These kinetic models have the common feature that the time, space and particle velocity are discretized in particular ways. 
The essential differences among them lie in how the collision term is simplified and how the discrete (equilibrium) distribution functions are calculated, which leads to different capabilities of the models. 
In the pioneering discrete velocity model (DVM) that aims to solve the Boltzmann equation \cite{Broadwell1964POF}, there are only six molecular velocities and the collision term is written as the gain minus the loss (in a nonlinear quadratic form). But the model is too simple to describe a real physical system \cite{Broadwell1964POF}. With the Gauss-Hermite or Newton-Cotes rule used in the discrete ordinate method, the DBM (for BGK or ES equation) is applicable to rarefied gas flows over a wide range of Mach and Knudsen numbers \cite{Yang1995JCP}. Furthermore, the conservation laws and entropy dissipation are obeyed, as the discrete equilibrium distribution functions are expressed by an exponential function with the introduction of a discrete-velocity grid \cite{Mieussens2000JCP}. Although the set of allowable velocities becomes finite in the DVM, the computational cost is still often too expensive to perform satisfactory simulations \cite{Yang1995JCP,Mieussens2000JCP}. 
In contrast, the DBM is designed to accurately predict fluid flows with HNE and TNE. To this end, a list of moment relations of discrete equilibrium distribution functions is required in the DBM. The collision term (including several relaxation times) and discrete equilibrium distribution functions (with the total number $16$ in this work) are calculated through the matrix inversion method, which is physically accurate, computationally efficient and numerically robust \cite{Lin2019PRE}. 	

\section{Verification and validation}\label{SecIII}

For practical calculations, it is convenient and useful to use dimensionless variables. In this work, physical quantities are expressed in nondimensional forms using the following references, i.e., the molar mass $m_0$, molar number density $n_0$, length $L_0$, temperature $T_0$, and universal gas constant $R$. For example, 
\begin{center}
	\begin{tabular}{lll}
		Distribution functions: $f_{i}^{\sigma }$ & by & $n_0$\\
		Mass density: $\rho^{\sigma}$, $\rho$ & by & $m_0 n_0$ \\
		Speed and velocity: $v_s^{\sigma }$, $\eta _{i}^{\sigma }$, $\mathbf{u}^{\sigma}$, $\mathbf{u}$ & by & $\sqrt{R T_0/{m_0}}$ \\
		Energy density: $E^{\sigma}$, $E$ & by & $n_0 R T_0$ \\
		Pressure: $p^{\sigma}$, $p$ & by & $n_0 R T_0$ \\
		Temperature: $T^{\sigma}$, $T$ & by & $T_0$ \\
		Coordinate: $x$, $y$ & by & $L_0$ \\
		Time: $t$ & by & $L_0/\sqrt{R T_0/{m_0}} $ \\
	\end{tabular}
\end{center}

In the following are three subsections. The first part is for the three-component diffusion, which is to demonstrate the capacity of the present DBM in dealing with the interaction among various nonpremixed chemical species. The second subsection is to use the thermal Couette flow to validate that our DBM is suitable for fluid flows where both Prandtl number and specific heat ratio are flexible. Finally, the Sod shock tube is simulated to show that this model has the capability of describing the shock wave with a high Mach number (as well as the rarefaction wave). 

\subsection{Three-component diffusion}

Diffusion is the net movement of molecules driven by a gradient in chemical potential of fluid species \cite{Cussler2000,Bird2002AMR}. As one of the most important and fundamental transport processes, it has received great attention due to its significance in chemical process and biological engineering \cite{Cussler2000,Bird2002AMR,Law2006}, etc. 

\begin{figure}
	\begin{center}
		\includegraphics[bbllx=0pt,bblly=0pt,bburx=158pt,bbury=107pt,width=0.35\textwidth]{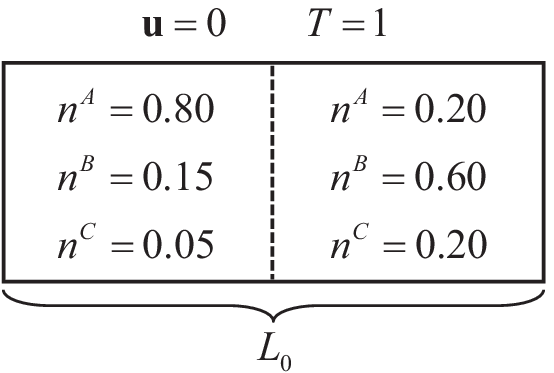}
	\end{center}
	\caption{Initial configuration of the three-component diffusion.}
	\label{Fig02}
\end{figure}

To demonstrate that the DBM could describe the interaction among various chemical species, we carry out the simulation of multicomponent diffusion. As shown in Fig. \ref{Fig02}, the initial configuration is
\begin{equation}
	\left\{
	\begin{array}{l}
		{{\left( {{n}^{A}},{{n}^{B}},{{n}^{C}} \right)}_{L}}=\left( 0.80,0.15,0.05 \right) 	\tt{,} \\
		{{\left( {{n}^{A}},{{n}^{B}},{{n}^{C}} \right)}_{R}}=\left( 0.20,0.60,0.20 \right) 	\tt{,}
	\end{array}
	\right.
\end{equation}
where the subscripts $L$ and $R$ indicate $0 < x \le L_0 / 2$ and $L_0 / 2 < x \le L_0$, respectively, with $L_0 = 0.1$. The superscripts $A$, $B$, and $C$ represent three chemical species, respectively. For simplicity, the molar mass is chosen as $m^{\sigma}=1$. The average velocity and temperature are $\mathbf{u} = 0$ and $T = 1$. The pressure on the two sides equals $p = 1$, hence the interface remains rest. In the horizontal direction the quantities on the ghost nodes outside the boundary are replaced by the neighbouring ones \cite{Qu2007PRE,Gan2018PRE}, while the boundary conditions are periodic in the vertical direction. In fact, this case is a $1$-D problem as the physical field is the same in the $y$ direction. Hence, the mesh is chosen as $N_x \times N_y = N_x \times 1$. The spatial step is $\Delta x = \Delta y = L_{0} / N_x$, the temporal step $\Delta t = 4 \times 10^{-4}$, the relaxation parameters $S_i = 10^{3}$, the extra degrees of freedom $I^{\sigma} = 3$, and the parameters ($v^{\sigma}_{a}$, $v^{\sigma}_{b}$, ${\eta}^{\sigma}_{a}$, ${\eta}^{\sigma}_{a}$) $=$ ($0.01$, $2$, $2.7$, $2.55$). 

\begin{figure}
	\begin{center}
		\includegraphics[bbllx=0pt,bblly=0pt,bburx=423pt,bbury=153pt,width=0.8\textwidth]{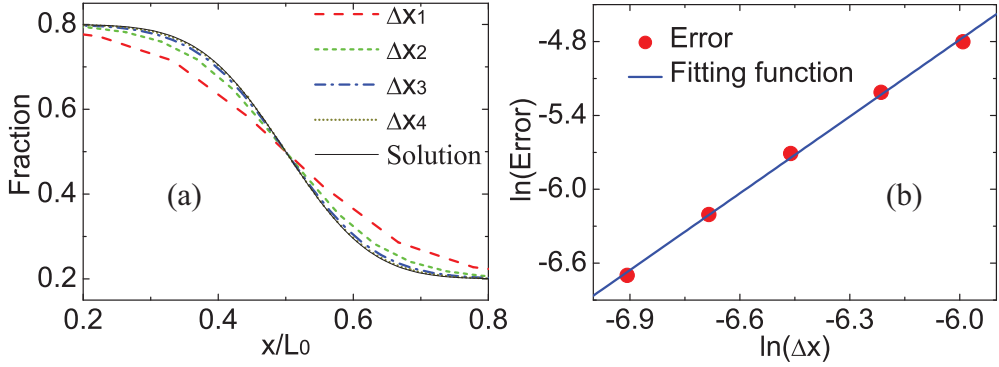}
	\end{center}
	\caption{Grid convergence analysis: (a) the horizontal distribution of mole fractions $X^{A}$ at the time $t = 0.05$, (b) relative errors under various spatial steps.}
	\label{Fig03}
\end{figure}

First of all, let us perform a grid convergence analysis, which is an important issue for numerical models. To this end, we carried out some simulations under various spatial steps $\Delta x_{1} = L_{0}/10$, $\Delta x_{2} = L_{0}/20$, $\Delta x_{3} = L_{0}/40$, and $\Delta x_{4} = L_{0}/80$, respectively. Figure \ref{Fig03} (a) shows the mole fraction of species $A$. The long-dashed, short-dashed, dash-dotted and short-dotted lines stand for DBM results under $\Delta x_{1}$, $\Delta x_{2}$, $\Delta x_{3}$, and $\Delta x_{4}$, respectively. The solid line denotes the analytical solution  \cite{Cussler2000,Bird2002AMR},
\begin{equation}
	X^{\sigma}=\frac{X^{\sigma}_{L}+X^{\sigma}_{R}}{2}-\frac{X^{\sigma}_{L}-X^{\sigma}_{R}}{2} {\rm Erf} \left(\frac{x-x_0}{\sqrt{4 D t}} \right) 
	\tt{,}
\end{equation}
where ${\rm Erf}$ is the complementary error function, $x_0 = L_0/2$ is the location of the interface, $D = 10^{-3}$ is the diffusivity. It can be found that, with decreasing spatial steps (i.e., increasing resolution), the numerical results converge towards the analytical solution. Particularly, the results with spatial step $\Delta x_{4}$ are quite close to the solution, which is satisfactory.

For the purpose of a quantitative analysis, Fig. \ref{Fig03} (b) gives relative errors versus spatial steps. The relative error takes the form 
\begin{equation}
	{\rm Error} (\phi)=\sqrt{\frac{\sum_{(x, y)}\left|\phi_{a}(x, y, t)-\phi_{n}(x, y, t)\right|^{2}}{\sum_{(x, y)}\left|\phi_{a}(x, y, t)\right|^{2}}}
	\tt{,}
\end{equation}
where $\phi_{a}$ and $\phi_{n}$ denote the analytical and numerical results of the variable $\phi$ (e.g., the mole fraction $X^{A}$). The circles represent the DBM results and the line stand for the fitting function, ${\rm ln (Error)} = 2.079 \ {\rm ln} (\Delta x) + 7.6887$. Clearly, the slope of the fitting function is close to $2.0$, which indicates that the current model has a second-order convergence rate in space. 

\begin{figure}
	\begin{center}
		\includegraphics[bbllx=0pt,bblly=0pt,bburx=168pt,bbury=309pt,width=0.4\textwidth]{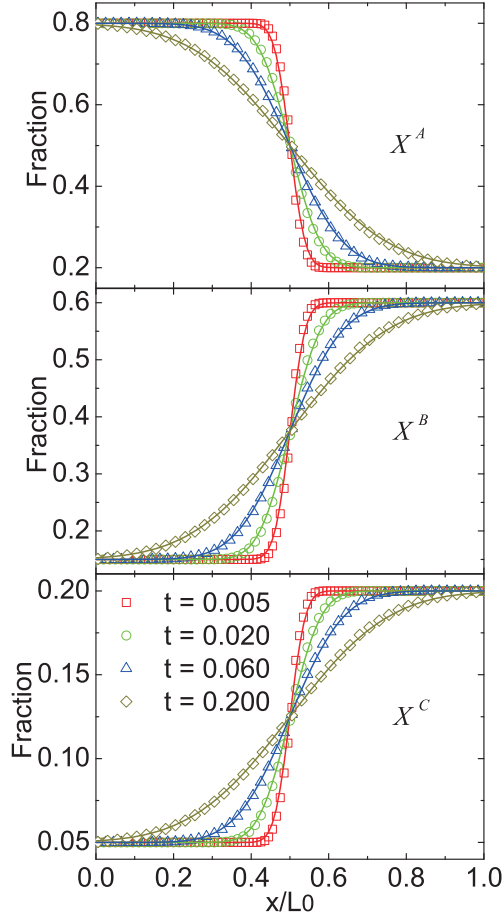}
	\end{center}
	\caption{Molar fractions in the diffusion process: $X^A$ (top), $X^B$ (middle), and $X^C$ (bottom). Squares, circles, triangles, and diamonds denote DBM results at time constants $t = 0.005$, $0.02$, $0.06$, and $0.2$, respectively. Solid lines stand for the corresponding analytical solutions.}
	\label{Fig04}
\end{figure}

Figure \ref{Fig04} illustrates molar fractions, $X^{\sigma} = n^{\sigma}/n$, at various times in the diffusion process. The spatial step is $\Delta x_4$, which is valided in Fig. \ref{Fig03} (a). Symbols denote numerical results at various times $t = 0.005$ (squares), $0.02$ (circles), $0.06$ (triangles), and $0.2$ (diamonds), respectively. Lines denote the analytical solutions. It is evident that the DBM results coincide with the analytical solutions in the evolution of the diffusion. 

\begin{figure}
	\begin{center}
		\includegraphics[bbllx=0pt,bblly=0pt,bburx=476pt,bbury=326pt,width=0.42\textwidth]{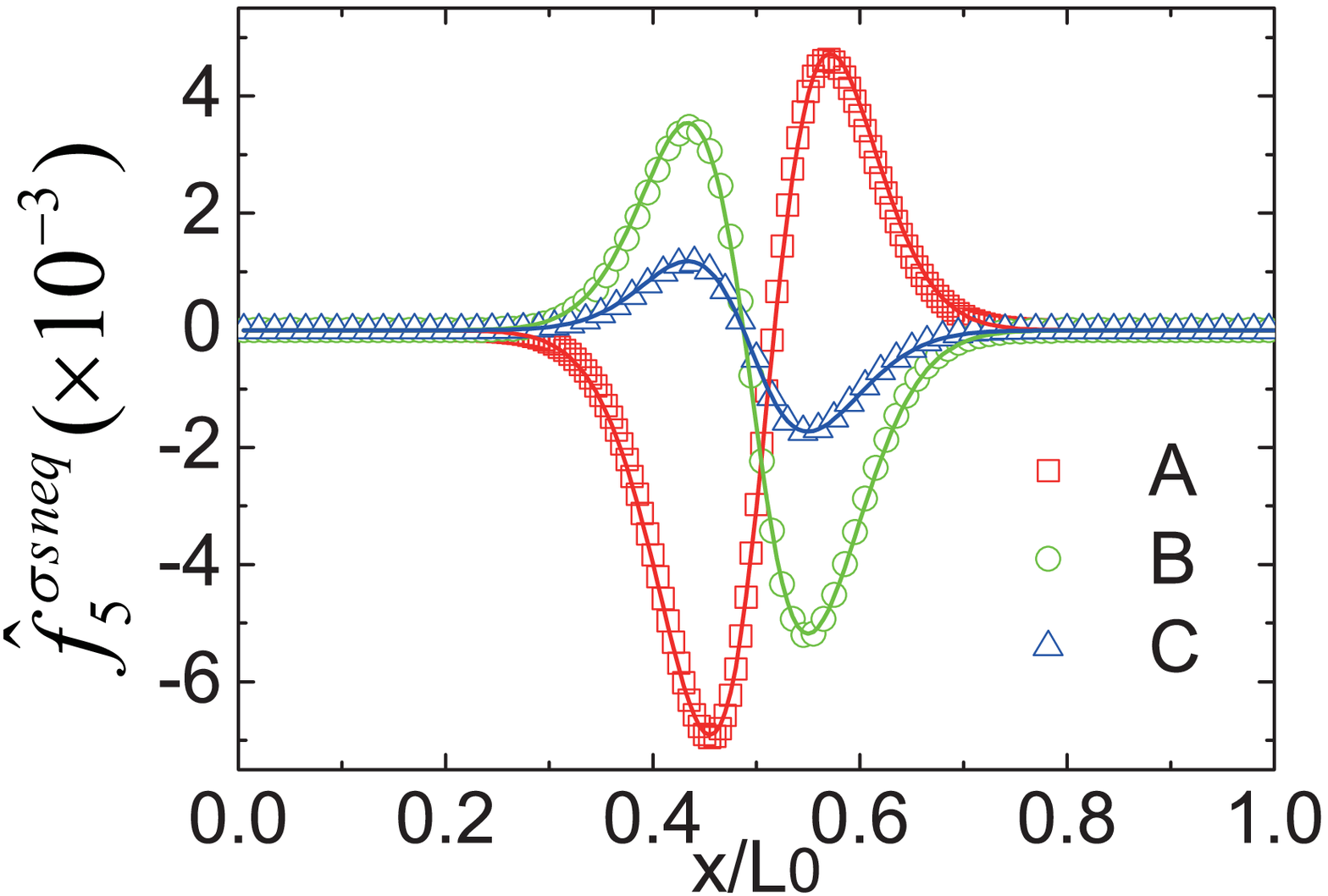}
	\end{center}
	\caption{Nonequilibrium quantities at time $t = 0.02$ in the diffusion process. Squares, circles, and triangles denote DBM results of $\hat{f}^{A sneq}_{5}$, $\hat{f}^{B sneq}_{5}$, and $\hat{f}^{C sneq}_{5}$, respectively. Solid lines stand for the corresponding analytical solutions.}
	\label{Fig05}
\end{figure}

Moreover, to further validate that the DBM has the capability of capturing nonequilibrium effects, Fig. \ref{Fig05} plots nonequilibrium quantities $\hat{f}^{\sigma sneq}_{5}$ at time $t = 0.02$ in the diffusion process. Symbols represent our DBM results, and lines represent the analytical solutions in Eq. (\ref{hatf5_sneq}). Obviously, our simulation results are in excellent agreement with the analytical solutions. Consequently, it is confirmed that the DBM can be used to probe and measure nonequilibrium manifestations. 

\subsection{Thermal Couette flow}

In fluid dynamics, thermal Couette flow is the flow of a viscous fluid between two surfaces with relative shear movement. It is a classical benchmark to test a model for compressible fluid flows where viscosity and heat transfer dominate \cite{Li2007PRE,Yang2016PRE}. Here we conduct simulations of the thermal Couette flow for two purposes. One aim is to verify that the DBM is suitable for various values of the specific heat ratio $\gamma$ and Prandtl number $\Pr$. The other aim is to verify the DBM for the case with premixed compressible fluid species. 

\begin{figure}
	\begin{center}
		\includegraphics[bbllx=0pt,bblly=0pt,bburx=110pt,bbury=90pt,width=0.3\textwidth]{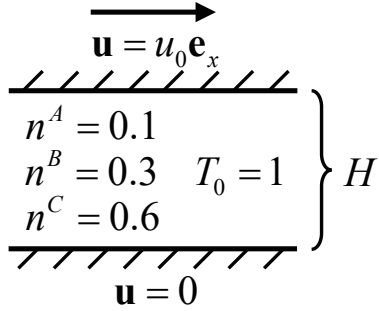}
	\end{center}
	\caption{Initial configuration of the thermal Couette flow.}
	\label{Fig06}
\end{figure}
\begin{table}
	\begin{tabular}{lccccc}
		\hline\hline
		Cases & $\Pr$ & $\gamma$ & $S^{\sigma}_{i = 5, 6, 7}$ & $S^{\sigma}_{i \ne 5, 6, 7}$ & $I^{\sigma}$ \\
		\hline
		Run I & $1.0$ & $1.3$ & $1000$ & $1000$ & $14/3$ \\
		\hline
		Run II & $1.0$ & $1.4$ & $1000$ & $1000$ & $3$ \\
		\hline
		Run III & $1.0$ & $1.5$ & $1000$ & $1000$ & $2$ \\
		\hline
		Run IV & $0.5$ & $1.4$ & $2000$ & $1000$ & $3$ \\
		\hline
		Run V & $2.0$ & $1.4$ & $500$ & $1000$ & $3$ \\
		\hline\hline
	\end{tabular}
	\caption{Parameters for the thermal Couette flow.}
	\label{TableI}
\end{table}

Figure \ref{Fig06} delineates the sketch of initial configuration for this problem. A premixed fluid flow with species, $\sigma = A$, $B$, $C$, is between two infinite parallel flat plates separated by a distance $H = 0.1$. The concentrations are ($n^{A}$, $n^{B}$, $n^{C}$) $=$ ($0.1$, $0.3$, $0.6$), the molar mass $m^{\sigma} = m_{0} = 1$, the temperature $T^{\sigma} = T_{0} = 1$, and the velocity $\mathbf{u}^{\sigma} = 0$. The upper plate moves horizontally at the speed $u_0 = 0.1$, while the lower plate keeps motionless. The nonequilibrium extrapolation scheme is imposed on the top and bottom, respectively \cite{Guo2002CP}. Periodic boundary conditions are applied for the left and right boundaries, respectively. Because the field is the same in the $y$ direction, the configuration is actually a $1$-D case. The mesh is chosen as $N_x \times N_y = 1 \times 200$, the spatial step $\Delta x = \Delta y = 5 \times 10^{-4}$, the temporal step $\Delta t = 2 \times 10^{-5}$, the parameters ($v^{\sigma}_{a}$, $v^{\sigma}_{b}$, ${\eta}^{\sigma}_{a}$, ${\eta}^{\sigma}_{a}$) $=$ ($1.5$, $1.8$, $1.6$, $2.5$), and the remaining parameters are listed in Table \ref{TableI}.

\begin{figure}
	\begin{center}
		\includegraphics[bbllx=0pt,bblly=0pt,bburx=483pt,bbury=616pt,width=0.4
		\textwidth]{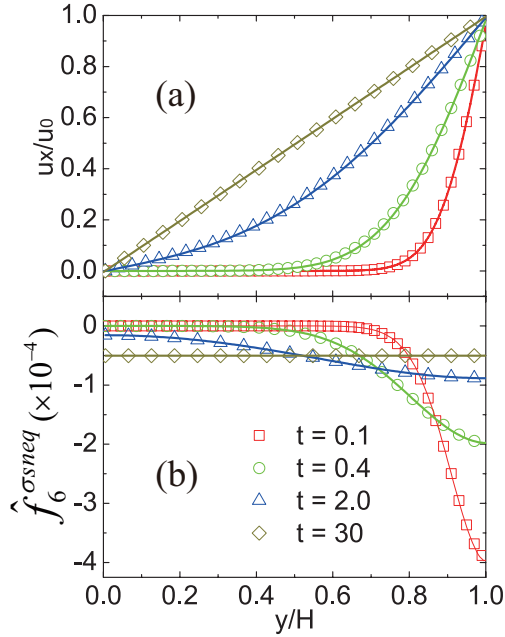}
	\end{center}
	\caption{Vertical distribution of the horizontal speed $u_x$ (a) and nonequilibrium quantity $\hat{f}_{6}^{A sneq}$ (b) in the thermal Couette flow. Squares, circles, triangles, and diamonds represent DBM results at time constants $t = 0.1$, $t = 0.4$, $t = 2.0$, and $t = 30$, respectively. Solid lines stand for the corresponding analytical solutions.}
	\label{Fig07}
\end{figure}

Five cases are under consideration with various values of the specific heat ratio and Prandtl number in Table \ref{TableI}. In the current DBM, the specific heat ratio of species $\sigma$ takes the form ${{\gamma }^{\sigma }}={\left( 4+{{I}^{\sigma }} \right)}/{\left( 2+{{I}^{\sigma }} \right)}$, and the Prandtl number of species $\sigma$ is 
${{\Pr }^{\sigma }}=S_{\kappa }^{\sigma }/S_{\mu }^{\sigma }$ under the conditions $S_{\mu }^{\sigma }=S_{5}^{\sigma }=S_{6}^{\sigma }=S_{7}^{\sigma }$ and $S_{\kappa }^{\sigma }=S_{8}^{\sigma }=S_{9}^{\sigma }$. Consequently, in terms of $I^{\sigma}$, $S^{\sigma}_{i = 5, 6, 7}$ and $S^{\sigma}_{i = 8, 9}$, we set $\Pr = 1.0$ and $\gamma = 1.3$, $1.4$, and $1.5$ for Runs I, II, and III, respectively. While the parameters are $\gamma = 1.4$ and $\Pr = 0.5$, $1.0$, and $2.0$ for Runs II, IV, and V, respectively. 

As an important parameter characterizing nonequilibrium flows, the Knudsen number is defined as ${\rm{Kn}} = \lambda / H$, where $H$ denotes the characteristic length scale, $\lambda = {v_s} \tau$ stands for the molecular mean-free-path, and $\tau = 1/{S^{\sigma}_{i \ne 5, 6, 7}}$ is the representative relaxation time. Hence, the Knudsen number is ${\rm{Kn}} = 0.0114$, $0.0118$, and $0.0122$ for ${\gamma} = 1.3$, $1.4$, and $1.5$, respectively. Moreover, the Mach number is defined as ${\rm{Ma}} = {u_0} / {v_s}$ with the sound speed ${v_s} = \sqrt{{\gamma} {T_0}/{m_0}}$. Thus, the Mach number is ${\rm{Ma}} = 0.0877$, $0.0845$, and $0.0816$ for ${\gamma} = 1.3$, $1.4$, and $1.5$, respectively. 

Firstly, we consider the case of Run IV in Table \ref{TableI}. Figure \ref{Fig07} (a) exhibits the comparisons between the numerical and analytical results of the horizontal speed along the $y$ axis at various time constants. Symbols represent numerical results, and lines represent the following analytical solutions \cite{Batchelor2000,Watari2003PRE},
\begin{equation}
	u=\frac{y}{H}u_{0}+\frac{2}{\pi}u_{0}\sum_{n=1}^{\infty}%
	\left[\frac{(-1)^{n}}{n}\exp \left(-n^{2}\pi^{2}\frac{\mu t}{\rho H^{2}} \right)\sin \left(\frac{n\pi y}{H}\right)\right] 
	\tt{,}
\end{equation}
where $\mu$ is the viscosity coefficient. Clearly, we can find a good agreement between them in the evolution of the thermal Couette flow. To further demonstrate its capability of measuring nonequilibrium manifestations, Fig. \ref{Fig07} (b) plots the vertical distribution of the nonequilibrium quantity $\hat{f}_{6}^{\sigma sneq}$ of species $\sigma = A$. Via the Chapman-Enskog analysis, we can obtain the analytical solution in Eq. (\ref{hatf6_sneq}). Obviously, the DBM results are consistent with the analytical solution in the thermal Couette flow. 

\begin{figure}
	\begin{center}
		\includegraphics[bbllx=0pt,bblly=0pt,bburx=372pt,bbury=508pt,width=0.4
		\textwidth]{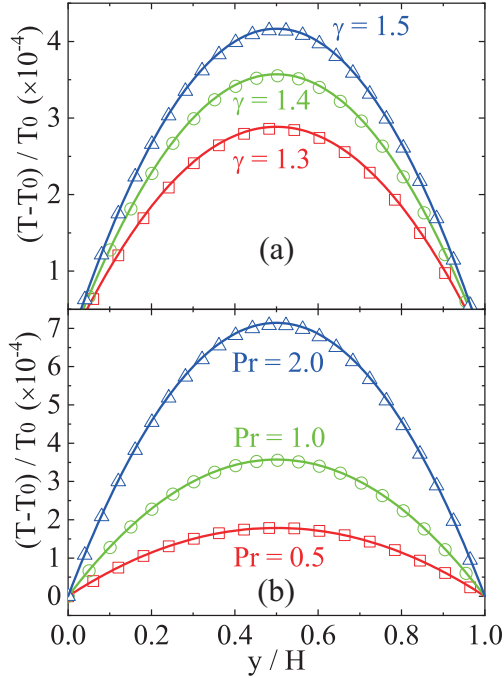}
	\end{center}
	\caption{Vertical distribution of the temperature in the steady Couette flow. (a) Cases with $\Pr = 1.0$ and $\gamma = 1.3$, $1.4$, and $1.5$, respectively. (b) Cases with $\gamma = 1.4$ and $\Pr  = 0.5$, $1.0$, and $2.0$, respectively.}
	\label{Fig08}
\end{figure}

Figure \ref{Fig08} shows the vertical distribution of the temperature when the thermal Couette flow achieves its steady state. In theory, the analytical solution reads \cite{Batchelor2000,Watari2003PRE},
\begin{equation}
	T=T_{0}+\frac{\Pr}{2 c_p} u_{0}^2 \frac{y}{H}\left(1-\frac{y}{H}\right)
	\tt{,}
\end{equation}
where $T_{0}$ is the temperature of the top/bottom wall, $c_p = \gamma c_v$ the specific heat at constant pressure, $c_v$ the specific heat at constant volume. Temperature depends upon the specific heat ratio and Prandtl number. Figure \ref{Fig08} (a) is for the cases with fixed $\Pr = 1.0$ and various $\gamma = 1.3$, $1.4$, and $1.5$, respectively. Figure \ref{Fig08} (b) is for the cases with fixed $\gamma = 1.4$ and various $\Pr  = 0.5$, $1.0$, and $2.0$, respectively. It is clear that simulation results match the corresponding analytical solutions for all cases. 

\subsection{Sod shock tube}

To verify the DBM for high-speed compressible flows, we consider a typical benchmark, the Sod shock tube that includes abundant and complex characteristic structures \cite{Sod1978JCP}. It is worth mentioning that, compared with single-component models, the current DBM is applicable to the Sod shock tube that contains various species (with different molar mass and/or specific-heat ratios, etc.) in different locations. As shown in Fig. \ref{Fig10}, the initial field reads,
\begin{equation}
	\left\{ 
	\begin{array}{l}
		{\left( {{n}^{A}}, {{n}^{B}}, {{n}^{C}}, p \right)}_{L}  = \left( 1.25, 0, 0, 1 \right) \tt{,}  \\ [6pt]  
		{{\left( {{n}^{A}}, {{n}^{B}}, {{n}^{C}}, p \right)}_{R}} = \left( 0, 0.0625, 0.0521, 0.1 \right) \tt{,}
	\end{array} \right.
\end{equation}
where the subscripts $L$ and $R$ denote the left part $-L_{0}/2 \le x < 0$ and right part $0 \le x < L_{0}/2$, respectively, with $L_{0} = 1.0$. Both parts are initially at rest, i.e. $\mathbf{u} = 0$. The molar mass is (${{m}^{A}}$, ${{m}^{B}}$, ${{m}^{C}}$) = ($0.8$, $1$, $1.2$). Consequently, it is easy to obtain $(\rho_{L}, \rho_{R}) = (1, 0.125)$ and $(T_{L}, T_{R}) = (0.8, 0.87273)$ in terms of $\rho = \sum\nolimits_{\sigma }{{{m }^{\sigma }} {{n }^{\sigma }}}$ and $T = p/\sum\nolimits_{\sigma } {{n }^{\sigma }}$. The specific-heat ratios are ($\gamma^{A}$, $\gamma^{B}$, $\gamma^{C}$) $=$ ($1.4$, $1.5$, $1.5$), and the parameters $S^{\sigma }_{i} = 2 \times 10 ^{4}$, ($v^{\sigma}_{a}$, $v^{\sigma}_{b}$, ${\eta}^{\sigma}_{a}$, ${\eta}^{\sigma}_{a}$) $=$ ($1.5$, $3.3$, $1.1$, $3.9$). The boundary conditions are the same with those in Fig. \ref{Fig04}. 

As numerical accuracy and robustness should be under consideration, we carry out simulations of the Sod shock tube with various spatial and temporal steps. Figure \ref{Fig10} plots density profiles at a time constant $t=0.2$ in the Sod shock tube. In Fig. \ref{Fig10} (a), the dashed, dotted, dash-dotted, and solid lines represents numerical results under spatial steps 
$\Delta x_{1} = 4 \times 10^{-3}$, 
$\Delta x_{2} = 2 \times 10^{-3}$, 
$\Delta x_{3} = 10^{-3}$, and 
$\Delta x_{4} = 5 \times 10^{-4}$, respectively.
The corresponding meshes are 
$N_x \times N_y = 250 \times 1$, 
$500 \times 1$, 
$1000 \times 1$, and
$2000 \times 1$, respectively. Besides, in Fig. \ref{Fig10} (b), the dashed, dotted, dash-dotted, and solid lines represents numerical results in cases with temporal steps
$\Delta t_{1} = 5 \times 10^{-5}$, 
$\Delta t_{2} = 2.5 \times 10^{-5}$, 
$\Delta t_{3} = 1.25 \times 10^{-5}$, and 
$\Delta t_{4} = 6.25 \times 10^{-6}$, respectively. The inset maps in Figs. \ref{Fig10} (a) and (b) are the enlargements of portions within $0.186 \le x \le 0.195$. 
It indicates that simulation results start to converge with decreasing spatial and temporal steps. Moreover, it can be found that the spatial step $\Delta x = 5 \times 10^{-4}$ and temporal step $2.5 \times 10^{-5}$, which are used in Fig. \ref{Fig11}, are small enough to give satisfactory simulation results. 

\begin{figure}
	\begin{center}
		\includegraphics[bbllx=0pt,bblly=0pt,bburx=158pt,bbury=96pt,width=0.35\textwidth]{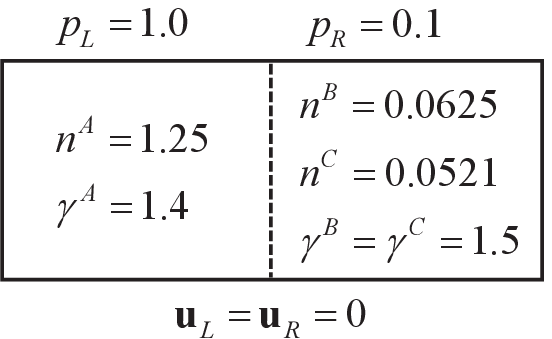}
	\end{center}
	\caption{Initial configuration of the Sod shock tube.}
	\label{Fig09}
\end{figure}
\begin{figure}
	\begin{center}
		\includegraphics[bbllx=22pt,bblly=7pt,bburx=570pt,bbury=210pt,width=0.99 \textwidth]{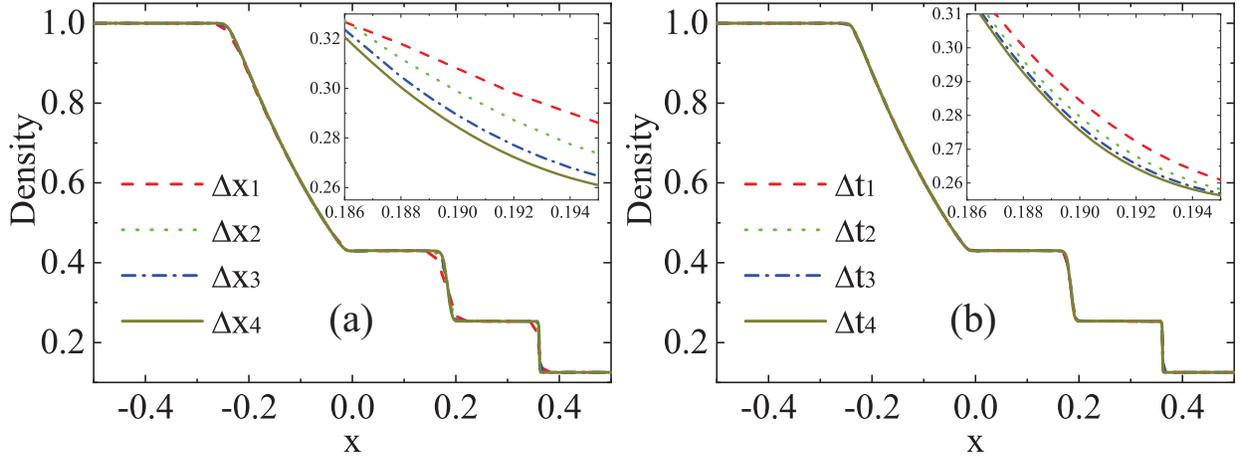}
	\end{center}
	\caption{Profiles of density at a time constant $t = 0.2$ in the Sod shock tube with various spatial steps (a) and temporal steps (b).}
	\label{Fig10}
\end{figure}
\begin{figure}
	\begin{center}
		\includegraphics[bbllx=0pt,bblly=0pt,bburx=395pt,bbury=263pt,width=0.9 \textwidth]{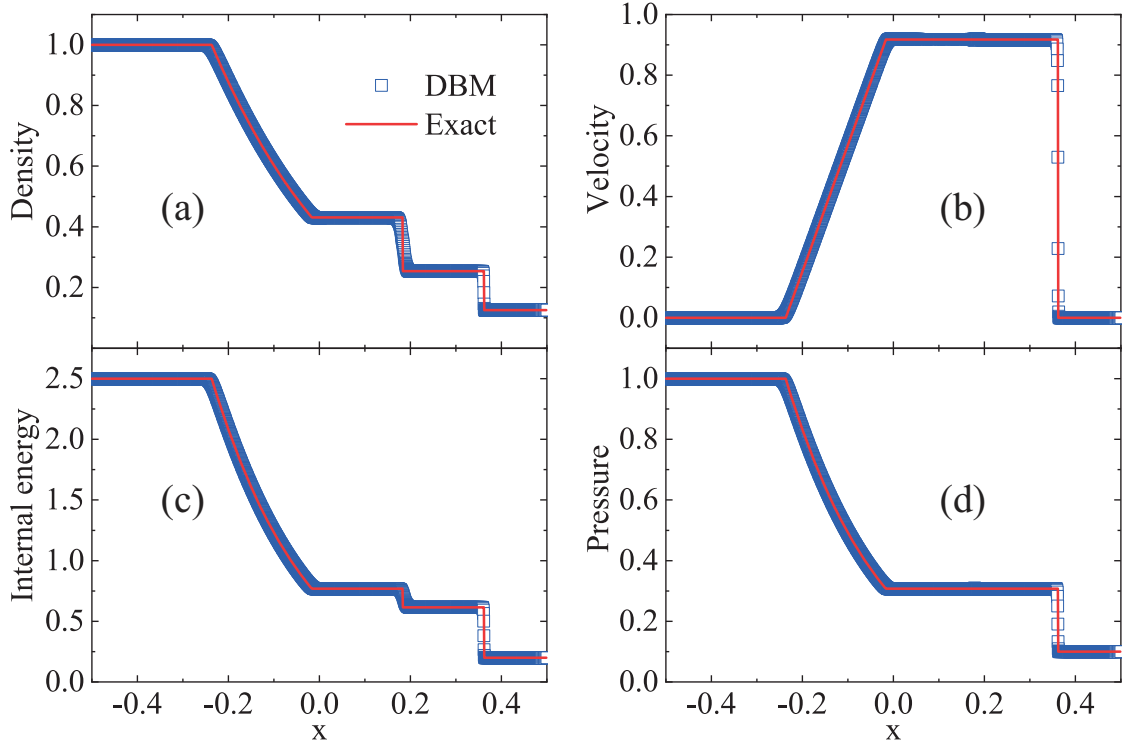}
	\end{center}
	\caption{Profiles of density (a), horizontal speed (b), temperature (c), and pressure (d) at a time constant $t = 0.2$ in the Sod shock tube. Symbols represent DBM results, and solid lines stand for Riemann solutions.}
	\label{Fig11}
\end{figure}

Figure \ref{Fig11} illustrates the density (a), horizontal speed (b), temperature (c), and pressure at a time constant $t = 0.2$ in the Sod shock tube. Symbols and lines stand for our DBM results and the Riemann solutions, respectively. As shown in Figs. \ref{Fig11} (a)-(d), the rarefaction wave (moving leftward), the contact discontinuity (between two media with different concentrations), and the (left-propagating) shock front are captured well. It is clear that the numerical and exact results coincide well with each other. 
For this problem, the Reynolds number is defined as ${\rm{Re}} = \rho_{c} u_{c} L_{c} / {\mu_{c}}$, where the characteristic density $\rho_{c} = 0.25340$, velocity $u_{c} = 0.91661$, dynamic viscosity $\mu_{c} = \sum\nolimits_{\sigma }{{{\mu }^{\sigma }}} = 0.30728 / (2 \times 10 ^{4})$ are behind the shock front, and the characteristic length equals the length of the shock tube $L_{c} = L_{0}$, hence ${\rm{Re}} =15117$. Besides, the Knudsen number is ${\rm{Kn}} = \lambda / L_{c} = 6.74 \times 10^{-5}$ in terms of the characteristic length scale $L_{c} = L_{0}$ and the molecular mean-free-path $\lambda = {v_s} \tau$, where $\tau = 1/{S^{\sigma }_{i}} = 5 \times 10^{-5}$ is the relaxation time and $v_s = 1.3487$ is the sound speed behind the shock wave. The Knudsen number is in the continuum regime (namely, the TNE is relatively weak), this is the physical reason why the DBM results (involving detailed TNE) agree with the exact solutions (without consideration of any TNE).

\section{Kelvin-Helmholtz instability} \label{SecIV}

As an essential physical mechanism in turbulence and fluids mixing process, the KHI has been studied extensively with experimental \cite{Wan2015PRL,Liu2015PNASUSA,Akula2017JFM}, theoretical \cite{Awasthi2014IJHMT,Liu2015IJHMT,Wang2017}, and computational \cite{Wang2009EPL,Wang2010POP,Lin2019CTP,Gan2019FOP} methods during the past decades. In this section, we further utilize the DBM to simulate and investigate the compressible KHI with both HNE and TNE. 

\begin{figure}
	\begin{center}
		\includegraphics[bbllx=0pt,bblly=0pt,bburx=257pt,bbury=136pt,width=0.35\textwidth]{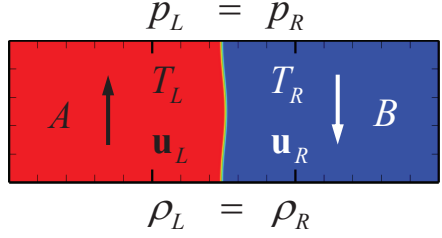}
	\end{center}
	\caption{Initial configuration of the KHI.}
	\label{Fig12}
\end{figure}

Figure \ref{Fig12} portrays the initial configuration for the KHI. The length and height of the calculation domain are $L_{x} = 1.5$ and $L_{y} = 0.5$, respectively. Initially, the left half part is occupied by upward-moving species $A$ with velocity $\mathbf{u}_{L} = 0.5 \mathbf{e}_{y}$, and the right is filled with $B$ travelling downwards with velocity $\mathbf{u}_{R} = - 0.5 \mathbf{e}_{y}$. To have an initial smooth interface, we impose a transition layer with width $W = L_{x}/300$ on the concentration and velocity fields across the interface. Moreover, to trigger the formation of the KHI, a sinusoidal perturbation, $w={{w}_{0}}\cos (2\pi y/{L}_{y})$, is imposed on the interface with an amplitude ${w}_{0} = L_{x}/200$. The concentration and velocity are expressed by, 
\[
\left\{
\begin{array}{l}
n=\frac{{n_{L}+n_{R}}}{2}-\frac{{n_{L}-n_{R}}}{2}\tanh (\frac{x-x_{0}+w}{W}) \tt{,}  \\ [6pt]  
\mathbf{u}=\frac{{\mathbf{u}_{L}+\mathbf{u}_{R}}}{2}-\frac{\mathbf{u}_{L}{-\mathbf{u}_{R}}}{2}\tanh (\frac{x-x_{0}+w}{W}) \tt{,} 
\end{array}
\right.
\]
where $x_{0}=L_{x}/2$ denotes the averaged $x$ position of the cosine-shaped interface, $n_{L}$ and $n_{R}$ are the concentrations in the left and right parts, respectively. Across the interface, pressure keeps homogeneous, i.e., $p_{L} = p_{R}$. The two species have an identical velocity and temperature at the same location. In addition, the specular reflection (periodic) boundary condition is used in the $x$ ($y$) direction. The time and space steps are as small as $\Delta t = 2.5 \times 10^{-5}$ and $\Delta x = \Delta y = 5 \times 10^{-4}$ to reduce numerical errors. Correspondingly, the mesh is $N_x \times N_y = 3000 \times 1000$. 

\begin{table}
	\begin{tabular}{lccc}
		\hline\hline
		Cases & (${T_L}$, ${T_R}$) & $\Pr^{\sigma}$ & (${\kappa}^{A}$, ${\kappa}^{B}$)  \\
		\hline
		Run I &($1$, $1$) & $0.25$ &  ($2.8$, $2.8$)$\times {10}^{-3}$  \\
		\hline
		Run II & ($1$, $1$) & $0.5$ & ($1.4$, $1.4$)$\times {10}^{-3}$  \\
		\hline
		Run III & ($1$, $1$) & $1.0$ & ($7.0$, $7.0$)$\times {10}^{-4}$  \\
		\hline
		Run IV & ($1$, $1$) & $2.0$ & ($3.5$, $3.5$)$\times {10}^{-4}$  \\
		\hline
		Run V & ($1$, $1$) & $4.0$ & ($1.75$, $1.75$)$\times {10}^{-4}$  \\
		\hline
		Run VI & ($1$, $2$) & $0.25$ & ($2.8$, $1.4$)$\times {10}^{-3}$  \\
		\hline
		Run VII & ($1$, $2$) & $0.5$ & ($1.4$, $0.7$)$\times {10}^{-3}$  \\
		\hline
		Run VIII & ($1$, $2$) & $1.0$ & ($7.0$, $3.5$)$\times {10}^{-4}$  \\
		\hline
		Run IX & ($1$, $2$) & $2.0$ & ($3.5$, $1.75$)$\times {10}^{-4}$  \\
		\hline
		Run X & ($1$, $2$) & $4.0$ & ($17.5$, $8.75$)$\times {10}^{-5}$  \\
		\hline\hline
	\end{tabular}
	\caption{Parameters for the KHI.}
	\label{TableII}
\end{table}

Next, let us study the influence of heat conduction upon the formation and evolution of the nonequilibrium KHI. To this end, ten representative cases are under consideration, see Table \ref{TableII}. 
For the first five cases, the temperatures in the two parts are equal, i.e., ${{T}_{L}} = {{T}_{R}} = 1.0$, the concentrations $n_{L}=n_{R}=1$, the molar mass $m^{\sigma} = 1$, and the parameters ($v^{\sigma}_{a}$, $v^{\sigma}_{b}$, ${\eta}^{\sigma}_{a}$, ${\eta}^{\sigma}_{a}$) $=$ ($2$, $3.7$, $1.5$, $5.5$). 
Moreover, the relaxation parameters are ${{S}_{8}} = {{S}_{9}} = 1.25 \times 10^{3}$, $2.5 \times 10^{3}$, $5.0 \times 10^{3}$, $1.0 \times 10^{4}$, and $2.0 \times 10^{4}$, respectively. The other relaxation parameters are ${{S}_{i}} = 5.0 \times 10^{3}$. The extra degrees of freedom $I^{\sigma} = 3$. Actually, in these cases, the initial dynamic viscosity is fixed, and the thermal conductivity is variable, i.e., ${\kappa}^{\sigma} = 2.8 \times {10}^{-3}$, $1.4 \times {10}^{-3}$, $7.0 \times {10}^{-4}$, $3.5 \times {10}^{-4}$, and $1.75\times {10}^{-4}$, respectively. In other words, the Prandtl number is variable in the five cases. 
In contrast, for the latter five cases, the temperatures in the two parts are different, namely, ${{T}_{L}} = 1.0$ and ${{T}_{R}} = 2.0$, the molar mass $m^{A} = 1$ and $m^{B} = 2$, and the parameters ($v^{\sigma}_{a}$, $v^{\sigma}_{b}$, ${\eta}^{\sigma}_{a}$, ${\eta}^{\sigma}_{a}$) $=$ ($1.4$, $2.8$, $5.0$, $2.5$). The particular thermal conductivity is (${\kappa}^{A}$, ${\kappa}^{B}$) $=$ ($2.8$, $1.4$)$\times {10}^{-3}$,
($1.4$, $0.7$)$\times {10}^{-3}$,
($7.0$, $3.5$)$\times {10}^{-4}$,
($3.5$, $1.75$)$\times {10}^{-4}$,
($17.5$, $8.75$)$\times {10}^{-5}$, respectively.
The other parameters in the latter five cases are the same with the former corresponding ones. Additionally, for all above cases, the density is homogeneous, i.e., ${\rho} = 1$ in the system, hence the Atwood number is a constant $At = ({\rho_{L}} - {\rho_{R}})/({\rho_{L}} + {\rho_{R}}) = 0$. 

\begin{figure}
	\begin{center}
		\includegraphics[bbllx=0pt,bblly=0pt,bburx=351pt,bbury=588pt,width=0.4\textwidth]{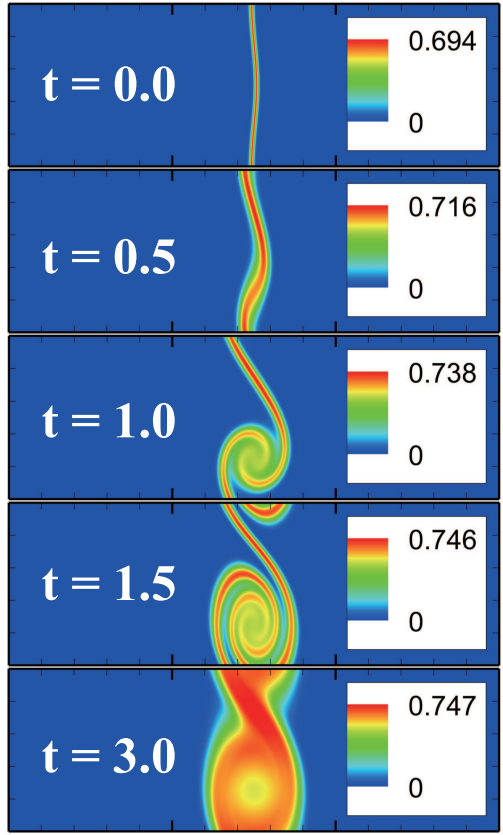}
	\end{center}
	\caption{Contours of the entropy of mixing at time constants $t = 0.0$, $0.5$, $1.0$, $1.5$, and $3.0$ in the evolution of KHI.}
	\label{Fig13}
\end{figure}

\begin{figure}
	\begin{center}
		\includegraphics[bbllx=0pt,bblly=0pt,bburx=488pt,bbury=285pt,width=0.99\textwidth]{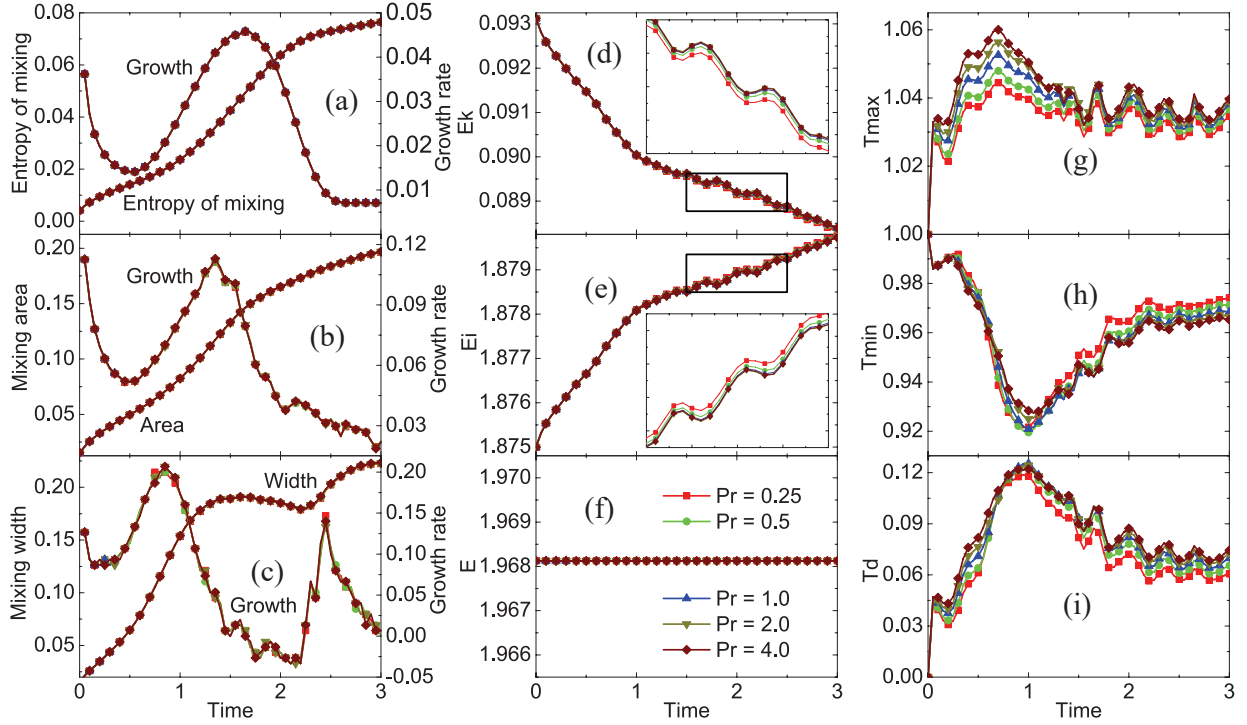}
	\end{center}
	\caption{Physical quantities in the evolution of KHI with ${T_L} = {T_R}$: (a) the entropy of mixing and its growth rate, (b) the mixing area and its growth rate, (c) the mixing width and its growth rate, (d) the kinetic energy, (e) the internal energy, (f) the total energy, (g) the maximum temperature, (h) the minimum temperature, (i) the temperature difference. The inserts in (d) and (e) correspond to the rectangles, respectively. The lines with squares, circles, upper triangles, lower triangles, and diamonds indicate $\Pr = 0.25$, $0.5$, $1.0$, $2.0$, and $4.0$, respectively.}
	\label{Fig14}
\end{figure}

\begin{figure}
	\begin{center}
		\includegraphics[bbllx=0pt,bblly=0pt,bburx=488pt,bbury=285pt,width=0.99\textwidth]{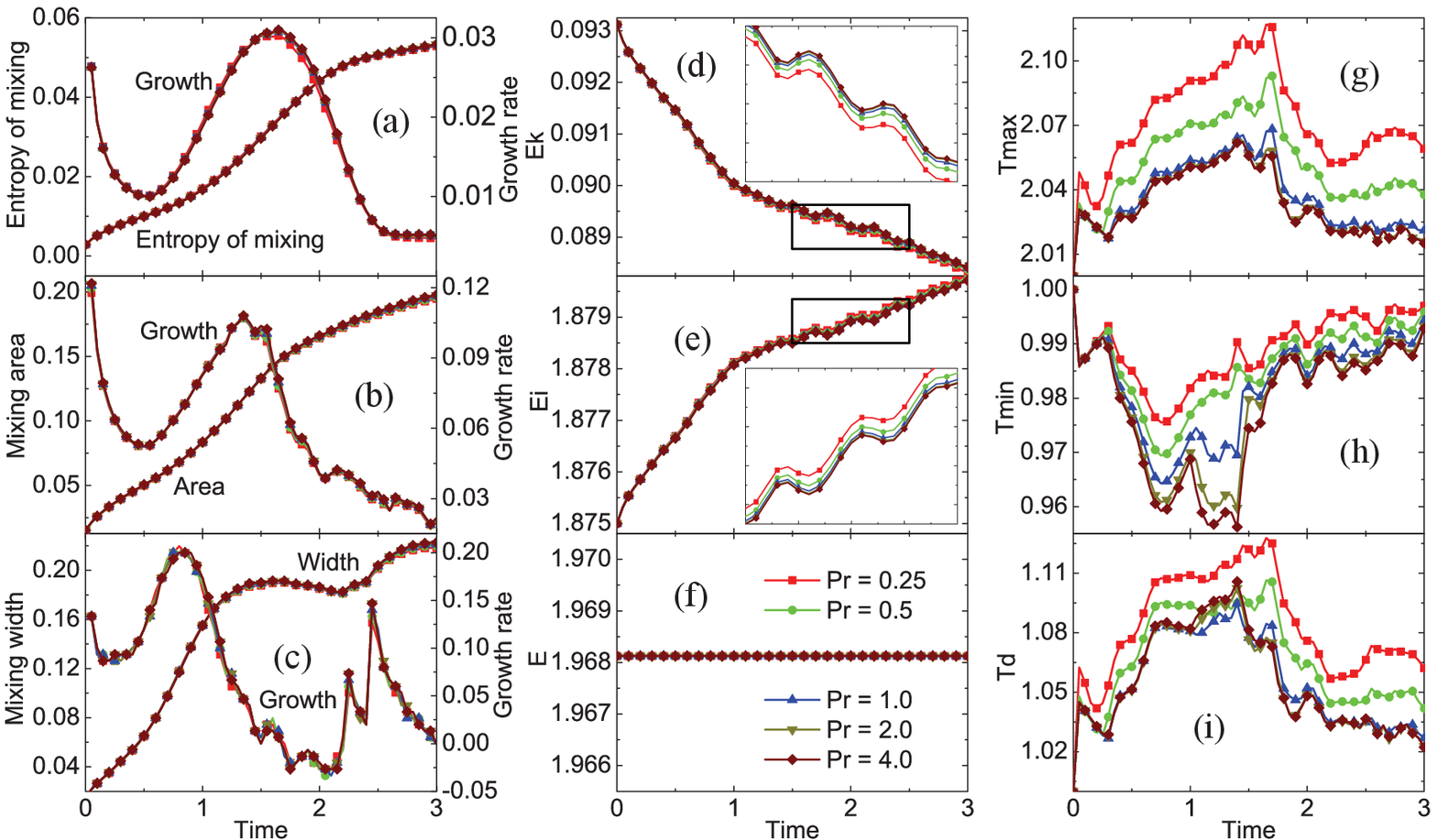}
	\end{center}
	\caption{Physical quantities in the evolution of KHI with ${T_L} \ne {T_R}$: (a) the entropy of mixing and its growth rate, (b) the mixing area and its growth rate, (c) the mixing width and its growth rate, (d) the kinetic energy, (e) the internal energy, (f) the total energy, (g) the maximum temperature, (h) the minimum temperature, (i) the temperature difference. The inserts in (d) and (e) correspond to the rectangles, respectively. The lines with squares, circles, upper triangles, lower triangles, and diamonds indicate $\Pr = 0.25$, $0.5$, $1.0$, $2.0$, and $4.0$, respectively.}
	\label{Fig15}
\end{figure}

To give an intuitive impression, we take Run I for example and depict the entropy of mixing in the evolution of KHI in Fig. \ref{Fig13}. From top to bottom are its contours at time constants $t = 0.0$, $0.5$, $1.0$, $1.5$, and $3.0$, respectively. It is clear to find a sequence of distinct evolutionary stages, namely, the initial linear growth period, then the nonlinear growth stage, the later time with a highly rolled-up vortex, and finally a 
sufficiently mixed phase with nonregular structures.
To be specific, firstly, the smooth interface starts to wiggle due to the initial perturbation and the velocity shear between the two layers. At the early stage, the perturbation grows exponentially in accordance with the linear stability theory	(see Figs. \ref{Fig14} and \ref{Fig15}), and the sinusoidal structure gradually becomes asymmetric.
Then, in the nonlinear stage, a braid-shape region is formed and a roughly circular vortex appears. Subsequently, the vortex becomes elliptical with its roll-up movement and it is further stretched in the vertical direction. In the final phase, 
with the development of the vortex, the rotating movements promote the mixing between the two parts until its saturation state (see Figs. \ref{Fig14} and \ref{Fig15}).

Figure \ref{Fig14} displays the evolution of physical quantities for the first five cases in Table \ref{TableII}. The lines with squares, circles, upper triangles, lower triangles, and diamonds stand for the Prandtl number $\Pr = 0.25$, $0.5$, $1.0$, $2.0$, and $4.0$, respectively. Figure \ref{Fig14} (a) shows the whole entropy of mixing $\int \int{{{S}_{M}}dxdy}$ and its growth rate. Here the integral is extended over the physical region $L_{x} \times L_{y}$. Figure \ref{Fig14} (b) exhibits the value of ${S_a}/({L_x}{L_y})$ and its growth rate, with the mixing area ${S_a}$ where the mass fraction of species $A$ is within the range $1 \% \le {{\lambda }^{A }} \le 99 \%$. Figure \ref{Fig14} (c) gives the value of ${L_M}/{L_x}$ and its growth rate. Here the mixing width ${L_M}$ is defined as the horizontal distance between the leftmost and rightmost points within the region $1 \% \le {{\lambda }^{A }} \le 99 \%$. It is clear in Figs. \ref{Fig14} (a)-(c) that the mixing degree, area, and width coincide well with each other in the five cases. 

With the definition of the kinetic energy ${{E}_{k}} = \frac{1}{2} \rho |\mathbf{u}|^{2}$, Fig. \ref{Fig14} (d) plots the whole kinetic energy $\int \int{{{E}_{k}}dxdy}$. With the introduction of the internal energy ${{E}_{i}}=\frac{1}{2}\sum\nolimits_{\sigma }{(D+{{I}^{\sigma }})}{{n}^{\sigma }}T$, we show the whole internal energy $\int \int{{{E}_{i}}dxdy}$ and its growth rate in Fig. \ref{Fig14} (e). The inserts in Figs. \ref{Fig14} (d) and (e) are enlargements of the portions in the corresponding rectangles. It can be found that the kinetic and internal energies in the five cases are almost the same with each other, and their differences are very small. The kinetic (internal) energy becomes only a little larger (smaller) with the increasing Prandtl number, i.e., the decreasing thermal conductivity. Figure \ref{Fig14} (f) plots the whole energy $\int \int{{E}dxdy}$ in terms of ${E} = {{E}_{k}} + {{E}_{i}}$. It is evident that the energy is a conserved quantity in the KHI process. For instance, in the first case, our DBM gives $\int \int{{E}dxdy} =  \int \int{{E_{k}}dxdy} + \int \int{{E_{i}}dxdy} = 0.0883230 + 1.8798020$ at the time $t = 3$, which equals its initial result $\int \int{{E}dxdy} =  0.0931250+ 1.8750000$. It is noteworthy that, apart from the energy conservation, the mass and momentum conservation is ensured by the DBM as well (which is not shown here). 

Figures \ref{Fig14} (g)-(i) are for the maximum temperature ${T_{max}}$, the minimum temperature ${T_{min}}$, and their difference ${T_{d}} = {T_{max}} - {T_{min}}$. On the whole, the maximum temperature is smaller for larger thermal conductivity. The minimum temperature with various Prandtl numbers competes with each other before the time $t = 1.5$, afterwards it is larger for larger thermal conductivity. Hence, the temperature difference becomes smaller with the increasing thermal conductivity that facilitates heat exchange. 

Figure \ref{Fig15} exhibits the evolution of physical quantities for the latter five cases in Table \ref{TableII}. In the following, comparison is made between Figs. \ref{Fig14} and \ref{Fig15}. The former is for the cases in an initial homogeneous temperature field, while the latter initially has a temperature difference between the left and right half parts of the physical domain. Some findings are listed as follows. 

(I) From Figs. \ref{Fig14} (a)-(c) and Figs. \ref{Fig15} (a)-(c), it is apparent that the whole entropy of mixing, the mixing area, the mixing width, and their growth rates for various Prandtl numbers basically coincide with each other. That is to say, the heat conduction has a weak effect on the formation and evolution of the KHI for the parameter range here we considered.

(II) It can be found in Figs. \ref{Fig14} (d)-(e) and Figs. \ref{Fig15} (d)-(e) that, the kinetic and internal energies have slight differences for various Prandtl numbers. The inserts show that, for either ${T_L} = {T_R}$ or ${T_L} \ne {T_R}$, the kinetic (internal) energy becomes only a bit smaller (larger) with the reducing Prandtl number, i.e., the increasing thermal conductivity. 

(III) The energy conservation is held in the DBM simulation, which is validated in Fig. \ref{Fig14} (f) and Fig. \ref{Fig15} (f). Take Run X in Table \ref{TableII} for instance, the simulation result remains $\int \int{{E}dxdy} = 1.96813$, which is exactly equal to its exact solution $1.96813$. Actually, the mass and momentum conservation is also obeyed by the DBM (which is not shown here). 

(IV) Comparison between Figs. \ref{Fig14} (g)-(i) and Figs. \ref{Fig15} (g)-(i) shows that the maximum and minimum temperatures and their differences for ${T_L} = {T_R}$ are quite different from those for ${T_L} \ne {T_R}$. In Figs. \ref{Fig15} (g)-(i), both maximum and minimum temperatures, and their differences on the whole are larger for a larger thermal conductivity. 

(V) Although the evolutionary temperature fields are quite different for various Prandtl numbers, the mixing process is almost the same for homogeneous or inhomogeneous initial temperature configuration. Consequently, the temperature plays a nonessential role in the formation and evolution of the KHI. 

\begin{figure}
	\begin{center}
		\includegraphics[bbllx=0pt,bblly=0pt,bburx=424pt,bbury=323pt,width=0.99\textwidth]{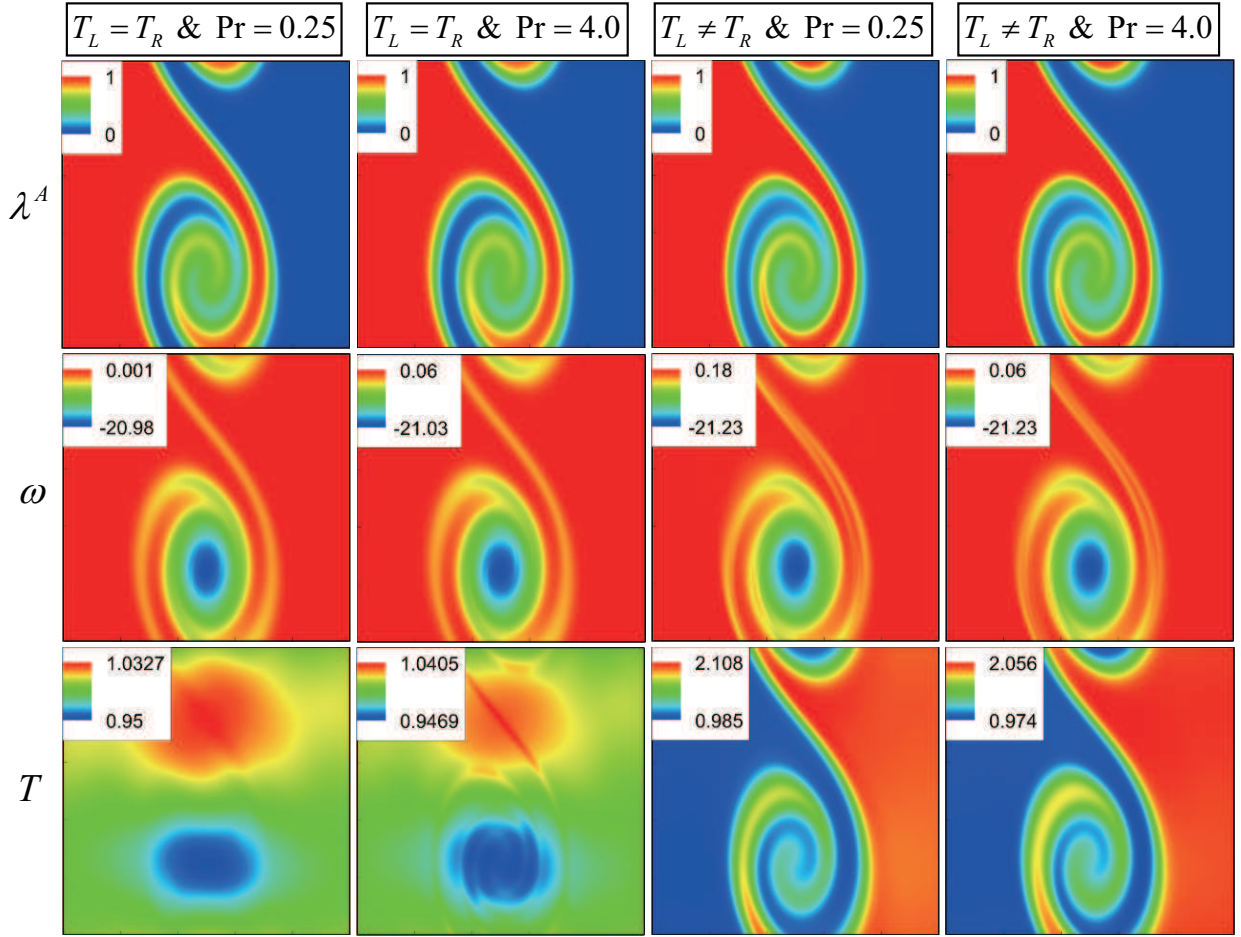}
	\end{center}
	\caption{Comparison of physical fields at the time $t = 1.5$ in the KHI process. From top to bottom are the mass fraction of species $A$, vorticity, and temperature in the three rows, respectively. From left to right are the cases (${T_{L}} = {T_{R}}$ and $\Pr = 0.25$), (${T_{L}} = {T_{R}}$ and $\Pr = 4.0$), (${T_{L}} \ne {T_{R}}$ and $\Pr = 0.25$), and (${T_{L}} \ne {T_{R}}$ and $\Pr = 4.0$) in the four columns, respectively. Only a part of horizontal range $0.5 \le x \le 1.0$ is shown in each subfigure.}
	\label{Fig16}
\end{figure}

Finally, for the sake of validating above conclusions again, let us compare the specific KHI fields in four representative cases, i.e., 
Run I (${T_{L}} = {T_{R}}$ and $\Pr = 0.25$), 
Run V (${T_{L}} = {T_{R}}$ and $\Pr = 4.0$), 
Run VI (${T_{L}} \ne {T_{R}}$ and $\Pr = 0.25$), and
Run X (${T_{L}} \ne {T_{R}}$ and $\Pr = 4.0$), respectively. Figure \ref{Fig16} depict the contours of physical fields at a time constant $t = 1.5$ in the KHI process. The four cases are shown from left to right columns, respectively. The mass fraction (${\lambda}^{A}$), the vorticity ($\omega = \partial _x u_y - \partial_y u_x$), and the temperature ($T$) are plotted from top to bottom rows, respectively. Only a part of the physical domain $0.5 \le x \le 1.0$ and $0 \le y \le 0.5$ is shown in each subfigure. 
Obviously, the fields of mass fraction and vorticity are almost the same (with negligible differences) for all cases. Their shapes and sizes are very similar, despite few differences of the vorticity maxima and minima in the four cases. On the contrary, the contours of temperature fields are similar for the same initial configurations, and are distinguishable for different initial configurations. It is further confirmed that neither temperature nor thermal conductivity has a strong influence on the mass fraction and vorticity in the KHI process. From the point view of mixing state (such as mixing area and degree) and flow state (including the vortex shapes and sizes), the temperature and thermal conductivity play inessential roles in the spatio-temporal  evolution of the KHI.

\section{Conclusions and Discussions}\label{SecV}

We presented an MRT DBM for compressible multicomponent mixtures with both HNE and TNE. Physically, the DBM formulation is not only consistent with the NS equations, Fick's law and Stefan-Maxwell diffusion equation under corresponding conditions in the continuum limit, but also provides more detailed kinetic thermodynamic nonequilibrium information. Such a capability of the DBM allows the study of nonequilibrium processes like the entropy production. Mathematically, a set of uniform discrete Boltzmann equations are used to describe multicomponent mixtures, and the linear form of evolution equations makes it easy to code. Computationally, it can be implemented on massively parallel clusters with excellent scalability because all information transfer in DBM is local in time and space. 

In addition, several prototype problems, including the three-component diffusion, thermal Couette flow, and Sod shock tube, are simulated to verify and validate the model. It is demonstrated that the present DBM is suitable for both low and high speed compressible nonequilibrium flows, with premixed or nonpremixed chemical species, whose specific heat ratio and Prandtl number are adjustable. Various detailed TNE in complex fluid flows can be captured, measured, and predicted effectively by the current versatile kinetic model. 

Furthermore, the current model is utilized to investigate the compressible KHI with TNE. Ten cases with various values of thermal conductivity and initial temperature configurations are compared and analyzed. It is found that the mixing state (such as the mixing area and degree) and flow state (including the vortex shapes and sizes) are quite similar for all cases in the dynamic KHI process, although the temperature is similar for the same initial configurations and is distinguishable for different initial configurations. The whole kinetic (internal) energy becomes only a bit smaller (larger) with the increasing thermal conductivity. It is concluded that both heat conduction and temperature exert slight influences on the formation and evolution of the KHI, which is absolutely different from previous studies for single component fluids \cite{Awasthi2014IJHMT,Liu2015IJHMT,Wang2009EPL,Wang2010POP,Gan2019FOP}. 

Moreover, the temperature field shows different trends in cases with or without spatial variation of temperature across the material interface in an initial configuration. To be specific, for the initial homogeneous temperature, the maximum temperature is smaller for larger thermal conductivity as a whole, while the minimum temperature with various Prandtl numbers competes with each other in the early stage and is larger for larger thermal conductivity afterwards. For the initial inhomogeneous temperature, both maximum and minimum temperatures, and their differences on the whole are larger for larger thermal conductivity. 

\section*{Data Availability Statement}
The data that support the findings of this study are available from the corresponding author upon reasonable request.

\begin{acknowledgments}
	This work is supported by the Natural Science Foundation of China (NSFC) under Grant Nos. 51806116, 91441120, 11772064 and 11875001, CAEP Foundation under Grant No. CX2019033, the opening project of State Key Laboratory of Explosion Science and Technology (Beijing Institute of Technology) under Grant No. KFJJ19-01M, and the Natural Science Foundation of Fujian Provinces under Grant No. 2018J01654. Support from the UK Engineering and Physical Sciences Research Council under the project “UK Consortium on Mesoscale Engineering Sciences (UKCOMES)” (Grant No. EP/R029598/1) is also gratefully acknowledged. 
\end{acknowledgments}

\appendix

\section{}\label{APPENDIXA}

In essence, to choose the discretization of velocities (e.g., Fig.\ref{Fig01}) is a process of determining the calculation of discrete (equilibrium) distribution functions, wherein the order of physical accuracy is specified. Actually, the physical accuracy is directly related to the kinetic moment relations. (The Boltzmann equation is equivalent to an infinite list of coupled moment equations \cite{Struchtrup2005Book}.) The more the moment relations, the higher the physical accuracy. 

In the current work, there are $16$ moment relations satisfied by the discrete equilibrium distribution functions ${f_{i}^{\sigma eq}}$ as below,
\begin{equation}
	\sum\nolimits_{i}{f_{i}^{\sigma eq}}=\int{\int{{{f}^{\sigma eq}}d\mathbf{v}d\eta }}
	\label{feqM1}
	\tt{,}
\end{equation}
\begin{equation}
	\sum\nolimits_{i}{f_{i}^{\sigma eq}v_{i\alpha }^{\sigma }}=\int{\int{{{f}^{\sigma eq}}{{v}_{\alpha }}d\mathbf{v}d\eta }}
	\label{feqM2}
	\tt{,}
\end{equation}
\begin{equation}
	\sum\nolimits_{i}{f_{i}^{\sigma eq}\left( v_{i}^{\sigma 2}+\eta _{i}^{\sigma 2} \right)}=\int{\int{{{f}^{\sigma eq}}\left( {{v}^{2}}+{{\eta }^{2}} \right)d\mathbf{v}d\eta }}
	\label{feqM3}
	\tt{,}
\end{equation}
\begin{equation}
	\sum\nolimits_{i}{f_{i}^{\sigma eq}v_{i\alpha }^{\sigma }v_{i\beta }^{\sigma }}=\int{\int{{{f}^{\sigma eq}}{{v}_{\alpha }}{{v}_{\beta }}d\mathbf{v}d\eta }}
	\label{feqM4}
	\tt{,}
\end{equation}
\begin{equation}
	\sum\nolimits_{i}{f_{i}^{\sigma eq}\left( v_{i}^{\sigma 2}+\eta _{i}^{\sigma 2} \right)v_{i\alpha }^{\sigma }}=\int{\int{{{f}^{\sigma eq}}\left( {{v}^{2}}+{{\eta }^{2}} \right){{v}_{\alpha }}d\mathbf{v}d\eta }}
	\label{feqM5}
	\tt{,}
\end{equation}
\begin{equation}
	\sum\nolimits_{i}{f_{i}^{\sigma eq}v_{i\alpha }^{\sigma }v_{i\beta }^{\sigma }v_{i\chi }^{\sigma }}=\int{\int{{{f}^{\sigma eq}}{{v}_{\alpha }}{{v}_{\beta }}{{v}_{\chi }}d\mathbf{v}d\eta }}
	\label{feqM6}
	\tt{,}
\end{equation}
\begin{equation}
	\sum\nolimits_{i}{f_{i}^{\sigma eq}\left( v_{i}^{\sigma 2}+\eta _{i}^{\sigma 2} \right)v_{i\alpha }^{\sigma }v_{i\beta }^{\sigma }}=\int{\int{{{f}^{\sigma eq}}\left( {{v}^{2}}+{{\eta }^{2}} \right){{v}_{\alpha }}{{v}_{\beta }}d\mathbf{v}d\eta }}
	\label{feqM7}
	\tt{,}
\end{equation}
where the equilibrium distribution function reads
\begin{equation}
	{{f}^{\sigma eq}}={{n}^{\sigma }}{{\left( \frac{{{m}^{\sigma }}}{2\pi T} \right)}^{D/2}}{{\left( \frac{{{m}^{\sigma }}}{2\pi {{I}^{\sigma }}T} \right)}^{1/2}}\exp \left[ -\frac{{{m}^{\sigma }}{{\left| \mathbf{v}-\mathbf{u} \right|}^{2}}}{2T}-\frac{{{m}^{\sigma }}{{\eta }^{2}}}{2{{I}^{\sigma }}T} \right]
	\tt{.}
\end{equation}
Mathematically, Eqs. (\ref{feqM1}) - (\ref{feqM7}) can be expressed in a uniform form (\ref{Moment_feq}), which leads to the solution of the discrete equilibrium distribution functions,
${{\mathbf{f}}^{\sigma eq}}=
{{{\mathbf{M}}^{\sigma }}^{-1}}
{{\mathbf{\hat{f}}}^{\sigma eq}}$.

The square matrix ${{\mathbf{M}}^{\sigma }}$ has $16 \times 16$ elements: ${{M}^{\sigma}_{1i}}=1$, $M_{2i}^{\sigma }=v_{ix}^{\sigma }$, $M_{3i}^{\sigma }=v_{iy}^{\sigma }$, $M_{4i}^{\sigma }=v_{i}^{\sigma 2}+\eta _{i}^{\sigma 2}$, $M_{5i}^{\sigma }=v_{ix}^{\sigma 2}$, 
$M_{6i}^{\sigma }=v_{ix}^{\sigma }v_{iy}^{\sigma }$, ${{M}^{\sigma}_{7i}}=v_{iy}^{2}$, $M_{8i}^{\sigma }=\left( v_{i}^{\sigma 2}+\eta _{i}^{\sigma 2} \right)v_{ix}^{\sigma }$, $M_{9i}^{\sigma }=\left( v_{i}^{\sigma 2}+\eta _{i}^{\sigma 2} \right)v_{iy}^{\sigma }$, 
$M_{10i}^{\sigma }=v_{ix}^{\sigma 3}$, $M_{11i}^{\sigma }=v_{ix}^{\sigma 2}v_{iy}^{\sigma }$, $M_{12i}^{\sigma }=v_{ix}^{\sigma }v_{iy}^{\sigma 2}$, $M_{13i}^{\sigma }=v_{iy}^{\sigma 3}$, 
$M_{14i}^{\sigma }=\left( v_{i}^{\sigma 2}+\eta _{i}^{\sigma 2} \right)v_{ix}^{\sigma 2}$, $M_{15i}^{\sigma }=\left( v_{i}^{\sigma 2}+\eta _{i}^{\sigma 2} \right)v_{ix}^{\sigma }v_{iy}^{\sigma }$, $M_{16i}^{\sigma }=\left( v_{i}^{\sigma 2}+\eta _{i}^{\sigma 2} \right)v_{iy}^{\sigma 2}$.

The column matrix ${{\mathbf{\hat{f}}}^{\sigma eq}}$ has $16$ elements: $\hat{f}_{1}^{\sigma eq}={{n}^{\sigma }}$, $\hat{f}_{2}^{\sigma eq}={{n}^{\sigma }}{{u}_{x}}$, $\hat{f}_{3}^{\sigma eq}={{n}^{\sigma }}{{u}_{y}}$, $\hat{f}_{4}^{\sigma eq}={{n}^{\sigma }}\left[ \left( D+{{I}^{\sigma }} \right){T}/{{{m}^{\sigma }}}\;+{{u}^{2}} \right]$, 
$\hat{f}_{5}^{\sigma eq}={{n}^{\sigma }}\left( {T}/{{{m}^{\sigma }}}\;+u_{x}^{2} \right)$, $\hat{f}_{6}^{\sigma eq}={{n}^{\sigma }}{{u}_{x}}{{u}_{y}}$, $\hat{f}_{7}^{\sigma eq}={{n}^{\sigma }}\left( {T}/{{{m}^{\sigma }}}\;+u_{y}^{2} \right)$,  
$\hat{f}_{8}^{\sigma eq}={{n}^{\sigma }}{{\xi }^{\sigma }}{{u}_{x}}$, $\hat{f}_{9}^{\sigma eq}={{n}^{\sigma }}{{\xi }^{\sigma }}{{u}_{y}}$, 
$\hat{f}_{10}^{\sigma eq}=3{{n}^{\sigma }}{{u}_{x}}{T}/{{{m}^{\sigma }}}\;+{{n}^{\sigma }}u_{x}^{3}$, $\hat{f}_{11}^{\sigma eq}={{n}^{\sigma }}{{u}_{y}}{T}/{{{m}^{\sigma }}}\;+{{n}^{\sigma }}u_{x}^{2}{{u}_{y}}$, 
$\hat{f}_{12}^{\sigma eq}={{n}^{\sigma }}{{u}_{x}}{T}/{{{m}^{\sigma }}}\;+{{n}^{\sigma }}{{u}_{x}}u_{y}^{2}$, $\hat{f}_{13}^{\sigma eq}=3{{n}^{\sigma }}{{u}_{y}}{T}/{{{m}^{\sigma }}}\;+{{n}^{\sigma }}u_{y}^{3}$, 
$\hat{f}_{14}^{\sigma eq}={{n}^{\sigma }}{{\xi }^{\sigma }}{T}/{{{m}^{\sigma }}}\;+{{n}^{\sigma }}u_{x}^{2}\left( {{\xi }^{\sigma }}+2{T}/{{{m}^{\sigma }}}\; \right)$, $\hat{f}_{15}^{\sigma eq}={{n}^{\sigma }}{{u}_{x}}{{u}_{y}}\left( {{\xi }^{\sigma }}+2{T}/{{{m}^{\sigma }}}\; \right)$, 
$\hat{f}_{16}^{\sigma eq}={{n}^{\sigma }}{{\xi }^{\sigma }}{T}/{{{m}^{\sigma }}}\;+{{n}^{\sigma }}u_{y}^{2}\left( {{\xi }^{\sigma }}+2{T}/{{{m}^{\sigma }}}\; \right)$, with ${{\xi }^{\sigma }}=\left( D+{{I}^{\sigma }}+2 \right){T}/{{{m}^{\sigma }}}\;+{{u}^{2}}$. 

Moreover, the expression and moment relations of ${f_{i}^{\sigma seq}}$ are obtained in a similar way (which is not shown here for brevity).  
The column matrix ${{\mathbf{\hat{f}}}^{\sigma seq}}$ has $16$ elements: $\hat{f}_{1}^{\sigma seq}={{n}^{\sigma }}$, $\hat{f}_{2}^{\sigma seq}={{n}^{\sigma }}{{u}_{x}^{\sigma}}$, $\hat{f}_{3}^{\sigma seq}={{n}^{\sigma }}{{u}_{y}^{\sigma}}$, $\hat{f}_{4}^{\sigma seq}={{n}^{\sigma }}\left[ \left( D+{{I}^{\sigma }} \right){{T}^{\sigma}}/{{{m}^{\sigma }}}\;+{{u}^{\sigma 2}} \right]$, 
$\hat{f}_{5}^{\sigma seq}={{n}^{\sigma }}\left( {{T}^{\sigma}}/{{{m}^{\sigma }}}\;+u_{x}^{\sigma 2} \right)$, $\hat{f}_{6}^{\sigma seq}={{n}^{\sigma }}{{u}_{x}^{\sigma}}{{u}_{y}^{\sigma}}$, $\hat{f}_{7}^{\sigma seq}={{n}^{\sigma }}\left( {{T}^{\sigma}}/{{{m}^{\sigma }}}\;+u_{y}^{\sigma 2} \right)$,  
$\hat{f}_{8}^{\sigma seq}={{n}^{\sigma }}{{\xi }^{\sigma s}}{{u}_{x}^{\sigma}}$, $\hat{f}_{9}^{\sigma seq}={{n}^{\sigma }}{{\xi }^{\sigma s}}{{u}_{y}^{\sigma}}$, 
$\hat{f}_{10}^{\sigma seq}=3{{n}^{\sigma }}{{u}_{x}^{\sigma}}{{T}^{\sigma}}/{{{m}^{\sigma }}}\;+{{n}^{\sigma }}u_{x}^{\sigma 3}$, $\hat{f}_{11}^{\sigma seq}={{n}^{\sigma }}{{u}_{y}^{\sigma}}{{T}^{\sigma}}/{{{m}^{\sigma }}}\;+{{n}^{\sigma }}u_{x}^{\sigma 2}{{u}_{y}^{\sigma}}$, 
$\hat{f}_{12}^{\sigma seq}={{n}^{\sigma }}{{u}_{x}^{\sigma}}{{T}^{\sigma}}/{{{m}^{\sigma }}}\;+{{n}^{\sigma }}{{u}_{x}^{\sigma}}u_{y}^{\sigma 2}$, $\hat{f}_{13}^{\sigma seq}=3{{n}^{\sigma }}{{u}_{y}^{\sigma}}{{T}^{\sigma}}/{{{m}^{\sigma }}}\;+{{n}^{\sigma }}u_{y}^{\sigma 3}$, 
$\hat{f}_{14}^{\sigma seq}={{n}^{\sigma }}{{\xi }^{\sigma s}}{{T}^{\sigma}}/{{{m}^{\sigma }}}\;+{{n}^{\sigma }}u_{x}^{\sigma 2}\left( {{\xi }^{\sigma s}}+2{{T}^{\sigma}}/{{{m}^{\sigma }}}\; \right)$, $\hat{f}_{15}^{\sigma seq}={{n}^{\sigma }}{{u}_{x}^{\sigma}}{{u}_{y}^{\sigma}}\left( {{\xi }^{\sigma s}}+2{{T}^{\sigma}}/{{{m}^{\sigma }}}\; \right)$, 
$\hat{f}_{16}^{\sigma seq}={{n}^{\sigma }}{{\xi }^{\sigma s}}{{T}^{\sigma}}/{{{m}^{\sigma }}}\;+{{n}^{\sigma }}u_{y}^{\sigma 2}\left( {{\xi }^{\sigma s}}+2{{T}^{\sigma}}/{{{m}^{\sigma }}}\; \right)$, with ${{\xi }^{\sigma s}}=\left( D+{{I}^{\sigma }}+2 \right){{T}^{\sigma}}/{{{m}^{\sigma }}}\;+{{u}^{\sigma 2}}$. 

It is worth mentioning that there are $16$ discrete velocities and discrete (equilibrium) distribution functions. Correspondingly, there are only $16$ sets of discrete Boltzmann equations (\ref{DBEquation}). Obviously, this type of methodology is economic. To achieve the same order of physical accuracy (namely, to have the same moment relations), more discrete velocities, discrete (equilibrium) distribution functions, and discrete Boltzmann equations are required in other kinetic models. For example, there are $65$ discrete velocities in a finite difference LBM proposed by Watari\cite{Watari2007PA}, and much more are needed in the discrete velocity model \cite{Mieussens2000JCP}.

\section{}\label{APPENDIXB}

Let us give the NS equations recovered from the DBM in the continuum limit via the Chapman-Enskog analysis. The Einstein summation convention is adopted here. The NS equations of individual species take the form,
\begin{equation}
	{{\partial }_{t}}{{\rho }^{\sigma }}+{{\partial }_{\alpha }}J_{\alpha }^{\sigma }=0
	\label{NSequation1}
	\tt{,}
\end{equation}
\begin{equation}
	{{\partial }_{t}}{J_{\alpha }^{\sigma }}+{{\partial }_{\beta }}\left( {{\delta }_{\alpha \beta }}{{p}^{\sigma }}+{{\rho }^{\sigma }}u_{\alpha }^{\sigma }u_{\beta }^{\sigma }+P_{\alpha \beta }^{\sigma }+U_{\alpha \beta }^{\sigma } \right)=S_{J\alpha }^{\sigma }{{\rho }^{\sigma }}\left( {{u}_{\alpha }}-u_{\alpha }^{\sigma } \right)
	\label{NSequation2}
	\tt{,}
\end{equation}
\begin{eqnarray}
	{{\partial }_{t}}{{E}^{\sigma }} +
	{{\partial }_{\alpha }}\left( {{E}^{\sigma }}u_{\alpha }^{\sigma }+{{p}^{\sigma }}u_{\alpha }^{\sigma }-\kappa _{\alpha }^{\sigma }{{\partial }_{\alpha }}{{T}^{\sigma }}+u_{\beta }^{\sigma }P_{\alpha \beta }^{\sigma }+Y_{\alpha }^{\sigma } \right)  \nonumber \\
	=\frac{1}{2}S_{4}^{\sigma }{{\rho }^{\sigma }}\left[ \left( D+{{I}^{\sigma }} \right)\frac{T-{{T}^{\sigma }}}{{{m}^{\sigma }}}+{{u}^{2}}-{{u}^{\sigma 2}} \right] ,~~~~~~~~~~~~~~
	\label{NSequation3}
\end{eqnarray}
in terms of
\begin{equation}
	P_{\alpha \beta }^{\sigma }=\frac{{{p}^{\sigma }}}{S_{P\alpha \beta }^{\sigma }}\left( \frac{2{{\delta }_{\alpha \beta }}}{D+{{I}^{\sigma }}}{{\partial }_{\chi }}u_{\chi }^{\sigma }-{{\partial }_{\beta }}u_{\alpha }^{\sigma }-{{\partial }_{\alpha }}u_{\beta }^{\sigma } \right)
	\label{P_ab}
	\tt{,}
\end{equation}
\begin{eqnarray}
	U_{\alpha \beta }^{\sigma }={{\delta }_{\alpha \beta }}\frac{S_{4}^{\sigma }-S_{P\alpha \beta }^{\sigma }}{S_{P\alpha \beta }^{\sigma }}{{\rho }^{\sigma }}\frac{{{T}^{\sigma }}-T}{{{m}^{\sigma }}}+{{\rho }^{\sigma }}\left( {{u}_{\alpha }}{{u}_{\beta }}-u_{\alpha }^{\sigma }u_{\beta }^{\sigma } \right) \nonumber \\ 
	+\frac{S_{J\alpha }^{\sigma }}{S_{P\alpha \beta }^{\sigma }}{{\rho }^{\sigma }}\left( u_{\alpha }^{\sigma }u_{\beta }^{\sigma }-{{u}_{\alpha }}u_{\beta }^{\sigma } \right)+\frac{S_{J\beta }^{\sigma }}{S_{P\alpha \beta }^{\sigma }}{{\rho }^{\sigma }}\left( u_{\alpha }^{\sigma }u_{\beta }^{\sigma }-u_{\alpha }^{\sigma }{{u}_{\beta }} \right) \nonumber \\ 
	-{{\delta }_{\alpha \beta }}\frac{S_{4}^{\sigma }}{S_{P\alpha \beta }^{\sigma }}{{\rho }^{\sigma }}\frac{{{u}^{2}}+{{u}^{\sigma 2}}-2u_{\chi }^{\sigma }{{u}_{\chi }}}{D+{{I}^{\sigma }}}
	\tt{,} 
	~~~~~~~~~~~~~~~~~~~~~~~~~~~
\end{eqnarray}
\begin{eqnarray}
	Y_{\alpha }^{\sigma }=-\frac{S_{4}^{\sigma }}{S_{\kappa \alpha }^{\sigma }}\frac{{{\rho }^{\sigma }}u_{\alpha }^{\sigma }}{D+{{I}^{\sigma }}}{{\left( u_{\beta }^{\sigma }-{{u}_{\beta }} \right)}^{2}}+\frac{S_{J\alpha }^{\sigma }-S_{4}^{\sigma }}{S_{\kappa \alpha }^{\sigma }}{{\rho }^{\sigma }}u_{\alpha }^{\sigma }\left( {{u}^{\sigma 2}}-u_{\beta }^{\sigma }{{u}_{\beta }} \right) \nonumber \\ 
	+\frac{{{\rho }^{\sigma }}}{2S_{\kappa \alpha }^{\sigma }}\left( S_{4}^{\sigma }u_{\alpha }^{\sigma }-S_{\kappa \alpha }^{\sigma }{{u}_{\alpha }} \right)\left[ \left( D+{{I}^{\sigma }}+2 \right)\frac{{{T}^{\sigma }}-T}{{{m}^{\sigma }}}+{{u}^{\sigma 2}}-{{u}^{2}} \right] \nonumber ~~~~\\ 
	+\frac{S_{\kappa \alpha }^{\sigma }-S_{J\alpha }^{\sigma }}{S_{\kappa \alpha }^{\sigma }}\frac{{{\rho }^{\sigma }}}{2}\left( {{u}_{\alpha }}-u_{\alpha }^{\sigma } \right)\left[ \left( D+{{I}^{\sigma }}+2 \right)\frac{{{T}^{\sigma }}}{{{m}^{\sigma }}}+{{u}^{\sigma 2}} \right] 
	\label{Y_a}
	\tt{,} 
	~~~~~~~~~~~~~
\end{eqnarray}
where 
$S_{Jx}^{\sigma }=S_{2}^{\sigma }$, 
$S_{Jy}^{\sigma }=S_{3}^{\sigma }$, 
$S_{Pxx}^{\sigma }=S_{5}^{\sigma }$, 
$S_{Pxy}^{\sigma }=S_{6}^{\sigma }$, 
$S_{Pyy}^{\sigma }=S_{7}^{\sigma }$, 
$S_{\kappa x}^{\sigma }=S_{8}^{\sigma }$, 
$S_{\kappa y}^{\sigma }=S_{9}^{\sigma }$. 
The thermal conductivity is 
\begin{equation}
	\kappa _{\alpha }^{\sigma }=\frac{D+{{I}^{\sigma }}+2}{2S_{\kappa \alpha }^{\sigma }}\frac{{{p}^{\sigma }}}{{{m}^{\sigma }}}
	\tt{,}
\end{equation}
which is reduced to
\begin{equation}
	{{\kappa }^{\sigma }}=\frac{D+{{I}^{\sigma }}+2}{2S_{\kappa }^{\sigma }}\frac{{{p}^{\sigma }}}{{{m}^{\sigma }}}
	\tt{,}
\end{equation}
in the case $S_{8}^{\sigma }=S_{9}^{\sigma }=S_{\kappa }^{\sigma }$. 
Moreover, if $S_{5}^{\sigma }=S_{6}^{\sigma }=S_{7}^{\sigma }=S_{\mu }^{\sigma }$, Eq. (\ref{P_ab}) can be rewritten into
\begin{equation}
	P_{\alpha \beta }^{\sigma }={{\mu }^{\sigma }}\left( {{\delta }_{\alpha \beta }}\frac{2}{D}{{\partial }_{\chi }}u_{\chi }^{\sigma }-{{\partial }_{\beta }}u_{\alpha }^{\sigma }-{{\partial }_{\alpha }}u_{\beta }^{\sigma } \right)-{{\delta }_{\alpha \beta }}\mu _{B}^{\sigma }{{\partial }_{\chi }}u_{\chi }^{\sigma }
	\tt{,}
\end{equation}
with the dynamic, kinematic, and bulk viscosities 
\begin{equation}
	{{\mu }^{\sigma }}=\frac{{{p}^{\sigma }}}{S_{\mu }^{\sigma }}
	\tt{,}
\end{equation}
\begin{equation}
	{{\nu }^{\sigma }}=\frac{{{\mu }^{\sigma }}}{{{\rho }^{\sigma }}}=\frac{{{T}^{\sigma }}}{{{m}^{\sigma }}S_{\mu }^{\sigma }}
	\tt{,}
\end{equation}
and
\begin{equation}
	\mu _{B}^{\sigma }={{\mu }^{\sigma }}\left( \frac{2}{D}-\frac{2}{D+{{I}^{\sigma }}} \right)
	\tt{,}
\end{equation}
respectively. 

The specific heat at constant pressure and volume are, respectively,
\begin{equation}
	c_{p}^{\sigma }=\frac{D+{{I}^{\sigma }}+2}{2{{m}^{\sigma }}}
	\tt{,}
\end{equation}
\begin{equation}
	c_{v}^{\sigma }=\frac{D+{{I}^{\sigma }}}{2{{m}^{\sigma }}}
	\tt{,}
\end{equation}
hence the specific-heat ratio is 
\begin{equation}
	{{\gamma }^{\sigma }}=\frac{c_{p}^{\sigma }}{c_{v}^{\sigma }}=\frac{D+{{I}^{\sigma }}+2}{D+{{I}^{\sigma }}}
	\tt{.}
\end{equation}
The number of degrees of freedom is a tunable parameter, which leads to a flexible specific heat ratio. To take account of real-gas effects associated with the gradual excitation of vibrational degrees of freedom with increasing temperature, the extra degrees of freedom are a function of temperature, i.e., $I^{\sigma} = I^{\sigma} (T^{\sigma})$. 
In addition, the Prandtl number is 
\begin{equation}
	{{\Pr }^{\sigma }}=\frac{c_{p}^{\sigma }{{\mu }^{\sigma }}}{{{\kappa }^{\sigma }}}=\frac{S_{\kappa }^{\sigma }}{S_{\mu }^{\sigma }}
	\tt{.}
\end{equation}
Consequently, both the specific-heat ratio and Prandtl number are flexible.

Furthermore, summing Eqs. (\ref{NSequation1}) - (\ref{NSequation3}) over all species $\sigma$ results in the NS equations describing mixing fluids as bellow,
\begin{equation}
	{{\partial }_{t}}\rho +{{\partial }_{\alpha }}\left( \rho {{u}_{\alpha }} \right)=0
	\label{NS_Sum_Detail1}
	\tt{,}
\end{equation}
\begin{equation}
	{{\partial }_{t}}\left( \rho {{u}_{\alpha }} \right)+{{\partial }_{\beta }}\sum\nolimits_{\sigma }{\left( {{\delta }_{\alpha \beta }}{{p}^{\sigma }}+{{\rho }^{\sigma }}u_{\alpha }^{\sigma }u_{\beta }^{\sigma }+P_{\alpha \beta }^{\sigma }+U_{\alpha \beta }^{\sigma } \right)}=0
	\label{NS_Sum_Detail2}
	\tt{,}
\end{equation}
\begin{equation}
	{{\partial }_{t}}E+{{\partial }_{\alpha }}\sum\nolimits_{\sigma }{\left( {{E}^{\sigma }}u_{\alpha }^{\sigma }+{{p}^{\sigma }}u_{\alpha }^{\sigma }-\kappa _{\alpha }^{\sigma }{{\partial }_{\alpha }}{{T}^{\sigma }}+u_{\beta }^{\sigma }P_{\alpha \beta }^{\sigma }+Y_{\alpha }^{\sigma } \right)}=0
	\label{NS_Sum_Detail3}
	\tt{,}
\end{equation}
under the condition of momentum and energy conservation,
\begin{equation}
	\sum\nolimits_{\sigma }{{{\rho }^{\sigma }}\left( {{u}_{\alpha }}-u_{\alpha }^{\sigma } \right)}=0
	\label{Conservation_momentum}
	\tt{,}
\end{equation}
\begin{equation}
	\sum\nolimits_{\sigma }{{{\rho }^{\sigma }}\left( \frac{D+{{I}^{\sigma }}}{2}\frac{{{T}^{\sigma }}-T}{{{m}^{\sigma }}}+\frac{{{u}^{\sigma 2}}-{{u}^{2}}}{2} \right)}=0
	\label{Conservation_energy}
	\tt{.}
\end{equation}

In addition, if individual velocities and temperatures of various chemical species are quite close to each other, i.e., ${u}^{\sigma }_{\alpha} = {u}_{\alpha}$ and ${T}^{\sigma } = T$, Eqs. (\ref{NS_Sum_Detail1}) - (\ref{NS_Sum_Detail3}) are simplified as
\begin{equation}
	{{\partial }_{t}}\rho +{{\partial }_{\alpha }}\left( \rho {{u}_{\alpha }} \right)=0
	\label{NS_Sum_simplified1}
	\tt{,}
\end{equation}
\begin{equation}
	{{\partial }_{t}}\left( \rho {{u}_{\alpha }} \right)+{{\partial }_{\beta }}\left( {{\delta }_{\alpha \beta }}p+\rho {{u}_{\alpha }}{{u}_{\beta }}+{{P}_{\alpha \beta }} \right)=0
	\label{NS_Sum_simplified2}
	\tt{,}
\end{equation}
\begin{equation}
	{{\partial }_{t}}E+{{\partial }_{\alpha }}\left( E{{u}_{\alpha }}+p{{u}_{\alpha }}-{{\kappa }_{\alpha }}{{\partial }_{\alpha }}T+{{u}_{\beta }}{{P}_{\alpha \beta }} \right)=0
	\label{NS_Sum_simplified3}
	\tt{,}
\end{equation}
with
\begin{equation}
	{{P}_{\alpha \beta }}=\mu \left( \frac{2{{\delta }_{\alpha \beta }}}{D}{{\partial }_{\chi }}{{u}_{\chi }}-{{\partial }_{\beta }}{{u}_{\alpha }}-{{\partial }_{\alpha }}{{u}_{\beta }} \right)-{{\delta }_{\alpha \beta }}{{\mu }_{B}}{{\partial }_{\chi }}{{u}_{\chi }}
	\tt{,}
\end{equation}
\begin{equation}
	\mu =\sum\nolimits_{\sigma }{{{\mu }^{\sigma }}}
	\tt{,}
\end{equation}
\begin{equation}
	E=\sum\nolimits_{\sigma }{{{E}^{\sigma }}}
	\tt{,}
\end{equation}
\begin{equation}
	{{\kappa }_{\alpha }}=\sum\nolimits_{\sigma }{\kappa _{\alpha }^{\sigma }}
	\tt{.}
\end{equation}
Clearly, the coefficients of viscosity and thermal conductivity become 
\begin{equation}
	\mu =\frac{p}{S_{\mu }}
	\tt{,}
\end{equation}
\begin{equation}
	\kappa =\frac{D+I+2}{2S_{\kappa }}\frac{p}{m}
	\tt{,}
\end{equation}
where $S_{\mu }^{\sigma } = S_{\mu }$, $S_{\kappa \alpha }^{\sigma } = S_{\kappa }$ and ${m}^{\sigma } = m$. Moreover, in practical systems, the parameters $S_{i }^{\sigma }$ are a function of physical variables, such as the density and temperature \cite{Yang1995JCP,Mieussens2000JCP,Sofonea2001PA}. For simplicity, values of $S_{i }^{\sigma }$ are kept constant in this work.

In fact, the expressions of viscosity $\mu$ and thermal conductivity $\kappa$ depend upon a particular simplified kinetic model that is an approximation to the original complex Boltzmann equation. For example, the ellipsoidal statistical model \cite{Holway1966POF} gives
\begin{equation}
	\mu = \Pr \frac{p}{S}
	\tt{,}
\end{equation}
\begin{equation}
	\kappa = \frac{5}{2} \frac{p}{S}
	\tt{,}
\end{equation}
where $S$ denotes the collision frequency, and the Prandtl number is specified as $\Pr = 1$ in the BGK model \cite{BGK1954}. For power potentials for the interaction between the particles, they take the form
\begin{equation}
	\mu = {{\mu }_{0}}{{\left( \frac{T}{{{T}_{0}}} \right)}^{\omega }}
	\tt{,}
\end{equation}
\begin{equation}
	\kappa =\frac{5}{2}\frac{\mu }{\Pr }
	\tt{,}
\end{equation}
where ${{\mu }_{0}}$ represents the viscosity at a reference temperature $T_{0}$, and $\omega \in \left[ 0.5, 1 \right]$ is a parameter depending upon the interaction potential \cite{Struchtrup2005Book}.

\section{}\label{APPENDIXC}

In a similar way to previous works \cite{Xu2005PRE,Lin2016CNF}, it is easy to demonstrate that the NS equations (\ref{NSequation1})-(\ref{NSequation3}) lead to the following diffusion equations.

(I) Fick' first law
\begin{equation}
	\Phi _{\alpha }^{\sigma }=-{{D}^{\sigma }}{{\partial }_{\alpha }}{{\rho }^{\sigma }}
	\tt{,}
\end{equation}
where $\Phi _{\alpha }^{\sigma } = \rho^{\sigma} \left( u^{\sigma}_{\alpha} - u_{\alpha} \right)$ is the individual diffusion flux of mass in the $\alpha$ direction, and ${{D}^{\sigma }}={T}/{( {{m}^{\sigma }}S_{J\alpha }^{\sigma } )}\;$ is the individual diffusivity.

(II) Fick's second law
\begin{equation}
	{{\partial }_{t}}{{\lambda }^{\sigma }}={{D}^{\sigma }}{{\partial }_{\alpha }}\left( {{\partial }_{\alpha }}{{\lambda }^{\sigma }} \right)
	\tt{,}
\end{equation}
where ${{\lambda }^{\sigma }}={{{\rho }^{\sigma }}}/{\rho }$ represents the mass fraction. 

(III) Stefan-Maxwell diffusion equation
\begin{equation}
	{{\partial }_{\alpha }}{{X}^{j}}=\frac{S_{J\alpha }^{j}}{p}\sum\limits_{k\ne j}^{Ns}{\frac{{{\rho }^{j}}{{\rho }^{k}}}{\rho }\left( u_{\alpha }^{k}-u_{\alpha }^{j} \right)}+\frac{{{\lambda }^{j}}-{{X}^{j}}}{p}{{\partial }_{\alpha }}p
	\label{SMequationI}
	\tt{.}
\end{equation}

Additionally, comparing Eq. (\ref{SMequationI}) with the traditional Stefan-Maxwell diffusion equation
\begin{equation}
	{{\partial }_{\alpha }}{{X}^{j}}=\sum\limits_{k\ne j}^{Ns}{\frac{{{X}^{j}}{{X}^{k}}}{{{D}^{jk}}}\left( u_{\alpha }^{k}-u_{\alpha }^{j} \right)}+\frac{{{\lambda }^{j}}-{{X}^{j}}}{p}{{\partial }_{\alpha }}p
	,
\end{equation}
we get
\begin{equation}
	S_{J\alpha }^{j}=p\frac{\sum\limits_{k\ne j}^{Ns}{\frac{{{X}^{j}}{{X}^{k}}}{{{D}^{jk}}}\left( u_{\alpha }^{j}-u_{\alpha }^{k} \right)}}{\sum\limits_{k\ne j}^{Ns}{\frac{{{\rho }^{j}}{{\rho }^{k}}}{\rho }\left( u_{\alpha }^{j}-u_{\alpha }^{k} \right)}}
	,
\end{equation}
with ${{D}^{jk}}$ the binary diffusivity. With the assumption that the quantity $\left( u_{\alpha }^{j}-u_{\alpha }^{k} \right)$ is of the same order for all $j \ne k$, the above equation is reduced to 
\begin{equation}
	S_{J\alpha }^{j}=p\frac{\sum\limits_{k\ne j}^{Ns}{\frac{{{X}^{j}}{{X}^{k}}}{{{D}^{jk}}}}}{\sum\limits_{k\ne j}^{Ns}{\frac{{{\rho }^{j}}{{\rho }^{k}}}{\rho }}}=p\frac{{{X}^{j}}\sum\limits_{k=1}^{Ns}{\frac{{{X}^{k}}}{{{D}^{jk}}}}}{{{\rho }^{j}}\left( 1-{{\lambda }^{j}} \right)}=\frac{p{{X}^{j}}}{{{\rho }^{j}}{{{\bar{D}}}^{j}}}
	\label{Formula_SJ}
	,
\end{equation}
in terms of
\begin{equation}
	{{\bar{D}}^{j}}={{\left( \sum\limits_{k\ne j}^{Ns}{\frac{{{X}^{k}}}{{{D}^{jk}}}} \right)}^{-1}}\left( 1-{{\lambda }^{j}} \right)
	,
\end{equation}
which is the mixture-averaged diffusion coefficient of component $j$ \cite{Arcidiacono2007PRE}.

Note that substituting Eq. (\ref{Formula_SJ}) into (\ref{NSequation2}) may give a result in contradiction to Eq. (\ref{Conservation_momentum}). A solution to this problem is to set $S_{J\alpha }^{\sigma }={\sum\nolimits_{j}{S_{J\alpha }^{j}}}/{{{N}_{s}}}\;$. Namely, the condition of momentum conservation is satisfied if all individual parameters $S_{J\alpha }^{\sigma }$ are equal to each other \cite{Sofonea2001PA}. Another solution is to modify the right-hand side of discrete Boltzmann equation (\ref{DBEquation}) \cite{Hosseini2018PA}. Similarly, a way to overcome the inconsistency between Eqs. (\ref{NSequation3}) and (\ref{Conservation_energy}) is to set $S^{\sigma}_4$ as the same value, or to add a modified term to Eq. (\ref{DBEquation}). More discussion is out of this paper.


\bibliography{References}

\begin{thebibliography}{98}%
\makeatletter
\providecommand \@ifxundefined [1]{%
 \@ifx{#1\undefined}
}%
\providecommand \@ifnum [1]{%
 \ifnum #1\expandafter \@firstoftwo
 \else \expandafter \@secondoftwo
 \fi
}%
\providecommand \@ifx [1]{%
 \ifx #1\expandafter \@firstoftwo
 \else \expandafter \@secondoftwo
 \fi
}%
\providecommand \natexlab [1]{#1}%
\providecommand \enquote  [1]{``#1''}%
\providecommand \bibnamefont  [1]{#1}%
\providecommand \bibfnamefont [1]{#1}%
\providecommand \citenamefont [1]{#1}%
\providecommand \href@noop [0]{\@secondoftwo}%
\providecommand \href [0]{\begingroup \@sanitize@url \@href}%
\providecommand \@href[1]{\@@startlink{#1}\@@href}%
\providecommand \@@href[1]{\endgroup#1\@@endlink}%
\providecommand \@sanitize@url [0]{\catcode `\\12\catcode `\$12\catcode
  `\&12\catcode `\#12\catcode `\^12\catcode `\_12\catcode `\%12\relax}%
\providecommand \@@startlink[1]{}%
\providecommand \@@endlink[0]{}%
\providecommand \url  [0]{\begingroup\@sanitize@url \@url }%
\providecommand \@url [1]{\endgroup\@href {#1}{\urlprefix }}%
\providecommand \urlprefix  [0]{URL }%
\providecommand \Eprint [0]{\href }%
\providecommand \doibase [0]{https://doi.org/}%
\providecommand \selectlanguage [0]{\@gobble}%
\providecommand \bibinfo  [0]{\@secondoftwo}%
\providecommand \bibfield  [0]{\@secondoftwo}%
\providecommand \translation [1]{[#1]}%
\providecommand \BibitemOpen [0]{}%
\providecommand \bibitemStop [0]{}%
\providecommand \bibitemNoStop [0]{.\EOS\space}%
\providecommand \EOS [0]{\spacefactor3000\relax}%
\providecommand \BibitemShut  [1]{\csname bibitem#1\endcsname}%
\let\auto@bib@innerbib\@empty
\bibitem [{\citenamefont {Cussler}(2000)}]{Cussler2000}%
  \BibitemOpen
  \bibfield  {author} {\bibinfo {author} {\bibfnamefont {E.~L.}\ \bibnamefont
  {Cussler}},\ }\href@noop {} {\emph {\bibinfo {title} {Diffusion: mass
  transfer in fluid systems}}}\ (\bibinfo  {publisher} {Cambridge university
  press},\ \bibinfo {address} {Cambridge},\ \bibinfo {year} {2000})\BibitemShut
  {NoStop}%
\bibitem [{\citenamefont {Law}(2006)}]{Law2006}%
  \BibitemOpen
  \bibfield  {author} {\bibinfo {author} {\bibfnamefont {C.~K.}\ \bibnamefont
  {Law}},\ }\href@noop {} {\emph {\bibinfo {title} {Combustion physics}}}\
  (\bibinfo  {publisher} {Cambridge University Press, Cambridge},\ \bibinfo
  {year} {2006})\BibitemShut {NoStop}%
\bibitem [{\citenamefont {Bertevas}\ \emph {et~al.}(2019)\citenamefont
  {Bertevas}, \citenamefont {Tran-Duc}, \citenamefont {Le-Cao}, \citenamefont
  {Khoo},\ and\ \citenamefont {Phan-Thien}}]{Bertevas2019POF}%
  \BibitemOpen
  \bibfield  {author} {\bibinfo {author} {\bibfnamefont {E.}~\bibnamefont
  {Bertevas}}, \bibinfo {author} {\bibfnamefont {T.}~\bibnamefont {Tran-Duc}},
  \bibinfo {author} {\bibfnamefont {K.}~\bibnamefont {Le-Cao}}, \bibinfo
  {author} {\bibfnamefont {B.~C.}\ \bibnamefont {Khoo}},\ and\ \bibinfo
  {author} {\bibfnamefont {N.}~\bibnamefont {Phan-Thien}},\ }\bibfield  {title}
  {\bibinfo {title} {{A smoothed particle hydrodynamics (SPH) formulation of a
  two-phase mixture model and its application to turbulent sediment
  transport}},\ }\bibfield  {journal} {\bibinfo  {journal} {{Phys. Fluids}}\
  }\textbf {\bibinfo {volume} {{31}}},\ \href
  {https://doi.org/{10.1063/1.5122671}} {{10.1063/1.5122671}} (\bibinfo {year}
  {{2019}})\BibitemShut {NoStop}%
\bibitem [{\citenamefont {Zhao}\ \emph {et~al.}(2019)\citenamefont {Zhao},
  \citenamefont {Moat},\ and\ \citenamefont {Qin}}]{Zhao2019SR}%
  \BibitemOpen
  \bibfield  {author} {\bibinfo {author} {\bibfnamefont {Z.~C.}\ \bibnamefont
  {Zhao}}, \bibinfo {author} {\bibfnamefont {R.~J.}\ \bibnamefont {Moat}},\
  and\ \bibinfo {author} {\bibfnamefont {R.~S.}\ \bibnamefont {Qin}},\
  }\bibfield  {title} {\bibinfo {title} {{Mesoscale modelling of miscible and
  immiscible multicomponent fluids}},\ }\bibfield  {journal} {\bibinfo
  {journal} {{Sci. Rep.}}\ }\textbf {\bibinfo {volume} {{9}}},\ \href
  {https://doi.org/{10.1038/s41598-019-44745-8}} {{10.1038/s41598-019-44745-8}}
  (\bibinfo {year} {{2019}})\BibitemShut {NoStop}%
\bibitem [{\citenamefont {Yang}\ \emph {et~al.}(2020)\citenamefont {Yang},
  \citenamefont {Wang}, \citenamefont {Yang},\ and\ \citenamefont
  {Shu}}]{Yang2020CF}%
  \BibitemOpen
  \bibfield  {author} {\bibinfo {author} {\bibfnamefont {T.}~\bibnamefont
  {Yang}}, \bibinfo {author} {\bibfnamefont {J.}~\bibnamefont {Wang}}, \bibinfo
  {author} {\bibfnamefont {L.}~\bibnamefont {Yang}},\ and\ \bibinfo {author}
  {\bibfnamefont {C.}~\bibnamefont {Shu}},\ }\bibfield  {title} {\bibinfo
  {title} {{Development of multi-component generalized sphere function based
  gas-kinetic flux solver for simulation of compressible viscous reacting
  flows}},\ }\href@noop {} {\bibfield  {journal} {\bibinfo  {journal} {{Comput.
  Fluids}}\ }\textbf {\bibinfo {volume} {{197}}},\ \bibinfo {pages} {104382}
  (\bibinfo {year} {{2020}})}\BibitemShut {NoStop}%
\bibitem [{\citenamefont {Yang}\ and\ \citenamefont
  {Huang}(1995)}]{Yang1995JCP}%
  \BibitemOpen
  \bibfield  {author} {\bibinfo {author} {\bibfnamefont {J.}~\bibnamefont
  {Yang}}\ and\ \bibinfo {author} {\bibfnamefont {J.}~\bibnamefont {Huang}},\
  }\bibfield  {title} {\bibinfo {title} {{Rarefied flow computations using
  nonlinear model Boltzmann equations}},\ }\href@noop {} {\bibfield  {journal}
  {\bibinfo  {journal} {J. Comput. Phys.}\ }\textbf {\bibinfo {volume} {120}},\
  \bibinfo {pages} {323} (\bibinfo {year} {1995})}\BibitemShut {NoStop}%
\bibitem [{\citenamefont {Peng}\ \emph {et~al.}(2016)\citenamefont {Peng},
  \citenamefont {Li}, \citenamefont {Wu},\ and\ \citenamefont
  {Jiang}}]{Peng2016JCP}%
  \BibitemOpen
  \bibfield  {author} {\bibinfo {author} {\bibfnamefont {A.}~\bibnamefont
  {Peng}}, \bibinfo {author} {\bibfnamefont {Z.}~\bibnamefont {Li}}, \bibinfo
  {author} {\bibfnamefont {J.}~\bibnamefont {Wu}},\ and\ \bibinfo {author}
  {\bibfnamefont {X.}~\bibnamefont {Jiang}},\ }\bibfield  {title} {\bibinfo
  {title} {Implicit gas-kinetic unified algorithm based on multi-block docking
  grid for multi-body reentry flows covering all flow regimes},\ }\href@noop {}
  {\bibfield  {journal} {\bibinfo  {journal} {J. Comput. Phys.}\ }\textbf
  {\bibinfo {volume} {327}},\ \bibinfo {pages} {919} (\bibinfo {year}
  {2016})}\BibitemShut {NoStop}%
\bibitem [{\citenamefont {Celiberto}\ \emph {et~al.}(2016)\citenamefont
  {Celiberto}, \citenamefont {Armenise}, \citenamefont {Cacciatore},
  \citenamefont {Capitelli}, \citenamefont {Esposito}, \citenamefont {Gamallo},
  \citenamefont {Janev}, \citenamefont {Lagana}, \citenamefont {Laporta},
  \citenamefont {Laricchiuta}, \citenamefont {Lombardi}, \citenamefont
  {Rutigliano}, \citenamefont {Sayos}, \citenamefont {Tennyson},\ and\
  \citenamefont {Wadehra}}]{Celiberto2016PSST}%
  \BibitemOpen
  \bibfield  {author} {\bibinfo {author} {\bibfnamefont {R.}~\bibnamefont
  {Celiberto}}, \bibinfo {author} {\bibfnamefont {I.}~\bibnamefont {Armenise}},
  \bibinfo {author} {\bibfnamefont {M.}~\bibnamefont {Cacciatore}}, \bibinfo
  {author} {\bibfnamefont {M.}~\bibnamefont {Capitelli}}, \bibinfo {author}
  {\bibfnamefont {F.}~\bibnamefont {Esposito}}, \bibinfo {author}
  {\bibfnamefont {P.}~\bibnamefont {Gamallo}}, \bibinfo {author} {\bibfnamefont
  {R.~K.}\ \bibnamefont {Janev}}, \bibinfo {author} {\bibfnamefont
  {A.}~\bibnamefont {Lagana}}, \bibinfo {author} {\bibfnamefont
  {V.}~\bibnamefont {Laporta}}, \bibinfo {author} {\bibfnamefont
  {A.}~\bibnamefont {Laricchiuta}}, \bibinfo {author} {\bibfnamefont
  {A.}~\bibnamefont {Lombardi}}, \bibinfo {author} {\bibfnamefont
  {M.}~\bibnamefont {Rutigliano}}, \bibinfo {author} {\bibfnamefont
  {R.}~\bibnamefont {Sayos}}, \bibinfo {author} {\bibfnamefont
  {J.}~\bibnamefont {Tennyson}},\ and\ \bibinfo {author} {\bibfnamefont
  {J.~M.}\ \bibnamefont {Wadehra}},\ }\bibfield  {title} {\bibinfo {title}
  {{Atomic and molecular data for spacecraft re-entry plasmas}},\ }\bibfield
  {journal} {\bibinfo  {journal} {{Plasma Sources Sci. Technol.}}\ }\textbf
  {\bibinfo {volume} {{25}}},\ \href
  {https://doi.org/{10.1088/0963-0252/25/3/033004}}
  {{10.1088/0963-0252/25/3/033004}} (\bibinfo {year} {{2016}})\BibitemShut
  {NoStop}%
\bibitem [{\citenamefont {Lin}\ \emph {et~al.}(2014)\citenamefont {Lin},
  \citenamefont {Xu}, \citenamefont {Zhang}, \citenamefont {Li},\ and\
  \citenamefont {Succi}}]{Lin2014PRE}%
  \BibitemOpen
  \bibfield  {author} {\bibinfo {author} {\bibfnamefont {C.}~\bibnamefont
  {Lin}}, \bibinfo {author} {\bibfnamefont {A.}~\bibnamefont {Xu}}, \bibinfo
  {author} {\bibfnamefont {G.}~\bibnamefont {Zhang}}, \bibinfo {author}
  {\bibfnamefont {Y.}~\bibnamefont {Li}},\ and\ \bibinfo {author}
  {\bibfnamefont {S.}~\bibnamefont {Succi}},\ }\bibfield  {title} {\bibinfo
  {title} {{Polar-coordinate lattice Boltzmann modeling of compressible
  flows}},\ }\href@noop {} {\bibfield  {journal} {\bibinfo  {journal} {Phys.
  Rev. E}\ }\textbf {\bibinfo {volume} {89}},\ \bibinfo {pages} {013307}
  (\bibinfo {year} {2014})}\BibitemShut {NoStop}%
\bibitem [{\citenamefont {Ivanov}\ and\ \citenamefont
  {Gimelshein}(1998)}]{Ivanov1998ARFM}%
  \BibitemOpen
  \bibfield  {author} {\bibinfo {author} {\bibfnamefont {M.}~\bibnamefont
  {Ivanov}}\ and\ \bibinfo {author} {\bibfnamefont {S.}~\bibnamefont
  {Gimelshein}},\ }\bibfield  {title} {\bibinfo {title} {Computational
  hypersonic rarefied flows},\ }\href@noop {} {\bibfield  {journal} {\bibinfo
  {journal} {Annu. Rev. Fluid Mech.}\ }\textbf {\bibinfo {volume} {30}},\
  \bibinfo {pages} {469} (\bibinfo {year} {1998})}\BibitemShut {NoStop}%
\bibitem [{\citenamefont {Rapaport}(2004)}]{Rapaport2004Book}%
  \BibitemOpen
  \bibfield  {author} {\bibinfo {author} {\bibfnamefont {D.~C.}\ \bibnamefont
  {Rapaport}},\ }\href@noop {} {\emph {\bibinfo {title} {{The art of molecular
  dynamics simulation}}}}\ (\bibinfo  {publisher} {Cambridge university
  press},\ \bibinfo {address} {Cambridge},\ \bibinfo {year} {2004})\BibitemShut
  {NoStop}%
\bibitem [{\citenamefont {Rykov}(1975)}]{Rykov1975FD}%
  \BibitemOpen
  \bibfield  {author} {\bibinfo {author} {\bibfnamefont {V.}~\bibnamefont
  {Rykov}},\ }\bibfield  {title} {\bibinfo {title} {A model kinetic equation
  for a gas with rotational degrees of freedom},\ }\href@noop {} {\bibfield
  {journal} {\bibinfo  {journal} {Fluid Dyn.}\ }\textbf {\bibinfo {volume}
  {10}},\ \bibinfo {pages} {959} (\bibinfo {year} {1975})}\BibitemShut
  {NoStop}%
\bibitem [{\citenamefont {Liu}\ \emph {et~al.}(2016{\natexlab{a}})\citenamefont
  {Liu}, \citenamefont {Kang}, \citenamefont {Zhang}, \citenamefont {Zhang},
  \citenamefont {Duan},\ and\ \citenamefont {He}}]{Liu2016FP}%
  \BibitemOpen
  \bibfield  {author} {\bibinfo {author} {\bibfnamefont {H.}~\bibnamefont
  {Liu}}, \bibinfo {author} {\bibfnamefont {W.}~\bibnamefont {Kang}}, \bibinfo
  {author} {\bibfnamefont {Q.}~\bibnamefont {Zhang}}, \bibinfo {author}
  {\bibfnamefont {Y.}~\bibnamefont {Zhang}}, \bibinfo {author} {\bibfnamefont
  {H.}~\bibnamefont {Duan}},\ and\ \bibinfo {author} {\bibfnamefont {X.~T.}\
  \bibnamefont {He}},\ }\bibfield  {title} {\bibinfo {title} {Molecular
  dynamics simulations of microscopic structure of ultra strong shock waves in
  dense helium},\ }\href@noop {} {\bibfield  {journal} {\bibinfo  {journal}
  {Front. Phys.}\ }\textbf {\bibinfo {volume} {11}},\ \bibinfo {pages} {115206}
  (\bibinfo {year} {2016}{\natexlab{a}})}\BibitemShut {NoStop}%
\bibitem [{\citenamefont {Liu}\ \emph {et~al.}(2017)\citenamefont {Liu},
  \citenamefont {Zhang}, \citenamefont {Kang}, \citenamefont {Zhang},
  \citenamefont {Duan},\ and\ \citenamefont {He}}]{Liu2017PRE}%
  \BibitemOpen
  \bibfield  {author} {\bibinfo {author} {\bibfnamefont {H.}~\bibnamefont
  {Liu}}, \bibinfo {author} {\bibfnamefont {Y.}~\bibnamefont {Zhang}}, \bibinfo
  {author} {\bibfnamefont {W.}~\bibnamefont {Kang}}, \bibinfo {author}
  {\bibfnamefont {P.}~\bibnamefont {Zhang}}, \bibinfo {author} {\bibfnamefont
  {H.}~\bibnamefont {Duan}},\ and\ \bibinfo {author} {\bibfnamefont {X.~T.}\
  \bibnamefont {He}},\ }\bibfield  {title} {\bibinfo {title} {Molecular
  dynamics simulation of strong shock waves propagating in dense deuterium,
  taking into consideration effects of excited electrons},\ }\href@noop {}
  {\bibfield  {journal} {\bibinfo  {journal} {Phys. Rev. E}\ }\textbf {\bibinfo
  {volume} {95}},\ \bibinfo {pages} {023201} (\bibinfo {year}
  {2017})}\BibitemShut {NoStop}%
\bibitem [{\citenamefont {Murugesan}\ \emph {et~al.}(2019)\citenamefont
  {Murugesan}, \citenamefont {Sirmas},\ and\ \citenamefont
  {Radulescu}}]{Murugesan2019CNF}%
  \BibitemOpen
  \bibfield  {author} {\bibinfo {author} {\bibfnamefont {R.}~\bibnamefont
  {Murugesan}}, \bibinfo {author} {\bibfnamefont {N.}~\bibnamefont {Sirmas}},\
  and\ \bibinfo {author} {\bibfnamefont {M.}~\bibnamefont {Radulescu},
  \bibfnamefont {I}},\ }\bibfield  {title} {\bibinfo {title} {{Non-equilibrium
  effects on thermal ignition using hard sphere molecular dynamics}},\ }\href
  {https://doi.org/{10.1016/j.combustflame.2019.04.037}} {\bibfield  {journal}
  {\bibinfo  {journal} {{Combust. Flame}}\ }\textbf {\bibinfo {volume}
  {{205}}},\ \bibinfo {pages} {457} (\bibinfo {year} {2019})}\BibitemShut
  {NoStop}%
\bibitem [{\citenamefont {Sebastiao}\ \emph {et~al.}(2018)\citenamefont
  {Sebastiao}, \citenamefont {Qiao},\ and\ \citenamefont
  {Alexeenko}}]{Sebastiao2018CNF}%
  \BibitemOpen
  \bibfield  {author} {\bibinfo {author} {\bibfnamefont {I.~B.}\ \bibnamefont
  {Sebastiao}}, \bibinfo {author} {\bibfnamefont {L.}~\bibnamefont {Qiao}},\
  and\ \bibinfo {author} {\bibfnamefont {A.}~\bibnamefont {Alexeenko}},\
  }\bibfield  {title} {\bibinfo {title} {{Direct simulation Monte Carlo
  modeling of H-2-O-2 deflagration waves}},\ }\href
  {https://doi.org/{10.1016/j.combustflame.2018.09.001}} {\bibfield  {journal}
  {\bibinfo  {journal} {{Combust. Flame}}\ }\textbf {\bibinfo {volume} {198}},\
  \bibinfo {pages} {40} (\bibinfo {year} {2018})}\BibitemShut {NoStop}%
\bibitem [{\citenamefont {White}\ \emph {et~al.}(2018)\citenamefont {White},
  \citenamefont {Borg}, \citenamefont {Scanlon}, \citenamefont {Longshaw},
  \citenamefont {John}, \citenamefont {Emerson},\ and\ \citenamefont
  {Reese}}]{White2018CPC}%
  \BibitemOpen
  \bibfield  {author} {\bibinfo {author} {\bibfnamefont {C.}~\bibnamefont
  {White}}, \bibinfo {author} {\bibfnamefont {M.~K.}\ \bibnamefont {Borg}},
  \bibinfo {author} {\bibfnamefont {T.~J.}\ \bibnamefont {Scanlon}}, \bibinfo
  {author} {\bibfnamefont {S.~M.}\ \bibnamefont {Longshaw}}, \bibinfo {author}
  {\bibfnamefont {B.}~\bibnamefont {John}}, \bibinfo {author} {\bibfnamefont
  {D.~R.}\ \bibnamefont {Emerson}},\ and\ \bibinfo {author} {\bibfnamefont
  {J.~M.}\ \bibnamefont {Reese}},\ }\bibfield  {title} {\bibinfo {title}
  {{dsmcFoam plus : An OpenFOAM based direct simulation Monte Carlo solver}},\
  }\href {https://doi.org/{10.1016/j.cpc.2017.09.030}} {\bibfield  {journal}
  {\bibinfo  {journal} {{Comput. Phys. Commun.}}\ }\textbf {\bibinfo {volume}
  {224}},\ \bibinfo {pages} {22} (\bibinfo {year} {2018})}\BibitemShut
  {NoStop}%
\bibitem [{\citenamefont {Gimelshein}\ and\ \citenamefont
  {Wysong}(2019)}]{Gimelshein2019PRF}%
  \BibitemOpen
  \bibfield  {author} {\bibinfo {author} {\bibfnamefont {S.~F.}\ \bibnamefont
  {Gimelshein}}\ and\ \bibinfo {author} {\bibfnamefont {I.~J.}\ \bibnamefont
  {Wysong}},\ }\bibfield  {title} {\bibinfo {title} {{Nonequilibrium air flow
  predictions with a high-fidelity direct simulation Monte Carlo approach}},\
  }\bibfield  {journal} {\bibinfo  {journal} {Phys. Rev. Fluids}\ }\textbf
  {\bibinfo {volume} {4}},\ \href
  {https://doi.org/{10.1103/PhysRevFluids.4.033405}}
  {{10.1103/PhysRevFluids.4.033405}} (\bibinfo {year} {2019})\BibitemShut
  {NoStop}%
\bibitem [{\citenamefont {Mieussens}(2000)}]{Mieussens2000JCP}%
  \BibitemOpen
  \bibfield  {author} {\bibinfo {author} {\bibfnamefont {L.}~\bibnamefont
  {Mieussens}},\ }\bibfield  {title} {\bibinfo {title} {{Discrete-velocity
  models and numerical schemes for the Boltzmann-BGK equation in plane and
  axisymmetric geometries}},\ }\href@noop {} {\bibfield  {journal} {\bibinfo
  {journal} {J. Comput. Phys.}\ }\textbf {\bibinfo {volume} {162}},\ \bibinfo
  {pages} {429} (\bibinfo {year} {2000})}\BibitemShut {NoStop}%
\bibitem [{\citenamefont {Succi}(2001)}]{SucciBook}%
  \BibitemOpen
  \bibfield  {author} {\bibinfo {author} {\bibfnamefont {S.}~\bibnamefont
  {Succi}},\ }\href@noop {} {\emph {\bibinfo {title} {{The Lattice Boltzmann
  Equation for Fluid Dynamics and Beyond}}}}\ (\bibinfo  {publisher} {Oxford
  University Press},\ \bibinfo {address} {New York},\ \bibinfo {year}
  {2001})\BibitemShut {NoStop}%
\bibitem [{\citenamefont {Wu}\ \emph {et~al.}(2013)\citenamefont {Wu},
  \citenamefont {White}, \citenamefont {Scanlon}, \citenamefont {Reese},\ and\
  \citenamefont {Zhang}}]{Wu2013JCP}%
  \BibitemOpen
  \bibfield  {author} {\bibinfo {author} {\bibfnamefont {L.}~\bibnamefont
  {Wu}}, \bibinfo {author} {\bibfnamefont {C.}~\bibnamefont {White}}, \bibinfo
  {author} {\bibfnamefont {T.~J.}\ \bibnamefont {Scanlon}}, \bibinfo {author}
  {\bibfnamefont {J.~M.}\ \bibnamefont {Reese}},\ and\ \bibinfo {author}
  {\bibfnamefont {Y.}~\bibnamefont {Zhang}},\ }\bibfield  {title} {\bibinfo
  {title} {{Deterministic numerical solutions of the Boltzmann equation using
  the fast spectral method}},\ }\href@noop {} {\bibfield  {journal} {\bibinfo
  {journal} {J. Comput. Phys.}\ }\textbf {\bibinfo {volume} {250}},\ \bibinfo
  {pages} {27} (\bibinfo {year} {2013})}\BibitemShut {NoStop}%
\bibitem [{\citenamefont {Liu}\ and\ \citenamefont {Xu}(2017)}]{Liu2017CCP}%
  \BibitemOpen
  \bibfield  {author} {\bibinfo {author} {\bibfnamefont {C.}~\bibnamefont
  {Liu}}\ and\ \bibinfo {author} {\bibfnamefont {K.}~\bibnamefont {Xu}},\
  }\bibfield  {title} {\bibinfo {title} {{A Unified Gas Kinetic Scheme for
  Continuum and Rarefied Flows V: Multiscale and Multi-Component Plasma
  Transport}},\ }\href@noop {} {\bibfield  {journal} {\bibinfo  {journal}
  {Commun. Comput. Phys.}\ }\textbf {\bibinfo {volume} {22}},\ \bibinfo {pages}
  {1175} (\bibinfo {year} {2017})}\BibitemShut {NoStop}%
\bibitem [{\citenamefont {Zhang}\ \emph {et~al.}(2018)\citenamefont {Zhang},
  \citenamefont {Zhu}, \citenamefont {Wang},\ and\ \citenamefont
  {Guo}}]{Zhang2018PRE}%
  \BibitemOpen
  \bibfield  {author} {\bibinfo {author} {\bibfnamefont {Y.}~\bibnamefont
  {Zhang}}, \bibinfo {author} {\bibfnamefont {L.}~\bibnamefont {Zhu}}, \bibinfo
  {author} {\bibfnamefont {R.}~\bibnamefont {Wang}},\ and\ \bibinfo {author}
  {\bibfnamefont {Z.}~\bibnamefont {Guo}},\ }\bibfield  {title} {\bibinfo
  {title} {{Discrete unified gas kinetic scheme for all Knudsen number flows.
  III. Binary gas mixtures of Maxwell molecules}},\ }\href@noop {} {\bibfield
  {journal} {\bibinfo  {journal} {Phys. Rev. E}\ }\textbf {\bibinfo {volume}
  {97}},\ \bibinfo {pages} {053306} (\bibinfo {year} {2018})}\BibitemShut
  {NoStop}%
\bibitem [{\citenamefont {Xu}\ \emph {et~al.}(2012)\citenamefont {Xu},
  \citenamefont {Zhang}, \citenamefont {Gan}, \citenamefont {Chen},\ and\
  \citenamefont {Yu}}]{Xu2012FP}%
  \BibitemOpen
  \bibfield  {author} {\bibinfo {author} {\bibfnamefont {A.}~\bibnamefont
  {Xu}}, \bibinfo {author} {\bibfnamefont {G.}~\bibnamefont {Zhang}}, \bibinfo
  {author} {\bibfnamefont {Y.}~\bibnamefont {Gan}}, \bibinfo {author}
  {\bibfnamefont {F.}~\bibnamefont {Chen}},\ and\ \bibinfo {author}
  {\bibfnamefont {X.}~\bibnamefont {Yu}},\ }\bibfield  {title} {\bibinfo
  {title} {{Lattice Boltzmann modeling and simulation of compressible flows}},\
  }\href@noop {} {\bibfield  {journal} {\bibinfo  {journal} {Front. Phys.}\
  }\textbf {\bibinfo {volume} {7}},\ \bibinfo {pages} {582} (\bibinfo {year}
  {2012})}\BibitemShut {NoStop}%
\bibitem [{\citenamefont {Xu}\ \emph {et~al.}(2018)\citenamefont {Xu},
  \citenamefont {Zhang},\ and\ \citenamefont {Zhang}}]{Xu2018Book}%
  \BibitemOpen
  \bibfield  {author} {\bibinfo {author} {\bibfnamefont {A.}~\bibnamefont
  {Xu}}, \bibinfo {author} {\bibfnamefont {G.}~\bibnamefont {Zhang}},\ and\
  \bibinfo {author} {\bibfnamefont {Y.}~\bibnamefont {Zhang}},\ }\bibfield
  {title} {\bibinfo {title} {{Discrete Boltzmann Modeling of Compressible
  Flows}},\ }in\ \href@noop {} {\emph {\bibinfo {booktitle} {Kinetic
  Theory}}},\ \bibinfo {editor} {edited by\ \bibinfo {editor} {\bibfnamefont
  {G.~Z.}\ \bibnamefont {Kyzas}}\ and\ \bibinfo {editor} {\bibfnamefont
  {A.~C.}\ \bibnamefont {Mitropoulos}}}\ (\bibinfo  {publisher} {IntechOpen},\
  \bibinfo {address} {Rijeka},\ \bibinfo {year} {2018})\ Chap.~\bibinfo
  {chapter} {2}\BibitemShut {NoStop}%
\bibitem [{\citenamefont {Bhatnagar}\ \emph {et~al.}(1954)\citenamefont
  {Bhatnagar}, \citenamefont {Gross},\ and\ \citenamefont {Krook}}]{BGK1954}%
  \BibitemOpen
  \bibfield  {author} {\bibinfo {author} {\bibfnamefont {P.~L.}\ \bibnamefont
  {Bhatnagar}}, \bibinfo {author} {\bibfnamefont {E.~P.}\ \bibnamefont
  {Gross}},\ and\ \bibinfo {author} {\bibfnamefont {M.}~\bibnamefont {Krook}},\
  }\bibfield  {title} {\bibinfo {title} {{A model for collision processes in
  gases. I. Small amplitude processes in charged and neutral one-component
  systems}},\ }\href@noop {} {\bibfield  {journal} {\bibinfo  {journal} {Phys.
  Rev.}\ }\textbf {\bibinfo {volume} {94}},\ \bibinfo {pages} {511} (\bibinfo
  {year} {1954})}\BibitemShut {NoStop}%
\bibitem [{\citenamefont {Holway~Jr}(1966)}]{Holway1966POF}%
  \BibitemOpen
  \bibfield  {author} {\bibinfo {author} {\bibfnamefont {L.~H.}\ \bibnamefont
  {Holway~Jr}},\ }\bibfield  {title} {\bibinfo {title} {New statistical models
  for kinetic theory: methods of construction},\ }\href@noop {} {\bibfield
  {journal} {\bibinfo  {journal} {Phys. Fluids}\ }\textbf {\bibinfo {volume}
  {9}},\ \bibinfo {pages} {1658} (\bibinfo {year} {1966})}\BibitemShut
  {NoStop}%
\bibitem [{\citenamefont {Shakhov}(1968)}]{Shakhov1968FD}%
  \BibitemOpen
  \bibfield  {author} {\bibinfo {author} {\bibfnamefont {E.}~\bibnamefont
  {Shakhov}},\ }\bibfield  {title} {\bibinfo {title} {{Generalization of the
  Krook kinetic relaxation equation}},\ }\href@noop {} {\bibfield  {journal}
  {\bibinfo  {journal} {Fluid Dyn.}\ }\textbf {\bibinfo {volume} {3}},\
  \bibinfo {pages} {95} (\bibinfo {year} {1968})}\BibitemShut {NoStop}%
\bibitem [{\citenamefont {Struchtrup}(2005)}]{Struchtrup2005Book}%
  \BibitemOpen
  \bibfield  {author} {\bibinfo {author} {\bibfnamefont {H.}~\bibnamefont
  {Struchtrup}},\ }\bibfield  {title} {\bibinfo {title} {Macroscopic transport
  equations for rarefied gas flows},\ }in\ \href@noop {} {\emph {\bibinfo
  {booktitle} {Macroscopic Transport Equations for Rarefied Gas Flows}}}\
  (\bibinfo  {publisher} {Springer},\ \bibinfo {year} {2005})\ pp.\ \bibinfo
  {pages} {145--160}\BibitemShut {NoStop}%
\bibitem [{\citenamefont {Andries}\ \emph {et~al.}(2002)\citenamefont
  {Andries}, \citenamefont {Aoki},\ and\ \citenamefont
  {Perthame}}]{Andries2002JSP}%
  \BibitemOpen
  \bibfield  {author} {\bibinfo {author} {\bibfnamefont {P.}~\bibnamefont
  {Andries}}, \bibinfo {author} {\bibfnamefont {K.}~\bibnamefont {Aoki}},\ and\
  \bibinfo {author} {\bibfnamefont {B.}~\bibnamefont {Perthame}},\ }\bibfield
  {title} {\bibinfo {title} {{A consistent BGK-type model for gas mixtures}},\
  }\href@noop {} {\bibfield  {journal} {\bibinfo  {journal} {J. Stat. Phys.}\
  }\textbf {\bibinfo {volume} {106}},\ \bibinfo {pages} {993} (\bibinfo {year}
  {2002})}\BibitemShut {NoStop}%
\bibitem [{\citenamefont {Groppi}\ and\ \citenamefont
  {Spiga}(2004)}]{Groppi2004POF}%
  \BibitemOpen
  \bibfield  {author} {\bibinfo {author} {\bibfnamefont {M.}~\bibnamefont
  {Groppi}}\ and\ \bibinfo {author} {\bibfnamefont {G.}~\bibnamefont {Spiga}},\
  }\bibfield  {title} {\bibinfo {title} {{A Bhatnagar--Gross--Krook-type
  approach for chemically reacting gas mixtures}},\ }\href@noop {} {\bibfield
  {journal} {\bibinfo  {journal} {Phys. Fluids}\ }\textbf {\bibinfo {volume}
  {16}},\ \bibinfo {pages} {4273} (\bibinfo {year} {2004})}\BibitemShut
  {NoStop}%
\bibitem [{\citenamefont {Titarev}(2007)}]{Titarev2007CF}%
  \BibitemOpen
  \bibfield  {author} {\bibinfo {author} {\bibfnamefont {V.~A.}\ \bibnamefont
  {Titarev}},\ }\bibfield  {title} {\bibinfo {title} {Conservative numerical
  methods for model kinetic equations},\ }\href@noop {} {\bibfield  {journal}
  {\bibinfo  {journal} {Comput. Fluids}\ }\textbf {\bibinfo {volume} {36}},\
  \bibinfo {pages} {1446} (\bibinfo {year} {2007})}\BibitemShut {NoStop}%
\bibitem [{\citenamefont {Morinishi}(2006)}]{Morinishi2006CF}%
  \BibitemOpen
  \bibfield  {author} {\bibinfo {author} {\bibfnamefont {K.}~\bibnamefont
  {Morinishi}},\ }\bibfield  {title} {\bibinfo {title} {{Numerical simulation
  for gas microflows using Boltzmann equation}},\ }\href@noop {} {\bibfield
  {journal} {\bibinfo  {journal} {Comput. Fluids}\ }\textbf {\bibinfo {volume}
  {35}},\ \bibinfo {pages} {978} (\bibinfo {year} {2006})}\BibitemShut
  {NoStop}%
\bibitem [{\citenamefont {Kudryavtsev}\ and\ \citenamefont
  {Shershnev}(2013)}]{Kudryavtsev2013JSC}%
  \BibitemOpen
  \bibfield  {author} {\bibinfo {author} {\bibfnamefont {A.~N.}\ \bibnamefont
  {Kudryavtsev}}\ and\ \bibinfo {author} {\bibfnamefont {A.~A.}\ \bibnamefont
  {Shershnev}},\ }\bibfield  {title} {\bibinfo {title} {{A numerical method for
  simulation of microflows by solving directly kinetic equations with WENO
  schemes}},\ }\href@noop {} {\bibfield  {journal} {\bibinfo  {journal} {J.
  Sci. Comput}\ }\textbf {\bibinfo {volume} {57}},\ \bibinfo {pages} {42}
  (\bibinfo {year} {2013})}\BibitemShut {NoStop}%
\bibitem [{\citenamefont {Broadwell}(1964)}]{Broadwell1964POF}%
  \BibitemOpen
  \bibfield  {author} {\bibinfo {author} {\bibfnamefont {J.~E.}\ \bibnamefont
  {Broadwell}},\ }\bibfield  {title} {\bibinfo {title} {Shock structure in a
  simple discrete velocity gas},\ }\href@noop {} {\bibfield  {journal}
  {\bibinfo  {journal} {Phys. Fluids}\ }\textbf {\bibinfo {volume} {7}},\
  \bibinfo {pages} {1243} (\bibinfo {year} {1964})}\BibitemShut {NoStop}%
\bibitem [{\citenamefont {Guo}\ and\ \citenamefont {Shu}(2013)}]{GuoBook2013}%
  \BibitemOpen
  \bibfield  {author} {\bibinfo {author} {\bibfnamefont {Z.}~\bibnamefont
  {Guo}}\ and\ \bibinfo {author} {\bibfnamefont {C.}~\bibnamefont {Shu}},\
  }\href@noop {} {\emph {\bibinfo {title} {{Lattice Boltzmann method and its
  applications in engineering}}}}\ (\bibinfo  {publisher} {World Scientific},\
  \bibinfo {address} {Singapore},\ \bibinfo {year} {2013})\BibitemShut
  {NoStop}%
\bibitem [{\citenamefont {Qian}\ \emph {et~al.}(1992)\citenamefont {Qian},
  \citenamefont {D'Humieres},\ and\ \citenamefont {Lallemand}}]{Qian1992EL}%
  \BibitemOpen
  \bibfield  {author} {\bibinfo {author} {\bibfnamefont {Y.~H.}\ \bibnamefont
  {Qian}}, \bibinfo {author} {\bibfnamefont {D.}~\bibnamefont {D'Humieres}},\
  and\ \bibinfo {author} {\bibfnamefont {P.}~\bibnamefont {Lallemand}},\
  }\bibfield  {title} {\bibinfo {title} {{Lattice BGK models for Navier--Stokes
  equation}},\ }\href@noop {} {\bibfield  {journal} {\bibinfo  {journal}
  {{Europhys. Lett.}}\ }\textbf {\bibinfo {volume} {{17}}},\ \bibinfo {pages}
  {479} (\bibinfo {year} {{1992}})}\BibitemShut {NoStop}%
\bibitem [{\citenamefont {Meng}\ \emph {et~al.}(2011)\citenamefont {Meng},
  \citenamefont {Zhang},\ and\ \citenamefont {Shan}}]{Zhang2011PRE}%
  \BibitemOpen
  \bibfield  {author} {\bibinfo {author} {\bibfnamefont {J.}~\bibnamefont
  {Meng}}, \bibinfo {author} {\bibfnamefont {Y.}~\bibnamefont {Zhang}},\ and\
  \bibinfo {author} {\bibfnamefont {X.}~\bibnamefont {Shan}},\ }\bibfield
  {title} {\bibinfo {title} {{Multiscale lattice Boltzmann approach to modeling
  gas flows}},\ }\href@noop {} {\bibfield  {journal} {\bibinfo  {journal}
  {Phys. Rev. E}\ }\textbf {\bibinfo {volume} {83}},\ \bibinfo {pages} {046701}
  (\bibinfo {year} {2011})}\BibitemShut {NoStop}%
\bibitem [{\citenamefont {Zhang}(2011)}]{Zhang2011MN}%
  \BibitemOpen
  \bibfield  {author} {\bibinfo {author} {\bibfnamefont {J.}~\bibnamefont
  {Zhang}},\ }\bibfield  {title} {\bibinfo {title} {{Lattice Boltzmann method
  for microfluidics: models and applications}},\ }\href
  {https://doi.org/{10.1007/s10404-010-0624-1}} {\bibfield  {journal} {\bibinfo
   {journal} {{Microfluid. Nanofluid.}}\ }\textbf {\bibinfo {volume} {{10}}},\
  \bibinfo {pages} {1} (\bibinfo {year} {{2011}})}\BibitemShut {NoStop}%
\bibitem [{\citenamefont {Meng}\ \emph {et~al.}(2013)\citenamefont {Meng},
  \citenamefont {Zhang}, \citenamefont {Hadjiconstantinou}, \citenamefont
  {Radtke},\ and\ \citenamefont {Shan}}]{Zhang2013}%
  \BibitemOpen
  \bibfield  {author} {\bibinfo {author} {\bibfnamefont {J.}~\bibnamefont
  {Meng}}, \bibinfo {author} {\bibfnamefont {Y.}~\bibnamefont {Zhang}},
  \bibinfo {author} {\bibfnamefont {N.~G.}\ \bibnamefont {Hadjiconstantinou}},
  \bibinfo {author} {\bibfnamefont {G.~A.}\ \bibnamefont {Radtke}},\ and\
  \bibinfo {author} {\bibfnamefont {X.}~\bibnamefont {Shan}},\ }\bibfield
  {title} {\bibinfo {title} {{Lattice ellipsoidal statistical BGK model for
  thermal non-equilibrium flows}},\ }\href
  {https://doi.org/10.1017/jfm.2012.616} {\bibfield  {journal} {\bibinfo
  {journal} {J. Fluid Mech.}\ }\textbf {\bibinfo {volume} {718}},\ \bibinfo
  {pages} {347} (\bibinfo {year} {2013})}\BibitemShut {NoStop}%
\bibitem [{\citenamefont {Qin}(2015)}]{Qin2015}%
  \BibitemOpen
  \bibfield  {author} {\bibinfo {author} {\bibfnamefont {R.}~\bibnamefont
  {Qin}},\ }\bibfield  {title} {\bibinfo {title} {Thermodynamic properties of
  phase separation in shear flow},\ }\href@noop {} {\bibfield  {journal}
  {\bibinfo  {journal} {Comput. Fluids}\ }\textbf {\bibinfo {volume} {117}},\
  \bibinfo {pages} {11} (\bibinfo {year} {2015})}\BibitemShut {NoStop}%
\bibitem [{\citenamefont {Fakhari}\ and\ \citenamefont
  {Lee}(2013)}]{Fakhari2013PRE}%
  \BibitemOpen
  \bibfield  {author} {\bibinfo {author} {\bibfnamefont {A.}~\bibnamefont
  {Fakhari}}\ and\ \bibinfo {author} {\bibfnamefont {T.}~\bibnamefont {Lee}},\
  }\bibfield  {title} {\bibinfo {title} {{Multiple-relaxation-time lattice
  Boltzmann method for immiscible fluids at high Reynolds numbers}},\
  }\href@noop {} {\bibfield  {journal} {\bibinfo  {journal} {Phys. Rev. E}\
  }\textbf {\bibinfo {volume} {87}},\ \bibinfo {pages} {023304} (\bibinfo
  {year} {2013})}\BibitemShut {NoStop}%
\bibitem [{\citenamefont {Liang}\ \emph {et~al.}(2016)\citenamefont {Liang},
  \citenamefont {Shi},\ and\ \citenamefont {Chai}}]{Liang2016PRE}%
  \BibitemOpen
  \bibfield  {author} {\bibinfo {author} {\bibfnamefont {H.}~\bibnamefont
  {Liang}}, \bibinfo {author} {\bibfnamefont {B.~C.}\ \bibnamefont {Shi}},\
  and\ \bibinfo {author} {\bibfnamefont {Z.~H.}\ \bibnamefont {Chai}},\
  }\bibfield  {title} {\bibinfo {title} {{Lattice Boltzmann modeling of
  three-phase incompressible flows}},\ }\href
  {https://doi.org/10.1103/PhysRevE.93.013308} {\bibfield  {journal} {\bibinfo
  {journal} {Phys. Rev. E}\ }\textbf {\bibinfo {volume} {93}},\ \bibinfo
  {pages} {013308} (\bibinfo {year} {2016})}\BibitemShut {NoStop}%
\bibitem [{\citenamefont {Qin}\ \emph {et~al.}(2018)\citenamefont {Qin},
  \citenamefont {Mazloomi~Moqaddam}, \citenamefont {Kang}, \citenamefont
  {Derome},\ and\ \citenamefont {Carmeliet}}]{Qin2018POF}%
  \BibitemOpen
  \bibfield  {author} {\bibinfo {author} {\bibfnamefont {F.}~\bibnamefont
  {Qin}}, \bibinfo {author} {\bibfnamefont {A.}~\bibnamefont
  {Mazloomi~Moqaddam}}, \bibinfo {author} {\bibfnamefont {Q.}~\bibnamefont
  {Kang}}, \bibinfo {author} {\bibfnamefont {D.}~\bibnamefont {Derome}},\ and\
  \bibinfo {author} {\bibfnamefont {J.}~\bibnamefont {Carmeliet}},\ }\bibfield
  {title} {\bibinfo {title} {{Entropic multiple-relaxation-time multirange
  pseudopotential lattice Boltzmann model for two-phase flow}},\ }\href@noop {}
  {\bibfield  {journal} {\bibinfo  {journal} {Phys. Fluids}\ }\textbf {\bibinfo
  {volume} {30}},\ \bibinfo {pages} {032104} (\bibinfo {year}
  {2018})}\BibitemShut {NoStop}%
\bibitem [{\citenamefont {Chen}\ \emph
  {et~al.}(2018{\natexlab{a}})\citenamefont {Chen}, \citenamefont {Shu},
  \citenamefont {Tan}, \citenamefont {Niu},\ and\ \citenamefont
  {Li}}]{Chen2018PRE}%
  \BibitemOpen
  \bibfield  {author} {\bibinfo {author} {\bibfnamefont {Z.}~\bibnamefont
  {Chen}}, \bibinfo {author} {\bibfnamefont {C.}~\bibnamefont {Shu}}, \bibinfo
  {author} {\bibfnamefont {D.}~\bibnamefont {Tan}}, \bibinfo {author}
  {\bibfnamefont {X.~D.}\ \bibnamefont {Niu}},\ and\ \bibinfo {author}
  {\bibfnamefont {Q.~Z.}\ \bibnamefont {Li}},\ }\bibfield  {title} {\bibinfo
  {title} {{Simplified multiphase lattice Boltzmann method for simulating
  multiphase flows with large density ratios and complex interfaces}},\ }\href
  {https://doi.org/{10.1103/PhysRevE.98.063314}} {\bibfield  {journal}
  {\bibinfo  {journal} {Phys. Rev. E}\ }\textbf {\bibinfo {volume} {{98}}},\
  \bibinfo {pages} {063314} (\bibinfo {year}
  {{2018}}{\natexlab{a}})}\BibitemShut {NoStop}%
\bibitem [{\citenamefont {Fei}\ \emph {et~al.}(2019)\citenamefont {Fei},
  \citenamefont {Du}, \citenamefont {Luo}, \citenamefont {Succi}, \citenamefont
  {Lauricella}, \citenamefont {Montessori},\ and\ \citenamefont
  {Wang}}]{Fei2019POF}%
  \BibitemOpen
  \bibfield  {author} {\bibinfo {author} {\bibfnamefont {L.}~\bibnamefont
  {Fei}}, \bibinfo {author} {\bibfnamefont {J.}~\bibnamefont {Du}}, \bibinfo
  {author} {\bibfnamefont {K.~H.}\ \bibnamefont {Luo}}, \bibinfo {author}
  {\bibfnamefont {S.}~\bibnamefont {Succi}}, \bibinfo {author} {\bibfnamefont
  {M.}~\bibnamefont {Lauricella}}, \bibinfo {author} {\bibfnamefont
  {A.}~\bibnamefont {Montessori}},\ and\ \bibinfo {author} {\bibfnamefont
  {Q.}~\bibnamefont {Wang}},\ }\bibfield  {title} {\bibinfo {title} {{Modeling
  realistic multiphase flows using a non-orthogonal multiple-relaxation-time
  lattice Boltzmann method}},\ }\href@noop {} {\bibfield  {journal} {\bibinfo
  {journal} {Phys. Fluids}\ }\textbf {\bibinfo {volume} {31}},\ \bibinfo
  {pages} {042105} (\bibinfo {year} {2019})}\BibitemShut {NoStop}%
\bibitem [{\citenamefont {Wang}\ \emph {et~al.}(2019)\citenamefont {Wang},
  \citenamefont {Tan},\ and\ \citenamefont {Phan-Thien}}]{Wang2019POF}%
  \BibitemOpen
  \bibfield  {author} {\bibinfo {author} {\bibfnamefont {D.}~\bibnamefont
  {Wang}}, \bibinfo {author} {\bibfnamefont {D.}~\bibnamefont {Tan}},\ and\
  \bibinfo {author} {\bibfnamefont {N.}~\bibnamefont {Phan-Thien}},\ }\bibfield
   {title} {\bibinfo {title} {{A lattice Boltzmann method for simulating
  viscoelastic drops}},\ }\bibfield  {journal} {\bibinfo  {journal} {{Phys.
  Fluids}}\ }\textbf {\bibinfo {volume} {{31}}},\ \href
  {https://doi.org/{10.1063/1.5100327}} {{10.1063/1.5100327}} (\bibinfo {year}
  {{2019}})\BibitemShut {NoStop}%
\bibitem [{\citenamefont {Makhija}\ \emph {et~al.}(2012)\citenamefont
  {Makhija}, \citenamefont {Pingen}, \citenamefont {Yang},\ and\ \citenamefont
  {Maute}}]{Makhija2012CAF}%
  \BibitemOpen
  \bibfield  {author} {\bibinfo {author} {\bibfnamefont {D.}~\bibnamefont
  {Makhija}}, \bibinfo {author} {\bibfnamefont {G.}~\bibnamefont {Pingen}},
  \bibinfo {author} {\bibfnamefont {R.}~\bibnamefont {Yang}},\ and\ \bibinfo
  {author} {\bibfnamefont {K.}~\bibnamefont {Maute}},\ }\bibfield  {title}
  {\bibinfo {title} {{Topology optimization of multi-component flows using a
  multi-relaxation time lattice Boltzmann method}},\ }\href@noop {} {\bibfield
  {journal} {\bibinfo  {journal} {Comput. Fluids}\ }\textbf {\bibinfo {volume}
  {67}},\ \bibinfo {pages} {104} (\bibinfo {year} {2012})}\BibitemShut
  {NoStop}%
\bibitem [{\citenamefont {Chai}\ and\ \citenamefont
  {Zhao}(2012)}]{Chai2012AMS}%
  \BibitemOpen
  \bibfield  {author} {\bibinfo {author} {\bibfnamefont {Z.}~\bibnamefont
  {Chai}}\ and\ \bibinfo {author} {\bibfnamefont {T.}~\bibnamefont {Zhao}},\
  }\bibfield  {title} {\bibinfo {title} {{A pseudopotential-based
  multiple-relaxation-time lattice Boltzmann model for
  multicomponent/multiphase flows}},\ }\href@noop {} {\bibfield  {journal}
  {\bibinfo  {journal} {Acta Mech. Sin.}\ }\textbf {\bibinfo {volume} {28}},\
  \bibinfo {pages} {983} (\bibinfo {year} {2012})}\BibitemShut {NoStop}%
\bibitem [{\citenamefont {Liu}\ \emph {et~al.}(2016{\natexlab{b}})\citenamefont
  {Liu}, \citenamefont {Wu}, \citenamefont {Ba}, \citenamefont {Xi},\ and\
  \citenamefont {Zhang}}]{Liu2016JCP}%
  \BibitemOpen
  \bibfield  {author} {\bibinfo {author} {\bibfnamefont {H.}~\bibnamefont
  {Liu}}, \bibinfo {author} {\bibfnamefont {L.}~\bibnamefont {Wu}}, \bibinfo
  {author} {\bibfnamefont {Y.}~\bibnamefont {Ba}}, \bibinfo {author}
  {\bibfnamefont {G.}~\bibnamefont {Xi}},\ and\ \bibinfo {author}
  {\bibfnamefont {Y.}~\bibnamefont {Zhang}},\ }\bibfield  {title} {\bibinfo
  {title} {{A lattice Boltzmann method for axisymmetric multicomponent flows
  with high viscosity ratio}},\ }\href@noop {} {\bibfield  {journal} {\bibinfo
  {journal} {J. Comput. Phys.}\ }\textbf {\bibinfo {volume} {327}},\ \bibinfo
  {pages} {873} (\bibinfo {year} {2016}{\natexlab{b}})}\BibitemShut {NoStop}%
\bibitem [{\citenamefont {Chai}\ \emph {et~al.}(2019)\citenamefont {Chai},
  \citenamefont {Guo}, \citenamefont {Wang},\ and\ \citenamefont
  {Shi}}]{Chai2019PRE}%
  \BibitemOpen
  \bibfield  {author} {\bibinfo {author} {\bibfnamefont {Z.}~\bibnamefont
  {Chai}}, \bibinfo {author} {\bibfnamefont {X.}~\bibnamefont {Guo}}, \bibinfo
  {author} {\bibfnamefont {L.}~\bibnamefont {Wang}},\ and\ \bibinfo {author}
  {\bibfnamefont {B.}~\bibnamefont {Shi}},\ }\bibfield  {title} {\bibinfo
  {title} {{Maxwell--Stefan-theory-based lattice Boltzmann model for diffusion
  in multicomponent mixtures}},\ }\href@noop {} {\bibfield  {journal} {\bibinfo
   {journal} {Phys. Rev. E}\ }\textbf {\bibinfo {volume} {99}},\ \bibinfo
  {pages} {023312} (\bibinfo {year} {2019})}\BibitemShut {NoStop}%
\bibitem [{\citenamefont {Hosseini}\ \emph {et~al.}(2019)\citenamefont
  {Hosseini}, \citenamefont {Darabiha},\ and\ \citenamefont
  {Th{\'e}venin}}]{Hosseini2019IJHMT}%
  \BibitemOpen
  \bibfield  {author} {\bibinfo {author} {\bibfnamefont {S.~A.}\ \bibnamefont
  {Hosseini}}, \bibinfo {author} {\bibfnamefont {N.}~\bibnamefont {Darabiha}},\
  and\ \bibinfo {author} {\bibfnamefont {D.}~\bibnamefont {Th{\'e}venin}},\
  }\bibfield  {title} {\bibinfo {title} {{Lattice Boltzmann advection-diffusion
  model for conjugate heat transfer in heterogeneous media}},\ }\href@noop {}
  {\bibfield  {journal} {\bibinfo  {journal} {Int. J. Heat Mass Transfer}\
  }\textbf {\bibinfo {volume} {132}},\ \bibinfo {pages} {906} (\bibinfo {year}
  {2019})}\BibitemShut {NoStop}%
\bibitem [{\citenamefont {Fei}\ \emph {et~al.}(2018)\citenamefont {Fei},
  \citenamefont {Luo}, \citenamefont {Lin},\ and\ \citenamefont
  {Li}}]{Fei2018IJHMT}%
  \BibitemOpen
  \bibfield  {author} {\bibinfo {author} {\bibfnamefont {L.}~\bibnamefont
  {Fei}}, \bibinfo {author} {\bibfnamefont {K.~H.}\ \bibnamefont {Luo}},
  \bibinfo {author} {\bibfnamefont {C.}~\bibnamefont {Lin}},\ and\ \bibinfo
  {author} {\bibfnamefont {Q.}~\bibnamefont {Li}},\ }\bibfield  {title}
  {\bibinfo {title} {{Modeling incompressible thermal flows using a
  central-moments-based lattice Boltzmann method}},\ }\href@noop {} {\bibfield
  {journal} {\bibinfo  {journal} {Int. J. Heat Mass Transfer}\ }\textbf
  {\bibinfo {volume} {120}},\ \bibinfo {pages} {624} (\bibinfo {year}
  {2018})}\BibitemShut {NoStop}%
\bibitem [{\citenamefont {Chen}\ \emph {et~al.}(2015)\citenamefont {Chen},
  \citenamefont {Kang}, \citenamefont {Tang}, \citenamefont {Robinson},
  \citenamefont {He},\ and\ \citenamefont {Tao}}]{Chen2015IJHMT}%
  \BibitemOpen
  \bibfield  {author} {\bibinfo {author} {\bibfnamefont {L.}~\bibnamefont
  {Chen}}, \bibinfo {author} {\bibfnamefont {Q.}~\bibnamefont {Kang}}, \bibinfo
  {author} {\bibfnamefont {Q.}~\bibnamefont {Tang}}, \bibinfo {author}
  {\bibfnamefont {B.~A.}\ \bibnamefont {Robinson}}, \bibinfo {author}
  {\bibfnamefont {Y.}~\bibnamefont {He}},\ and\ \bibinfo {author}
  {\bibfnamefont {W.}~\bibnamefont {Tao}},\ }\bibfield  {title} {\bibinfo
  {title} {{Pore-scale simulation of multicomponent multiphase reactive
  transport with dissolution and precipitation}},\ }\href@noop {} {\bibfield
  {journal} {\bibinfo  {journal} {Int. J. Heat Mass Transfer}\ }\textbf
  {\bibinfo {volume} {85}},\ \bibinfo {pages} {935} (\bibinfo {year}
  {2015})}\BibitemShut {NoStop}%
\bibitem [{\citenamefont {Feng}\ \emph {et~al.}(2018)\citenamefont {Feng},
  \citenamefont {Tayyab},\ and\ \citenamefont {Boivin}}]{Feng2018CNF}%
  \BibitemOpen
  \bibfield  {author} {\bibinfo {author} {\bibfnamefont {Y.}~\bibnamefont
  {Feng}}, \bibinfo {author} {\bibfnamefont {M.}~\bibnamefont {Tayyab}},\ and\
  \bibinfo {author} {\bibfnamefont {P.}~\bibnamefont {Boivin}},\ }\bibfield
  {title} {\bibinfo {title} {{A Lattice-Boltzmann model for low-Mach reactive
  flows}},\ }\href@noop {} {\bibfield  {journal} {\bibinfo  {journal} {Combust.
  Flame}\ }\textbf {\bibinfo {volume} {196}},\ \bibinfo {pages} {249} (\bibinfo
  {year} {2018})}\BibitemShut {NoStop}%
\bibitem [{\citenamefont {Kang}\ \emph {et~al.}(2014)\citenamefont {Kang},
  \citenamefont {Prasianakis},\ and\ \citenamefont {Mantzaras}}]{Kang2014PRE}%
  \BibitemOpen
  \bibfield  {author} {\bibinfo {author} {\bibfnamefont {J.}~\bibnamefont
  {Kang}}, \bibinfo {author} {\bibfnamefont {N.~I.}\ \bibnamefont
  {Prasianakis}},\ and\ \bibinfo {author} {\bibfnamefont {J.}~\bibnamefont
  {Mantzaras}},\ }\bibfield  {title} {\bibinfo {title} {{Thermal multicomponent
  lattice Boltzmann model for catalytic reactive flows}},\ }\href
  {https://doi.org/10.1103/PhysRevE.89.063310} {\bibfield  {journal} {\bibinfo
  {journal} {Phys. Rev. E}\ }\textbf {\bibinfo {volume} {89}},\ \bibinfo
  {pages} {063310} (\bibinfo {year} {2014})}\BibitemShut {NoStop}%
\bibitem [{\citenamefont {Gan}\ \emph {et~al.}(2015)\citenamefont {Gan},
  \citenamefont {Xu}, \citenamefont {Zhang},\ and\ \citenamefont
  {Succi}}]{Gan2015SM}%
  \BibitemOpen
  \bibfield  {author} {\bibinfo {author} {\bibfnamefont {Y.}~\bibnamefont
  {Gan}}, \bibinfo {author} {\bibfnamefont {A.}~\bibnamefont {Xu}}, \bibinfo
  {author} {\bibfnamefont {G.}~\bibnamefont {Zhang}},\ and\ \bibinfo {author}
  {\bibfnamefont {S.}~\bibnamefont {Succi}},\ }\bibfield  {title} {\bibinfo
  {title} {{Discrete Boltzmann modeling of multiphase flows: hydrodynamic and
  thermodynamic non-equilibrium effects}},\ }\href@noop {} {\bibfield
  {journal} {\bibinfo  {journal} {Soft Matter}\ }\textbf {\bibinfo {volume}
  {11}},\ \bibinfo {pages} {5336} (\bibinfo {year} {2015})}\BibitemShut
  {NoStop}%
\bibitem [{\citenamefont {Zhang}\ \emph
  {et~al.}(2019{\natexlab{a}})\citenamefont {Zhang}, \citenamefont {Xu},
  \citenamefont {Zhang}, \citenamefont {Gan}, \citenamefont {Chen},\ and\
  \citenamefont {Succi}}]{Zhang2019SM}%
  \BibitemOpen
  \bibfield  {author} {\bibinfo {author} {\bibfnamefont {Y.}~\bibnamefont
  {Zhang}}, \bibinfo {author} {\bibfnamefont {A.}~\bibnamefont {Xu}}, \bibinfo
  {author} {\bibfnamefont {G.}~\bibnamefont {Zhang}}, \bibinfo {author}
  {\bibfnamefont {Y.}~\bibnamefont {Gan}}, \bibinfo {author} {\bibfnamefont
  {Z.}~\bibnamefont {Chen}},\ and\ \bibinfo {author} {\bibfnamefont
  {S.}~\bibnamefont {Succi}},\ }\bibfield  {title} {\bibinfo {title} {Entropy
  production in thermal phase separation: a kinetic-theory approach},\
  }\href@noop {} {\bibfield  {journal} {\bibinfo  {journal} {Soft Matter}\
  }\textbf {\bibinfo {volume} {15}},\ \bibinfo {pages} {2245} (\bibinfo {year}
  {2019}{\natexlab{a}})}\BibitemShut {NoStop}%
\bibitem [{\citenamefont {Xu}\ \emph {et~al.}(2015)\citenamefont {Xu},
  \citenamefont {Lin}, \citenamefont {Zhang},\ and\ \citenamefont
  {Li}}]{Lin2015PRE}%
  \BibitemOpen
  \bibfield  {author} {\bibinfo {author} {\bibfnamefont {A.}~\bibnamefont
  {Xu}}, \bibinfo {author} {\bibfnamefont {C.}~\bibnamefont {Lin}}, \bibinfo
  {author} {\bibfnamefont {G.}~\bibnamefont {Zhang}},\ and\ \bibinfo {author}
  {\bibfnamefont {Y.}~\bibnamefont {Li}},\ }\bibfield  {title} {\bibinfo
  {title} {{Multiple-relaxation-time lattice Boltzmann kinetic model for
  combustion}},\ }\href@noop {} {\bibfield  {journal} {\bibinfo  {journal}
  {Phys. Rev. E}\ }\textbf {\bibinfo {volume} {91}},\ \bibinfo {pages} {043306}
  (\bibinfo {year} {2015})}\BibitemShut {NoStop}%
\bibitem [{\citenamefont {Lin}\ \emph {et~al.}(2016)\citenamefont {Lin},
  \citenamefont {Xu}, \citenamefont {Zhang},\ and\ \citenamefont
  {Li}}]{Lin2016CNF}%
  \BibitemOpen
  \bibfield  {author} {\bibinfo {author} {\bibfnamefont {C.}~\bibnamefont
  {Lin}}, \bibinfo {author} {\bibfnamefont {A.}~\bibnamefont {Xu}}, \bibinfo
  {author} {\bibfnamefont {G.}~\bibnamefont {Zhang}},\ and\ \bibinfo {author}
  {\bibfnamefont {Y.}~\bibnamefont {Li}},\ }\bibfield  {title} {\bibinfo
  {title} {{Double-distribution-function discrete Boltzmann model for
  combustion}},\ }\href@noop {} {\bibfield  {journal} {\bibinfo  {journal}
  {Combust. Flame}\ }\textbf {\bibinfo {volume} {164}},\ \bibinfo {pages} {137}
  (\bibinfo {year} {2016})}\BibitemShut {NoStop}%
\bibitem [{\citenamefont {Zhang}\ \emph {et~al.}(2016)\citenamefont {Zhang},
  \citenamefont {Xu}, \citenamefont {Zhang}, \citenamefont {Zhu},\ and\
  \citenamefont {Lin}}]{Zhang2016CNF}%
  \BibitemOpen
  \bibfield  {author} {\bibinfo {author} {\bibfnamefont {Y.}~\bibnamefont
  {Zhang}}, \bibinfo {author} {\bibfnamefont {A.}~\bibnamefont {Xu}}, \bibinfo
  {author} {\bibfnamefont {G.}~\bibnamefont {Zhang}}, \bibinfo {author}
  {\bibfnamefont {C.}~\bibnamefont {Zhu}},\ and\ \bibinfo {author}
  {\bibfnamefont {C.}~\bibnamefont {Lin}},\ }\bibfield  {title} {\bibinfo
  {title} {Kinetic modeling of detonation and effects of negative temperature
  coefficient},\ }\href@noop {} {\bibfield  {journal} {\bibinfo  {journal}
  {Combust. Flame}\ }\textbf {\bibinfo {volume} {173}},\ \bibinfo {pages} {483}
  (\bibinfo {year} {2016})}\BibitemShut {NoStop}%
\bibitem [{\citenamefont {Lin}\ \emph {et~al.}(2017{\natexlab{a}})\citenamefont
  {Lin}, \citenamefont {Luo}, \citenamefont {Fei},\ and\ \citenamefont
  {Succi}}]{Lin2017SR}%
  \BibitemOpen
  \bibfield  {author} {\bibinfo {author} {\bibfnamefont {C.}~\bibnamefont
  {Lin}}, \bibinfo {author} {\bibfnamefont {K.~H.}\ \bibnamefont {Luo}},
  \bibinfo {author} {\bibfnamefont {L.}~\bibnamefont {Fei}},\ and\ \bibinfo
  {author} {\bibfnamefont {S.}~\bibnamefont {Succi}},\ }\bibfield  {title}
  {\bibinfo {title} {{A multi-component discrete Boltzmann model for
  nonequilibrium reactive flows}},\ }\href@noop {} {\bibfield  {journal}
  {\bibinfo  {journal} {Sci. Rep.}\ }\textbf {\bibinfo {volume} {7}},\ \bibinfo
  {pages} {14580} (\bibinfo {year} {2017}{\natexlab{a}})}\BibitemShut {NoStop}%
\bibitem [{\citenamefont {Lin}\ and\ \citenamefont
  {Luo}(2018{\natexlab{a}})}]{Lin2018CNF}%
  \BibitemOpen
  \bibfield  {author} {\bibinfo {author} {\bibfnamefont {C.}~\bibnamefont
  {Lin}}\ and\ \bibinfo {author} {\bibfnamefont {K.~H.}\ \bibnamefont {Luo}},\
  }\bibfield  {title} {\bibinfo {title} {{Mesoscopic simulation of
  nonequilibrium detonation with discrete Boltzmann method}},\ }\href@noop {}
  {\bibfield  {journal} {\bibinfo  {journal} {Combust. Flame}\ }\textbf
  {\bibinfo {volume} {198}},\ \bibinfo {pages} {356} (\bibinfo {year}
  {2018}{\natexlab{a}})}\BibitemShut {NoStop}%
\bibitem [{\citenamefont {Lai}\ \emph {et~al.}(2016)\citenamefont {Lai},
  \citenamefont {Xu}, \citenamefont {Zhang}, \citenamefont {Gan}, \citenamefont
  {Ying},\ and\ \citenamefont {Succi}}]{Lai2016PRE}%
  \BibitemOpen
  \bibfield  {author} {\bibinfo {author} {\bibfnamefont {H.}~\bibnamefont
  {Lai}}, \bibinfo {author} {\bibfnamefont {A.}~\bibnamefont {Xu}}, \bibinfo
  {author} {\bibfnamefont {G.}~\bibnamefont {Zhang}}, \bibinfo {author}
  {\bibfnamefont {Y.}~\bibnamefont {Gan}}, \bibinfo {author} {\bibfnamefont
  {Y.}~\bibnamefont {Ying}},\ and\ \bibinfo {author} {\bibfnamefont
  {S.}~\bibnamefont {Succi}},\ }\bibfield  {title} {\bibinfo {title}
  {{Nonequilibrium thermohydrodynamic effects on the Rayleigh--Taylor
  instability in compressible flows}},\ }\href@noop {} {\bibfield  {journal}
  {\bibinfo  {journal} {Phys. Rev. E}\ }\textbf {\bibinfo {volume} {94}},\
  \bibinfo {pages} {023106} (\bibinfo {year} {2016})}\BibitemShut {NoStop}%
\bibitem [{\citenamefont {Lin}\ \emph {et~al.}(2017{\natexlab{b}})\citenamefont
  {Lin}, \citenamefont {Xu}, \citenamefont {Zhang}, \citenamefont {Luo},\ and\
  \citenamefont {Li}}]{Lin2017PRE}%
  \BibitemOpen
  \bibfield  {author} {\bibinfo {author} {\bibfnamefont {C.}~\bibnamefont
  {Lin}}, \bibinfo {author} {\bibfnamefont {A.}~\bibnamefont {Xu}}, \bibinfo
  {author} {\bibfnamefont {G.}~\bibnamefont {Zhang}}, \bibinfo {author}
  {\bibfnamefont {K.~H.}\ \bibnamefont {Luo}},\ and\ \bibinfo {author}
  {\bibfnamefont {Y.}~\bibnamefont {Li}},\ }\bibfield  {title} {\bibinfo
  {title} {{Discrete Boltzmann modeling of Rayleigh--Taylor instability in
  two-component compressible flows}},\ }\href@noop {} {\bibfield  {journal}
  {\bibinfo  {journal} {Phys. Rev. E}\ }\textbf {\bibinfo {volume} {96}},\
  \bibinfo {pages} {053305} (\bibinfo {year} {2017}{\natexlab{b}})}\BibitemShut
  {NoStop}%
\bibitem [{\citenamefont {Chen}\ \emph
  {et~al.}(2018{\natexlab{b}})\citenamefont {Chen}, \citenamefont {Xu},\ and\
  \citenamefont {Zhang}}]{Chen2018POF}%
  \BibitemOpen
  \bibfield  {author} {\bibinfo {author} {\bibfnamefont {F.}~\bibnamefont
  {Chen}}, \bibinfo {author} {\bibfnamefont {A.}~\bibnamefont {Xu}},\ and\
  \bibinfo {author} {\bibfnamefont {G.}~\bibnamefont {Zhang}},\ }\bibfield
  {title} {\bibinfo {title} {{Collaboration and competition between
  Richtmyer--Meshkov instability and Rayleigh--Taylor instability}},\
  }\href@noop {} {\bibfield  {journal} {\bibinfo  {journal} {Phys. Fluids}\
  }\textbf {\bibinfo {volume} {30}},\ \bibinfo {pages} {102105} (\bibinfo
  {year} {2018}{\natexlab{b}})}\BibitemShut {NoStop}%
\bibitem [{\citenamefont {Lin}\ and\ \citenamefont
  {Luo}(2018{\natexlab{b}})}]{Lin2018CAF}%
  \BibitemOpen
  \bibfield  {author} {\bibinfo {author} {\bibfnamefont {C.}~\bibnamefont
  {Lin}}\ and\ \bibinfo {author} {\bibfnamefont {K.~H.}\ \bibnamefont {Luo}},\
  }\bibfield  {title} {\bibinfo {title} {{MRT discrete Boltzmann method for
  compressible exothermic reactive flows}},\ }\href@noop {} {\bibfield
  {journal} {\bibinfo  {journal} {Comput. Fluids}\ }\textbf {\bibinfo {volume}
  {166}},\ \bibinfo {pages} {176} (\bibinfo {year}
  {2018}{\natexlab{b}})}\BibitemShut {NoStop}%
\bibitem [{\citenamefont {Lin}\ and\ \citenamefont {Luo}(2019)}]{Lin2019PRE}%
  \BibitemOpen
  \bibfield  {author} {\bibinfo {author} {\bibfnamefont {C.}~\bibnamefont
  {Lin}}\ and\ \bibinfo {author} {\bibfnamefont {K.~H.}\ \bibnamefont {Luo}},\
  }\bibfield  {title} {\bibinfo {title} {{Discrete Boltzmann modeling of
  unsteady reactive flows with nonequilibrium effects}},\ }\href@noop {}
  {\bibfield  {journal} {\bibinfo  {journal} {Phys. Rev. E}\ }\textbf {\bibinfo
  {volume} {99}},\ \bibinfo {pages} {012142} (\bibinfo {year}
  {2019})}\BibitemShut {NoStop}%
\bibitem [{\citenamefont {Wang}\ \emph {et~al.}(2007)\citenamefont {Wang},
  \citenamefont {He}, \citenamefont {Zhao}, \citenamefont {Tang},\ and\
  \citenamefont {Tao}}]{Wang2007IJMPC}%
  \BibitemOpen
  \bibfield  {author} {\bibinfo {author} {\bibfnamefont {Y.}~\bibnamefont
  {Wang}}, \bibinfo {author} {\bibfnamefont {Y.~L.}\ \bibnamefont {He}},
  \bibinfo {author} {\bibfnamefont {T.~S.}\ \bibnamefont {Zhao}}, \bibinfo
  {author} {\bibfnamefont {G.~H.}\ \bibnamefont {Tang}},\ and\ \bibinfo
  {author} {\bibfnamefont {W.~Q.}\ \bibnamefont {Tao}},\ }\bibfield  {title}
  {\bibinfo {title} {{Implicit-explicit finite-difference lattice boltzmann
  method for compressible flows}},\ }\href
  {https://doi.org/{10.1142/S0129183107011868}} {\bibfield  {journal} {\bibinfo
   {journal} {Int. J. Mod. Phys. C}\ }\textbf {\bibinfo {volume} {18}},\
  \bibinfo {pages} {1961} (\bibinfo {year} {2007})}\BibitemShut {NoStop}%
\bibitem [{\citenamefont {Gan}\ \emph {et~al.}(2019)\citenamefont {Gan},
  \citenamefont {Xu}, \citenamefont {Zhang}, \citenamefont {Lin}, \citenamefont
  {Lai},\ and\ \citenamefont {Liu}}]{Gan2019FOP}%
  \BibitemOpen
  \bibfield  {author} {\bibinfo {author} {\bibfnamefont {Y.}~\bibnamefont
  {Gan}}, \bibinfo {author} {\bibfnamefont {A.}~\bibnamefont {Xu}}, \bibinfo
  {author} {\bibfnamefont {G.}~\bibnamefont {Zhang}}, \bibinfo {author}
  {\bibfnamefont {C.}~\bibnamefont {Lin}}, \bibinfo {author} {\bibfnamefont
  {H.}~\bibnamefont {Lai}},\ and\ \bibinfo {author} {\bibfnamefont
  {Z.}~\bibnamefont {Liu}},\ }\bibfield  {title} {\bibinfo {title}
  {{Nonequilibrium and morphological characterizations of Kelvin--Helmholtz
  instability in compressible flows}},\ }\href@noop {} {\bibfield  {journal}
  {\bibinfo  {journal} {Front. Phys.}\ }\textbf {\bibinfo {volume} {14}},\
  \bibinfo {pages} {43602} (\bibinfo {year} {2019})}\BibitemShut {NoStop}%
\bibitem [{\citenamefont {Ye}\ \emph {et~al.}(2020)\citenamefont {Ye},
  \citenamefont {Lai}, \citenamefont {Li}, \citenamefont {Gan}, \citenamefont
  {Lin}, \citenamefont {Chen},\ and\ \citenamefont {Xu}}]{Ye2020Entropy}%
  \BibitemOpen
  \bibfield  {author} {\bibinfo {author} {\bibfnamefont {H.}~\bibnamefont
  {Ye}}, \bibinfo {author} {\bibfnamefont {H.}~\bibnamefont {Lai}}, \bibinfo
  {author} {\bibfnamefont {D.}~\bibnamefont {Li}}, \bibinfo {author}
  {\bibfnamefont {Y.}~\bibnamefont {Gan}}, \bibinfo {author} {\bibfnamefont
  {C.}~\bibnamefont {Lin}}, \bibinfo {author} {\bibfnamefont {L.}~\bibnamefont
  {Chen}},\ and\ \bibinfo {author} {\bibfnamefont {A.}~\bibnamefont {Xu}},\
  }\bibfield  {title} {\bibinfo {title} {{Knudsen Number Effects on
  Two-Dimensional Rayleigh--Taylor Instability in Compressible Fluid: Based on
  a Discrete Boltzmann Method}},\ }\href@noop {} {\bibfield  {journal}
  {\bibinfo  {journal} {Entropy}\ }\textbf {\bibinfo {volume} {22}},\ \bibinfo
  {pages} {500} (\bibinfo {year} {2020})}\BibitemShut {NoStop}%
\bibitem [{\citenamefont {Zhang}\ \emph
  {et~al.}(2019{\natexlab{b}})\citenamefont {Zhang}, \citenamefont {Xu},
  \citenamefont {Zhang}, \citenamefont {Chen},\ and\ \citenamefont
  {Wang}}]{Zhang2019CPC}%
  \BibitemOpen
  \bibfield  {author} {\bibinfo {author} {\bibfnamefont {Y.}~\bibnamefont
  {Zhang}}, \bibinfo {author} {\bibfnamefont {A.}~\bibnamefont {Xu}}, \bibinfo
  {author} {\bibfnamefont {G.}~\bibnamefont {Zhang}}, \bibinfo {author}
  {\bibfnamefont {Z.}~\bibnamefont {Chen}},\ and\ \bibinfo {author}
  {\bibfnamefont {P.}~\bibnamefont {Wang}},\ }\bibfield  {title} {\bibinfo
  {title} {{Discrete Boltzmann method for non-equilibrium flows: Based on
  Shakhov model}},\ }\href@noop {} {\bibfield  {journal} {\bibinfo  {journal}
  {Comput. Phys. Commun.}\ }\textbf {\bibinfo {volume} {238}},\ \bibinfo
  {pages} {50} (\bibinfo {year} {2019}{\natexlab{b}})}\BibitemShut {NoStop}%
\bibitem [{\citenamefont {Batchelor}(2000)}]{Batchelor2000}%
  \BibitemOpen
  \bibfield  {author} {\bibinfo {author} {\bibfnamefont {C.~K.}\ \bibnamefont
  {Batchelor}},\ }\href@noop {} {\emph {\bibinfo {title} {An introduction to
  fluid dynamics}}}\ (\bibinfo  {publisher} {Cambridge university press},\
  \bibinfo {address} {Cambridge},\ \bibinfo {year} {2000})\BibitemShut
  {NoStop}%
\bibitem [{\citenamefont {Umeda}(2020)}]{Umeda2020PP}%
  \BibitemOpen
  \bibfield  {author} {\bibinfo {author} {\bibfnamefont {T.}~\bibnamefont
  {Umeda}},\ }\bibfield  {title} {\bibinfo {title} {{Evaluating higher moments
  in the transverse Kelvin--Helmholtz instability by full kinetic
  simulation}},\ }\bibfield  {journal} {\bibinfo  {journal} {{Phys. Plasmas}}\
  }\textbf {\bibinfo {volume} {{27}}},\ \href
  {https://doi.org/{10.1063/1.5139442}} {{10.1063/1.5139442}} (\bibinfo {year}
  {{2020}})\BibitemShut {NoStop}%
\bibitem [{\citenamefont {Hoshoudy}\ and\ \citenamefont
  {Awasthi}(2020)}]{Hoshoudy2020EPJP}%
  \BibitemOpen
  \bibfield  {author} {\bibinfo {author} {\bibfnamefont {G.~A.}\ \bibnamefont
  {Hoshoudy}}\ and\ \bibinfo {author} {\bibfnamefont {M.~K.}\ \bibnamefont
  {Awasthi}},\ }\bibfield  {title} {\bibinfo {title} {{Compressibility effects
  on the Kelvin--Helmholtz and Rayleigh--Taylor instabilities between two
  immiscible fluids flowing through a porous medium}},\ }\bibfield  {journal}
  {\bibinfo  {journal} {{Eur. Phys. J. Plus}}\ }\textbf {\bibinfo {volume}
  {{135}}},\ \href {https://doi.org/{10.1140/epjp/s13360-020-00160-x}}
  {{10.1140/epjp/s13360-020-00160-x}} (\bibinfo {year} {{2020}})\BibitemShut
  {NoStop}%
\bibitem [{\citenamefont {Awasthi}\ \emph {et~al.}(2014)\citenamefont
  {Awasthi}, \citenamefont {Asthana},\ and\ \citenamefont
  {Agrawal}}]{Awasthi2014IJHMT}%
  \BibitemOpen
  \bibfield  {author} {\bibinfo {author} {\bibfnamefont {M.~K.}\ \bibnamefont
  {Awasthi}}, \bibinfo {author} {\bibfnamefont {R.}~\bibnamefont {Asthana}},\
  and\ \bibinfo {author} {\bibfnamefont {G.}~\bibnamefont {Agrawal}},\
  }\bibfield  {title} {\bibinfo {title} {{Viscous correction for the viscous
  potential flow analysis of Kelvin--Helmholtz instability of cylindrical flow
  with heat and mass transfer}},\ }\href@noop {} {\bibfield  {journal}
  {\bibinfo  {journal} {Int. J. Heat Mass Transfer}\ }\textbf {\bibinfo
  {volume} {78}},\ \bibinfo {pages} {251} (\bibinfo {year} {2014})}\BibitemShut
  {NoStop}%
\bibitem [{\citenamefont {Liu}\ \emph {et~al.}(2015{\natexlab{a}})\citenamefont
  {Liu}, \citenamefont {Wang}, \citenamefont {Zang},\ and\ \citenamefont
  {Zhao}}]{Liu2015IJHMT}%
  \BibitemOpen
  \bibfield  {author} {\bibinfo {author} {\bibfnamefont {G.}~\bibnamefont
  {Liu}}, \bibinfo {author} {\bibfnamefont {Y.}~\bibnamefont {Wang}}, \bibinfo
  {author} {\bibfnamefont {G.}~\bibnamefont {Zang}},\ and\ \bibinfo {author}
  {\bibfnamefont {H.}~\bibnamefont {Zhao}},\ }\bibfield  {title} {\bibinfo
  {title} {{Viscous Kelvin--Helmholtz instability analysis of liquid-vapor
  two-phase stratified flow for condensation in horizontal tubes}},\
  }\href@noop {} {\bibfield  {journal} {\bibinfo  {journal} {Int. J. Heat Mass
  Transfer}\ }\textbf {\bibinfo {volume} {84}},\ \bibinfo {pages} {592}
  (\bibinfo {year} {2015}{\natexlab{a}})}\BibitemShut {NoStop}%
\bibitem [{\citenamefont {Wang}\ \emph {et~al.}(2009)\citenamefont {Wang},
  \citenamefont {Ye},\ and\ \citenamefont {Li}}]{Wang2009EPL}%
  \BibitemOpen
  \bibfield  {author} {\bibinfo {author} {\bibfnamefont {L.}~\bibnamefont
  {Wang}}, \bibinfo {author} {\bibfnamefont {W.}~\bibnamefont {Ye}},\ and\
  \bibinfo {author} {\bibfnamefont {Y.}~\bibnamefont {Li}},\ }\bibfield
  {title} {\bibinfo {title} {{Numerical investigation on the ablative
  Kelvin--Helmholtz instability}},\ }\href@noop {} {\bibfield  {journal}
  {\bibinfo  {journal} {Europhys. Lett.}\ }\textbf {\bibinfo {volume} {87}},\
  \bibinfo {pages} {54005} (\bibinfo {year} {2009})}\BibitemShut {NoStop}%
\bibitem [{\citenamefont {Wang}\ \emph {et~al.}(2010)\citenamefont {Wang},
  \citenamefont {Ye}, \citenamefont {Don}, \citenamefont {Sheng}, \citenamefont
  {Li},\ and\ \citenamefont {He}}]{Wang2010POP}%
  \BibitemOpen
  \bibfield  {author} {\bibinfo {author} {\bibfnamefont {L.}~\bibnamefont
  {Wang}}, \bibinfo {author} {\bibfnamefont {W.}~\bibnamefont {Ye}}, \bibinfo
  {author} {\bibfnamefont {W.}~\bibnamefont {Don}}, \bibinfo {author}
  {\bibfnamefont {Z.}~\bibnamefont {Sheng}}, \bibinfo {author} {\bibfnamefont
  {Y.}~\bibnamefont {Li}},\ and\ \bibinfo {author} {\bibfnamefont
  {X.}~\bibnamefont {He}},\ }\bibfield  {title} {\bibinfo {title} {{Formation
  of large-scale structures in ablative Kelvin--Helmholtz instability}},\
  }\href@noop {} {\bibfield  {journal} {\bibinfo  {journal} {Phys. Plasmas}\
  }\textbf {\bibinfo {volume} {17}},\ \bibinfo {pages} {122308} (\bibinfo
  {year} {2010})}\BibitemShut {NoStop}%
\bibitem [{\citenamefont {Lin}\ \emph {et~al.}(2019)\citenamefont {Lin},
  \citenamefont {Luo}, \citenamefont {Gan},\ and\ \citenamefont
  {Liu}}]{Lin2019CTP}%
  \BibitemOpen
  \bibfield  {author} {\bibinfo {author} {\bibfnamefont {C.}~\bibnamefont
  {Lin}}, \bibinfo {author} {\bibfnamefont {K.~H.}\ \bibnamefont {Luo}},
  \bibinfo {author} {\bibfnamefont {Y.}~\bibnamefont {Gan}},\ and\ \bibinfo
  {author} {\bibfnamefont {Z.}~\bibnamefont {Liu}},\ }\bibfield  {title}
  {\bibinfo {title} {{Kinetic Simulation of Nonequilibrium Kelvin--Helmholtz
  Instability}},\ }\href@noop {} {\bibfield  {journal} {\bibinfo  {journal}
  {Commun. Theor. Phys.}\ }\textbf {\bibinfo {volume} {71}},\ \bibinfo {pages}
  {132} (\bibinfo {year} {2019})}\BibitemShut {NoStop}%
\bibitem [{\citenamefont {Zhang}\ and\ \citenamefont
  {Zhuang}(1991)}]{Zhang1991NND}%
  \BibitemOpen
  \bibfield  {author} {\bibinfo {author} {\bibfnamefont {H.}~\bibnamefont
  {Zhang}}\ and\ \bibinfo {author} {\bibfnamefont {F.}~\bibnamefont {Zhuang}},\
  }\bibfield  {title} {\bibinfo {title} {{NND schemes and their applications to
  numerical simulation of two- and three-dimensional flows}},\ }\href@noop {}
  {\bibfield  {journal} {\bibinfo  {journal} {Adv. Appl. Mech.}\ }\textbf
  {\bibinfo {volume} {29}},\ \bibinfo {pages} {193} (\bibinfo {year}
  {1991})}\BibitemShut {NoStop}%
\bibitem [{\citenamefont {Bird}(2002)}]{Bird2002AMR}%
  \BibitemOpen
  \bibfield  {author} {\bibinfo {author} {\bibfnamefont {R.~B.}\ \bibnamefont
  {Bird}},\ }\bibfield  {title} {\bibinfo {title} {Transport phenomena},\
  }\href@noop {} {\bibfield  {journal} {\bibinfo  {journal} {Appl. Mech. Rev.}\
  }\textbf {\bibinfo {volume} {55}},\ \bibinfo {pages} {R1} (\bibinfo {year}
  {2002})}\BibitemShut {NoStop}%
\bibitem [{\citenamefont {Qu}\ \emph {et~al.}(2007)\citenamefont {Qu},
  \citenamefont {Shu},\ and\ \citenamefont {Chew}}]{Qu2007PRE}%
  \BibitemOpen
  \bibfield  {author} {\bibinfo {author} {\bibfnamefont {K.}~\bibnamefont
  {Qu}}, \bibinfo {author} {\bibfnamefont {C.}~\bibnamefont {Shu}},\ and\
  \bibinfo {author} {\bibfnamefont {Y.~T.}\ \bibnamefont {Chew}},\ }\bibfield
  {title} {\bibinfo {title} {{Alternative method to construct equilibrium
  distribution functions in lattice-Boltzmann method simulation of inviscid
  compressible flows at high Mach number}},\ }\href
  {https://doi.org/10.1103/PhysRevE.75.036706} {\bibfield  {journal} {\bibinfo
  {journal} {Phys. Rev. E}\ }\textbf {\bibinfo {volume} {75}},\ \bibinfo
  {pages} {036706} (\bibinfo {year} {2007})}\BibitemShut {NoStop}%
\bibitem [{\citenamefont {Gan}\ \emph {et~al.}(2018)\citenamefont {Gan},
  \citenamefont {Xu}, \citenamefont {Zhang}, \citenamefont {Zhang},\ and\
  \citenamefont {Succi}}]{Gan2018PRE}%
  \BibitemOpen
  \bibfield  {author} {\bibinfo {author} {\bibfnamefont {Y.}~\bibnamefont
  {Gan}}, \bibinfo {author} {\bibfnamefont {A.}~\bibnamefont {Xu}}, \bibinfo
  {author} {\bibfnamefont {G.}~\bibnamefont {Zhang}}, \bibinfo {author}
  {\bibfnamefont {Y.}~\bibnamefont {Zhang}},\ and\ \bibinfo {author}
  {\bibfnamefont {S.}~\bibnamefont {Succi}},\ }\bibfield  {title} {\bibinfo
  {title} {{Discrete Boltzmann trans-scale modeling of high-speed compressible
  flows}},\ }\href {https://doi.org/10.1103/PhysRevE.97.053312} {\bibfield
  {journal} {\bibinfo  {journal} {Phys. Rev. E}\ }\textbf {\bibinfo {volume}
  {97}},\ \bibinfo {pages} {053312} (\bibinfo {year} {2018})}\BibitemShut
  {NoStop}%
\bibitem [{\citenamefont {Li}\ \emph {et~al.}(2007)\citenamefont {Li},
  \citenamefont {He}, \citenamefont {Wang},\ and\ \citenamefont
  {Tao}}]{Li2007PRE}%
  \BibitemOpen
  \bibfield  {author} {\bibinfo {author} {\bibfnamefont {Q.}~\bibnamefont
  {Li}}, \bibinfo {author} {\bibfnamefont {Y.~L.}\ \bibnamefont {He}}, \bibinfo
  {author} {\bibfnamefont {Y.}~\bibnamefont {Wang}},\ and\ \bibinfo {author}
  {\bibfnamefont {W.~Q.}\ \bibnamefont {Tao}},\ }\bibfield  {title} {\bibinfo
  {title} {{Coupled double-distribution-function lattice Boltzmann method for
  the compressible Navier--Stokes equations}},\ }\href@noop {} {\bibfield
  {journal} {\bibinfo  {journal} {Phys. Rev. E}\ }\textbf {\bibinfo {volume}
  {76}},\ \bibinfo {pages} {056705} (\bibinfo {year} {2007})}\BibitemShut
  {NoStop}%
\bibitem [{\citenamefont {Yang}\ \emph {et~al.}(2016)\citenamefont {Yang},
  \citenamefont {Shu},\ and\ \citenamefont {Wang}}]{Yang2016PRE}%
  \BibitemOpen
  \bibfield  {author} {\bibinfo {author} {\bibfnamefont {L.~M.}\ \bibnamefont
  {Yang}}, \bibinfo {author} {\bibfnamefont {C.}~\bibnamefont {Shu}},\ and\
  \bibinfo {author} {\bibfnamefont {Y.}~\bibnamefont {Wang}},\ }\bibfield
  {title} {\bibinfo {title} {Development of a discrete gas-kinetic scheme for
  simulation of two-dimensional viscous incompressible and compressible
  flows},\ }\href@noop {} {\bibfield  {journal} {\bibinfo  {journal} {Phys.
  Rev. E}\ }\textbf {\bibinfo {volume} {93}},\ \bibinfo {pages} {033311}
  (\bibinfo {year} {2016})}\BibitemShut {NoStop}%
\bibitem [{\citenamefont {Guo}\ \emph {et~al.}(2002)\citenamefont {Guo},
  \citenamefont {Zheng},\ and\ \citenamefont {Shi}}]{Guo2002CP}%
  \BibitemOpen
  \bibfield  {author} {\bibinfo {author} {\bibfnamefont {Z.}~\bibnamefont
  {Guo}}, \bibinfo {author} {\bibfnamefont {C.}~\bibnamefont {Zheng}},\ and\
  \bibinfo {author} {\bibfnamefont {B.}~\bibnamefont {Shi}},\ }\bibfield
  {title} {\bibinfo {title} {{Non-equilibrium extrapolation method for velocity
  and pressure boundary conditions in the lattice Boltzmann method}},\ }\href
  {https://doi.org/10.1088/1009-1963/11/4/310} {\bibfield  {journal} {\bibinfo
  {journal} {Chin. Phys.}\ }\textbf {\bibinfo {volume} {11}},\ \bibinfo {pages}
  {366} (\bibinfo {year} {2002})}\BibitemShut {NoStop}%
\bibitem [{\citenamefont {Watari}\ and\ \citenamefont
  {Tsutahara}(2003)}]{Watari2003PRE}%
  \BibitemOpen
  \bibfield  {author} {\bibinfo {author} {\bibfnamefont {M.}~\bibnamefont
  {Watari}}\ and\ \bibinfo {author} {\bibfnamefont {M.}~\bibnamefont
  {Tsutahara}},\ }\bibfield  {title} {\bibinfo {title} {{Two-dimensional
  thermal model of the finite-difference lattice Boltzmann method with high
  spatial isotropy}},\ }\href@noop {} {\bibfield  {journal} {\bibinfo
  {journal} {Phys. Rev. E}\ }\textbf {\bibinfo {volume} {67}},\ \bibinfo
  {pages} {036306} (\bibinfo {year} {2003})}\BibitemShut {NoStop}%
\bibitem [{\citenamefont {Sod}(1978)}]{Sod1978JCP}%
  \BibitemOpen
  \bibfield  {author} {\bibinfo {author} {\bibfnamefont {G.~A.}\ \bibnamefont
  {Sod}},\ }\bibfield  {title} {\bibinfo {title} {A survey of several finite
  difference methods for systems of nonlinear hyperbolic conservation laws},\
  }\href@noop {} {\bibfield  {journal} {\bibinfo  {journal} {J. Comput. Phys.}\
  }\textbf {\bibinfo {volume} {27}},\ \bibinfo {pages} {1} (\bibinfo {year}
  {1978})}\BibitemShut {NoStop}%
\bibitem [{\citenamefont {Wan}\ \emph {et~al.}(2015)\citenamefont {Wan},
  \citenamefont {Malamud}, \citenamefont {Shimony}, \citenamefont {Di~Stefano},
  \citenamefont {Trantham}, \citenamefont {Klein}, \citenamefont {Shvarts},
  \citenamefont {Kuranz},\ and\ \citenamefont {Drake}}]{Wan2015PRL}%
  \BibitemOpen
  \bibfield  {author} {\bibinfo {author} {\bibfnamefont {W.~C.}\ \bibnamefont
  {Wan}}, \bibinfo {author} {\bibfnamefont {G.}~\bibnamefont {Malamud}},
  \bibinfo {author} {\bibfnamefont {A.}~\bibnamefont {Shimony}}, \bibinfo
  {author} {\bibfnamefont {C.~A.}\ \bibnamefont {Di~Stefano}}, \bibinfo
  {author} {\bibfnamefont {M.~R.}\ \bibnamefont {Trantham}}, \bibinfo {author}
  {\bibfnamefont {S.~R.}\ \bibnamefont {Klein}}, \bibinfo {author}
  {\bibfnamefont {D.}~\bibnamefont {Shvarts}}, \bibinfo {author} {\bibfnamefont
  {C.~C.}\ \bibnamefont {Kuranz}},\ and\ \bibinfo {author} {\bibfnamefont
  {R.~P.}\ \bibnamefont {Drake}},\ }\bibfield  {title} {\bibinfo {title}
  {{Observation of Single-Mode, Kelvin--Helmholtz Instability in a Supersonic
  Flow}},\ }\href {https://doi.org/10.1103/PhysRevLett.115.145001} {\bibfield
  {journal} {\bibinfo  {journal} {Phys. Rev. Lett.}\ }\textbf {\bibinfo
  {volume} {115}},\ \bibinfo {pages} {145001} (\bibinfo {year}
  {2015})}\BibitemShut {NoStop}%
\bibitem [{\citenamefont {Liu}\ \emph {et~al.}(2015{\natexlab{b}})\citenamefont
  {Liu}, \citenamefont {Tan},\ and\ \citenamefont {Xu}}]{Liu2015PNASUSA}%
  \BibitemOpen
  \bibfield  {author} {\bibinfo {author} {\bibfnamefont {Y.}~\bibnamefont
  {Liu}}, \bibinfo {author} {\bibfnamefont {P.}~\bibnamefont {Tan}},\ and\
  \bibinfo {author} {\bibfnamefont {L.}~\bibnamefont {Xu}},\ }\bibfield
  {title} {\bibinfo {title} {{Kelvin--Helmholtz instability in an ultrathin air
  film causes drop splashing on smooth surfaces}},\ }\href@noop {} {\bibfield
  {journal} {\bibinfo  {journal} {Proc. Natl. Acad. Sci. U.S.A.}\ }\textbf
  {\bibinfo {volume} {112}},\ \bibinfo {pages} {3280} (\bibinfo {year}
  {2015}{\natexlab{b}})}\BibitemShut {NoStop}%
\bibitem [{\citenamefont {Akula}\ \emph {et~al.}(2017)\citenamefont {Akula},
  \citenamefont {Suchandra}, \citenamefont {Mikhaeil},\ and\ \citenamefont
  {Ranjan}}]{Akula2017JFM}%
  \BibitemOpen
  \bibfield  {author} {\bibinfo {author} {\bibfnamefont {B.}~\bibnamefont
  {Akula}}, \bibinfo {author} {\bibfnamefont {P.}~\bibnamefont {Suchandra}},
  \bibinfo {author} {\bibfnamefont {M.}~\bibnamefont {Mikhaeil}},\ and\
  \bibinfo {author} {\bibfnamefont {D.}~\bibnamefont {Ranjan}},\ }\bibfield
  {title} {\bibinfo {title} {{Dynamics of unstably stratified free shear flows:
  an experimental investigation of coupled Kelvin--Helmholtz and
  Rayleigh--Taylor instability}},\ }\href@noop {} {\bibfield  {journal}
  {\bibinfo  {journal} {J. Fluid Mech.}\ }\textbf {\bibinfo {volume} {816}},\
  \bibinfo {pages} {619} (\bibinfo {year} {2017})}\BibitemShut {NoStop}%
\bibitem [{\citenamefont {Wang}\ \emph {et~al.}(2017)\citenamefont {Wang},
  \citenamefont {Ye}, \citenamefont {He}, \citenamefont {Wu}, \citenamefont
  {Fan}, \citenamefont {Xue}, \citenamefont {Guo}, \citenamefont {Miao},
  \citenamefont {Yuan}, \citenamefont {Dong}, \citenamefont {Jia},
  \citenamefont {Zhang}, \citenamefont {Li}, \citenamefont {Liu}, \citenamefont
  {Wang}, \citenamefont {Ding},\ and\ \citenamefont {Zhang}}]{Wang2017}%
  \BibitemOpen
  \bibfield  {author} {\bibinfo {author} {\bibfnamefont {L.}~\bibnamefont
  {Wang}}, \bibinfo {author} {\bibfnamefont {W.}~\bibnamefont {Ye}}, \bibinfo
  {author} {\bibfnamefont {X.}~\bibnamefont {He}}, \bibinfo {author}
  {\bibfnamefont {J.}~\bibnamefont {Wu}}, \bibinfo {author} {\bibfnamefont
  {Z.}~\bibnamefont {Fan}}, \bibinfo {author} {\bibfnamefont {C.}~\bibnamefont
  {Xue}}, \bibinfo {author} {\bibfnamefont {H.}~\bibnamefont {Guo}}, \bibinfo
  {author} {\bibfnamefont {W.}~\bibnamefont {Miao}}, \bibinfo {author}
  {\bibfnamefont {Y.}~\bibnamefont {Yuan}}, \bibinfo {author} {\bibfnamefont
  {J.}~\bibnamefont {Dong}}, \bibinfo {author} {\bibfnamefont {G.}~\bibnamefont
  {Jia}}, \bibinfo {author} {\bibfnamefont {J.}~\bibnamefont {Zhang}}, \bibinfo
  {author} {\bibfnamefont {Y.}~\bibnamefont {Li}}, \bibinfo {author}
  {\bibfnamefont {J.}~\bibnamefont {Liu}}, \bibinfo {author} {\bibfnamefont
  {M.}~\bibnamefont {Wang}}, \bibinfo {author} {\bibfnamefont {Y.}~\bibnamefont
  {Ding}},\ and\ \bibinfo {author} {\bibfnamefont {W.}~\bibnamefont {Zhang}},\
  }\bibfield  {title} {\bibinfo {title} {Theoretical and simulation research of
  hydrodynamic instabilities in inertial-confinement fusion implosions},\
  }\href@noop {} {\bibfield  {journal} {\bibinfo  {journal} {Sci. China-Phys.
  Mech. Astron.}\ }\textbf {\bibinfo {volume} {60}},\ \bibinfo {pages} {055201}
  (\bibinfo {year} {2017})}\BibitemShut {NoStop}%
\bibitem [{\citenamefont {Watari}(2007)}]{Watari2007PA}%
  \BibitemOpen
  \bibfield  {author} {\bibinfo {author} {\bibfnamefont {M.}~\bibnamefont
  {Watari}},\ }\bibfield  {title} {\bibinfo {title} {{Finite difference lattice
  Boltzmann method with arbitrary specific heat ratio applicable to supersonic
  flow simulations}},\ }\href {https://doi.org/{10.1016/j.physa.2007.03.037}}
  {\bibfield  {journal} {\bibinfo  {journal} {{Physica A}}\ }\textbf {\bibinfo
  {volume} {{382}}},\ \bibinfo {pages} {502} (\bibinfo {year}
  {2007})}\BibitemShut {NoStop}%
\bibitem [{\citenamefont {Sofonea}\ and\ \citenamefont
  {Sekerka}(2001)}]{Sofonea2001PA}%
  \BibitemOpen
  \bibfield  {author} {\bibinfo {author} {\bibfnamefont {V.}~\bibnamefont
  {Sofonea}}\ and\ \bibinfo {author} {\bibfnamefont {R.~F.}\ \bibnamefont
  {Sekerka}},\ }\bibfield  {title} {\bibinfo {title} {{BGK models for diffusion
  in isothermal binary fluid systems}},\ }\href@noop {} {\bibfield  {journal}
  {\bibinfo  {journal} {Physica A}\ }\textbf {\bibinfo {volume} {299}},\
  \bibinfo {pages} {494} (\bibinfo {year} {2001})}\BibitemShut {NoStop}%
\bibitem [{\citenamefont {Xu}(2005)}]{Xu2005PRE}%
  \BibitemOpen
  \bibfield  {author} {\bibinfo {author} {\bibfnamefont {A.}~\bibnamefont
  {Xu}},\ }\bibfield  {title} {\bibinfo {title} {{Finite-difference
  lattice-Boltzmann methods for binary fluids}},\ }\href@noop {} {\bibfield
  {journal} {\bibinfo  {journal} {Phys. Rev. E}\ }\textbf {\bibinfo {volume}
  {71}},\ \bibinfo {pages} {066706} (\bibinfo {year} {2005})}\BibitemShut
  {NoStop}%
\bibitem [{\citenamefont {Arcidiacono}\ \emph {et~al.}(2007)\citenamefont
  {Arcidiacono}, \citenamefont {Karlin}, \citenamefont {Mantzaras},\ and\
  \citenamefont {Frouzakis}}]{Arcidiacono2007PRE}%
  \BibitemOpen
  \bibfield  {author} {\bibinfo {author} {\bibfnamefont {S.}~\bibnamefont
  {Arcidiacono}}, \bibinfo {author} {\bibfnamefont {I.~V.}\ \bibnamefont
  {Karlin}}, \bibinfo {author} {\bibfnamefont {J.}~\bibnamefont {Mantzaras}},\
  and\ \bibinfo {author} {\bibfnamefont {C.~E.}\ \bibnamefont {Frouzakis}},\
  }\bibfield  {title} {\bibinfo {title} {{Lattice Boltzmann model for the
  simulation of multicomponent mixtures}},\ }\href@noop {} {\bibfield
  {journal} {\bibinfo  {journal} {Phys. Rev. E}\ }\textbf {\bibinfo {volume}
  {76}},\ \bibinfo {pages} {046703} (\bibinfo {year} {2007})}\BibitemShut
  {NoStop}%
\bibitem [{\citenamefont {Hosseini}\ \emph {et~al.}(2018)\citenamefont
  {Hosseini}, \citenamefont {Darabiha},\ and\ \citenamefont
  {Th{\'e}venin}}]{Hosseini2018PA}%
  \BibitemOpen
  \bibfield  {author} {\bibinfo {author} {\bibfnamefont {S.~A.}\ \bibnamefont
  {Hosseini}}, \bibinfo {author} {\bibfnamefont {N.}~\bibnamefont {Darabiha}},\
  and\ \bibinfo {author} {\bibfnamefont {D.}~\bibnamefont {Th{\'e}venin}},\
  }\bibfield  {title} {\bibinfo {title} {{Mass-conserving advection-diffusion
  Lattice Boltzmann model for multi-species reacting flows}},\ }\href@noop {}
  {\bibfield  {journal} {\bibinfo  {journal} {Physica A}\ }\textbf {\bibinfo
  {volume} {499}},\ \bibinfo {pages} {40} (\bibinfo {year} {2018})}\BibitemShut
  {NoStop}%
\end{thebibliography}%

\end{document}